\pdfoutput=1
\documentclass[11pt,twoside,a4paper,cmspaper,final,collab]{cms-tdr}
\def\svnVersion{c5f488b-D}\def\svnDate{2021/01/30}\def\cmsCernNoTag{CERN-EP-2020-164}\def\cmsCernDate{\today}\def\cmsMessage{"Published in the Journal of High Energy Physics as \href{http://dx.doi.org/10.1007/JHEP01(2021)148}{\doi{10.1007/JHEP01(2021)148}.}"}
\begin{document}\cmsNoteHeader{HIG-19-006}

\RCS$Revision$
\RCS$HeadURL$
\RCS$Id$

\newcommand{\Zmm}{\ensuremath{\PZ \to \mu\mu}\xspace}
\newcommand{\Zjets}{\ensuremath{\PZ\text{+jets}}\xspace}
\newcommand{\Zjjew}{\ensuremath{\PZ\text{jj-EW}}\xspace}
\newcommand{\Gjets}{\ensuremath{\gamma\text{+jets}}\xspace}
\newcommand{\hmm}{\ensuremath{{\PH\to\mu^{+}\mu^{-}}}\xspace}
\newcommand{\brhmm}{\ensuremath{{\mathcal{B}(\PH \to \mu^{+}\mu^{-})}}\xspace}
\newcommand{\mh}{\ensuremath{m_{\PH}}\xspace}
\newcommand{\mjj}{\ensuremath{m_{\text{jj}}}\xspace}
\newcommand{\detajj}{\ensuremath{\Delta\eta_{\text{jj}}}\xspace}
\newcommand{\dphijj}{\ensuremath{\Delta\phi_{\text{jj}}}\xspace}
\newcommand{\HERWIGSeven}{{\HERWIG7}\xspace}

\newlength\cmsTabSkip\setlength{\cmsTabSkip}{0.5ex}
\providecommand{\cmsTable}[1]{\resizebox{\textwidth}{!}{#1}}

\ifthenelse{\boolean{cms@external}}{\providecommand{\cmsLeft}{upper\xspace}}{\providecommand{\cmsLeft}{left\xspace}}
\ifthenelse{\boolean{cms@external}}{\providecommand{\cmsRight}{lower\xspace}}{\providecommand{\cmsRight}{right\xspace}}

\cmsNoteHeader{HIG-19-006}

\title{Evidence for Higgs boson decay to a pair of muons}

\author*[inst1]{The CMS Collaboration}

\date{\today}

\abstract{Evidence for Higgs boson decay to a pair of muons is presented. This result combines searches in four exclusive categories targeting the production of the Higgs boson via gluon fusion, via vector boson fusion, in association with a vector boson, and in association with a top quark-antiquark pair. The analysis is performed using proton-proton collision data at $\sqrt{s}=13\TeV$, corresponding to an integrated luminosity of $137\fbinv$, recorded by the CMS experiment at the CERN LHC. An excess of events over the background expectation is observed in data with a significance of 3.0 standard deviations, where the expectation for the standard model (SM) Higgs boson with mass of 125.38\GeV is 2.5. The combination of this result with that from data recorded at $\sqrt{s} = 7$ and 8\TeV, corresponding to integrated luminosities of 5.1 and 19.7\fbinv, respectively, increases both the expected and observed significances by 1\%. The measured signal strength, relative to the SM prediction, is $1.19^{+0.40}_{-0.39}\stat^{+0.15}_{-0.14}\syst$. This result constitutes the first evidence for the decay of the Higgs boson to second generation fermions and is the most precise measurement of the Higgs boson coupling to muons reported to date.}

\hypersetup{
pdfauthor={CMS Collaboration},
pdftitle={Evidence for Higgs boson decay to a pair of muons},
pdfsubject={CMS},
pdfkeywords={CMS, physics, Higgs boson, symmetry breaking, muons}
}

\maketitle

\section{Introduction}\label{sec:introduction}

Since the discovery of the Higgs boson at the {CERN} {LHC} in 2012~\cite{Aad:2012tfa,Chatrchyan:2012xdj,Chatrchyan:2013lba}, various measurements of its interactions with standard model (SM) particles have been performed. The interactions of the Higgs boson with the electroweak gauge bosons and charged fermions belonging to the third generation of the SM have been observed, with coupling strengths consistent with the SM predictions~\cite{Sirunyan:2017khh,Sirunyan:2018hoz,Sirunyan:2018kst,Sirunyan:2018ouh,Sirunyan:2018egh,Sirunyan:2017exp,Sirunyan:2018koj,Aaboud:2018pen,Aaboud:2018zhk,Aaboud:2018urx,Aaboud:2018jqu,Aaboud:2017vzb,Aaboud:2018xdt,Aad:2019mbh}. The Yukawa couplings of the Higgs boson to fermions of the first and second generation, however, have yet to be established experimentally. The SM predicts that the strengths of the couplings of the Higgs boson to fermions are proportional to the fermion masses~\cite{PhysRevLett.13.321,HIGGS1964132,PhysRev.145.1156,Weinberg:1967tq}. Consequently, the branching fractions of the Higgs boson to fermions of the first and second generation are expected to be small, and their measurement at hadron colliders is challenging. The expected branching fraction for the decay of the Higgs boson with mass of 125\GeV to a pair of muons is ${\brhmm = 2.18 \times 10^{-4}}$~\cite{deFlorian:2016spz}. The study of \hmm decays is of particular importance since it is the most experimentally sensitive probe of the Higgs boson couplings to second-generation fermions at the {LHC}.

The {CMS} Collaboration performed a search for \hmm decays using a combination of proton-proton ($\Pp\Pp$) collision data collected at centre-of-mass energies of 7, 8, and 13\TeV, corresponding to integrated luminosities of 5.0, 19.7, and 35.9\fbinv, respectively. An observed (expected in absence of \hmm decays) upper limit of 2.9 (2.2) times the SM prediction was set at the 95\% confidence level (\CL) on the product of the Higgs boson production cross section and \brhmm~\cite{Sirunyan:2018hbu}. The corresponding signal strength, relative to the SM expectation, was $\mu = 1.0 \pm 1.0$. The {ATLAS} Collaboration has performed a search for \hmm decays using 13\TeV $\Pp\Pp$ collision data, corresponding to an integrated luminosity of 139\fbinv, resulting in an observed (expected for $\mu=0$) upper limit at 95\% \CL of 2.2 (1.1) times the SM prediction and a signal strength $\mu = 1.2 \pm 0.6$~\cite{Aad:2020xfq}.

This paper reports the first evidence for \hmm decays, obtained using $\Pp\Pp$ collision data collected by the {CMS} experiment at $\sqrt{s}=13\TeV$ and corresponding to a total integrated luminosity of 137\fbinv. The final states considered contain two prompt, isolated, and oppositely charged muons from the Higgs boson decay, with a narrow resonant invariant mass peak around the Higgs boson mass for signal events. The dimuon mass serves as a powerful discriminant against SM background processes. Events are separated into mutually exclusive categories targeting the main production modes of the Higgs boson at hadron colliders, namely gluon fusion ($\Pg\Pg\PH$), vector boson fusion (VBF), associated production with a vector boson ($\PV\PH$, where $\PV=\PW$ or $\PZ$), and associated production with a top quark-antiquark pair ($\ttbar\PH$). Results are given for ${\mh = 125.38 \pm 0.14 \GeV}$, corresponding to the most precise measurement of the Higgs boson mass to date~\cite{Sirunyan:2020xwk}.

The $\Pg\Pg\PH$ and VBF Higgs boson production modes have the largest cross sections at the LHC, and the event categories targeting these production modes are the most sensitive in this measurement. In the $\Pg\Pg\PH$ category, the final state may contain additional hadronic jets produced by initial-state (ISR) or final-state (FSR) radiation. The largest background in this category consists of Drell--Yan (DY) events in which an off-shell $\PZ$ boson decays to a pair of muons. Smaller background contaminations arise from $\ttbar$ and diboson ($\PW\PW$, $\PW\PZ$, $\PZ\PZ$) processes. In the VBF analysis, the final state contains two jets with a large pseudorapidity separation (\detajj) and large dijet invariant mass (\mjj). These characteristic features allow a significant suppression of the DY background, providing an expected sensitivity to \hmm decays that is better than that of the $\Pg\Pg\PH$ category, despite the smaller VBF production cross section. The VH signal events targeted by this analysis contain leptonic decays of the $\PW$ or $\PZ$ boson. This results in a final state with three or more charged leptons, with the dominant background from $\PW\PZ$ and $\PZ\PZ$ events. Finally, the $\ttbar\PH$ category contains the decays of a top quark-antiquark pair. Events in this category are therefore characterized by the presence of one or more $\cPqb$ quark jets, and may contain additional charged leptons. The dominant backgrounds in the $\ttbar\PH$ category are the $\ttbar$ and $\ttbar\PZ$ processes.

This paper is organized as follows: after a brief description of the CMS detector in Section~\ref{sec:cms_detector}, the event reconstruction, simulation, and selection are discussed in Sections~\ref{sec:event_reconstruction},~\ref{sec:event_simulation}, and~\ref{sec:event_selection}, respectively. Sections~\ref{sec:vbf_category},~\ref{sec:ggh_category},~\ref{sec:tth_category}, and~\ref{sec:vh_category} are dedicated to the description of the four exclusive event categories designed to target the VBF, $\Pg\Pg\PH$, $\ttbar\PH$, and $\PV\PH$ production modes, respectively. Finally, Section~\ref{sec:results} describes the main results and their combination which are then summarized in Section~\ref{sec:summary}.

\section{The CMS detector}\label{sec:cms_detector}

The central feature of the {CMS} apparatus is a superconducting solenoid of 6\unit{m} internal diameter, providing a magnetic field of 3.8\unit{T}. Within the solenoid volume are a silicon pixel and strip tracker, a lead tungstate crystal electromagnetic calorimeter (ECAL), and a brass and scintillator hadron calorimeter (HCAL), each composed of a barrel and two endcap sections. Forward calorimeters extend the pseudorapidity coverage provided by the barrel and endcap detectors. Muons are detected in gas-ionization chambers embedded in the steel flux-return yoke outside the solenoid. Events of interest are selected using a two-tiered trigger system~\cite{Khachatryan:2016bia}. The first level (L1) is composed of custom hardware processors, which use information from the calorimeters and muon detectors to select events at a rate of about 100\unit{kHz}. The second level, known as high-level trigger (HLT), is a software-based system which runs a version of the {CMS} full event reconstruction optimized for fast processing, reducing the event rate to about 1\unit{kHz}. A more detailed description of the {CMS} detector, together with a definition of the coordinate system used and the relevant kinematic variables, can be found in Ref.~\cite{Chatrchyan:2008zzk}. 

\section{Event reconstruction}\label{sec:event_reconstruction}

The particle-flow (PF) algorithm~\cite{Sirunyan:2017ulk} aims to reconstruct and identify each individual particle (PF candidate) in an event, with an optimized combination of information from the various elements of the {CMS} detector. The energy of photons is obtained from the ECAL measurement. The energy of electrons is determined from a combination of the electron momentum at the primary interaction vertex as determined by the silicon tracker, the energy of the corresponding ECAL cluster, and the energy sum of all bremsstrahlung photons spatially compatible with originating from the electron track. The energy of charged hadrons is determined from a combination of their momentum measured in the silicon tracker and the matching ECAL and HCAL energy deposits, corrected for the response function of the calorimeters to hadronic showers. The energy of neutral hadrons is obtained from the corresponding corrected ECAL and HCAL energies. Finally, the momentum of muons is obtained from the curvature of the corresponding track reconstructed in the silicon tracker as well as in the muon system. 

For each event, hadronic jets are clustered from these reconstructed particles using the infrared and collinear-safe anti-\kt algorithm~\cite{Cacciari:2008gp,Cacciari:2011ma} with a distance parameter of ${R = 0.4}$. The jet momentum is determined from the vectorial sum of the momenta of all particles in the jet, and is found from simulation to be, on average, within 5 to 10\% of the true transverse momentum over the whole \pt spectrum and detector acceptance. Additional $\Pp\Pp$ interactions within the same or nearby bunch crossings (pileup) can contribute additional tracks and calorimetric energy depositions to the jet momentum. To mitigate this effect, charged particles identified as originating from pileup vertices are discarded and an offset correction is applied to subtract the remaining contributions from neutral particles~\cite{Khachatryan:2016kdb}. Jet energy corrections are derived from simulation to bring, on average, the measured response of jets to that of particle-level jets. In situ measurements of the momentum balance in dijet, \Gjets, \Zjets, and multijet events are used to account for any residual differences in jet energy scale between data and simulation. The jet energy resolution amounts typically to 15--20\% at 30\GeV, 10\% at 100\GeV, and 5\% at 1\TeV~\cite{Khachatryan:2016kdb}. Additional selection criteria are applied to each jet to remove those potentially dominated by anomalous contributions from various subdetector components or reconstruction failures~\cite{Sirunyan:2020foa}. 

The missing transverse momentum vector \ptvecmiss is computed as the negative vector \pt sum of all the PF candidates in an event, and its magnitude is denoted as \ptmiss~\cite{Sirunyan:2019kia}. The \ptvecmiss is modified to account for corrections to the energy scale of the reconstructed jets in the event. Events with anomalously high-\ptmiss can arise from a variety of reconstruction failures, detector malfunctions, or noncollision backgrounds. Such events are rejected by event filters that are designed to identify more than 85--90\% of the spurious high-\ptmiss events with a mistagging rate smaller than 0.1\%~\cite{Sirunyan:2019kia}. 

Primary vertices are reconstructed from charged-particle tracks in the event.
The candidate vertex with the largest value of the sum of the $\pt^2$ of all associated physics objects is taken to be the primary $\Pp\Pp$ interaction vertex. In this sum, the physics objects are the jets, clustered using the jet finding algorithm~\cite{Cacciari:2008gp,Cacciari:2011ma} with the tracks assigned to candidate vertices as inputs, and the associated \ptmiss, taken as the negative vector \pt sum of those jets.

Jets originating from the hadronization of $\cPqb$ quarks are identified using a deep neural network (DeepCSV) that takes as input tracks displaced from the primary interaction vertex, identified secondary vertices, jet kinematic variables, and information related to the presence of soft leptons in the jet~\cite{Sirunyan:2017ezt}. Working points (WPs) that yield either a 1\% (medium WP) or a 10\% (loose WP) probability of misidentifying a light-flavour ($\cPqu\cPqd\cPqs\Pg$) jet with ${\pt > 30 \GeV}$ as a $\cPqb$ quark jet are used. The corresponding average efficiencies for the identification of the hadronization products of a bottom quark as a $\cPqb$ quark jet are about 70 and 85\%, respectively. 

Muon candidates, within the geometrical acceptance of the muon detectors (${\abs{\eta} < 2.4}$), are reconstructed by combining the information from the silicon tracker and the muon chambers~\cite{Sirunyan:2018fpa}. These candidates are required to satisfy a set of quality criteria based on the number of hits measured in the silicon tracker and in the muon system, the properties of the fitted muon track, and the impact parameters of the track with respect to the primary vertex of the event. Electron candidates within ${\abs{\eta} < 2.5}$ are reconstructed using an algorithm that associates fitted tracks in the silicon tracker with electromagnetic energy clusters in the ECAL~\cite{Khachatryan:2015hwa}. To reduce the misidentification rate, these candidates are required to satisfy identification criteria based on the shower shape of the energy deposit, the matching of the electron track to the ECAL energy cluster, the relative amount of energy deposited in the HCAL detector, and the consistency of the electron track with the primary vertex. Because of nonoptimal reconstruction performance, electron candidates in the transition region between the ECAL barrel and endcaps, ${1.44 < \abs{\eta} < 1.57}$, are discarded. Electron candidates identified as coming from photon conversions in the detector are also rejected. Identified muons and electrons are required to be isolated from hadronic activity in the event. The isolation sum is defined by summing the \pt of all the PF candidates in a cone of radius ${R = \sqrt{\smash[b]{(\Delta\eta)^{2}+(\Delta\phi)^{2}}} = 0.4~(0.3)}$ around the muon (electron) track, where $\phi$ is the azimuthal angle in radians, and is corrected for the contribution of neutral particles from pileup interactions ~\cite{Sirunyan:2018fpa,Khachatryan:2015hwa}. 

\section{Event simulation}\label{sec:event_simulation}

Simulated events from Monte Carlo (MC) event generators for the signal and dominant background processes are used to optimize the analysis strategy, evaluate the acceptance, and assess systematic uncertainties. The generated events are processed through a detailed simulation of the {CMS} detector based on \GEANTfour~\cite{AGOSTINELLI2003250} and are reconstructed with the same algorithms that are used for data. The effect of pileup interactions is modelled by overlaying simulated inelastic $\Pp\Pp$ collisions on the hard-scattering event. The MC simulated events are weighted to reproduce the distribution of the number of interactions per bunch crossing observed in data.

The $\Pg\Pg\PH$ signal process is simulated at next-to-leading order (NLO) accuracy in perturbative quantum chromodynamics (QCD), using both the \MGvATNLO~v2.4.2~\cite{Alwall:2014hca} and \POWHEG~v2.0~\cite{Nason:2004rx,Frixione:2007vw,Alioli:2010xd,Bagnaschi:2011tu} MC event generators. In the \MGvATNLO event generation, up to two additional partons in the final state are included in the matrix element (ME) calculation. The \pt distribution of the Higgs boson produced via gluon fusion is then reweighted to match the \POWHEG~\textsc{nnlops} predictions~\cite{Hamilton:2013fea,Hamilton:2015nsa}. The VBF, ${\Pq\Pq\to\PV\PH}$, and $\ttbar\PH$ processes are simulated with \POWHEG~v2.0~\cite{Nason:2009ai,Luisoni:2013kna,Hartanto:2015uka} at NLO precision in QCD. In addition to the four main production modes, the contributions due to Higgs boson production in association with a pair of $\cPqb$ quarks ($\bbbar\PH$), with a $\PZ$ boson through gluon fusion (${\Pg\Pg \to \PZ\PH}$), and with a single top quark and either a $\PW$ boson ($\cPqt\PH\PW$) or a quark ($\cPqt\PH\Pq$) are also considered. The $\bbbar\PH$ process is simulated at NLO precision in QCD with \POWHEG, while $\cPqt\PH\Pq$ and $\cPqt\PH\PW$ (${\Pg\Pg \to \PZ\PH}$) events are generated at leading order (LO) with the \MGvATNLO (\POWHEG) generator. Simulated signal events are generated, for each production mode, at $\mh$ values of 120, 125, and 130\GeV in order to interpolate signal models for every $\mh$ hypothesis in the $125\pm5\GeV$ range, following the procedure detailed in Section~\ref{sec:results}.

Expected signal yields are normalized to the production cross sections and \brhmm values taken from the recommendations of Ref.~\cite{deFlorian:2016spz}. The $\Pg\Pg\PH$ production cross section is computed at next-to-next-to-NLO ($\mathrm{N}^{3}\mathrm{LO}$) precision in QCD, and at NLO in electroweak (EW) theory~\cite{Anastasiou:2016cez}. The cross section of Higgs boson production in the VBF~\cite{Cacciari:2015jma} and ${\Pq\Pq\to\PV\PH}$~\cite{Brein:2003wg} modes is calculated at next-to-NLO (NNLO) in QCD, including NLO EW corrections, while the $\ttbar\PH$ cross section is computed at NLO in QCD and EW theory~\cite{Dawson:2003zu,Frixione:2014qaa}. The $\bbbar\PH$, $\cPqt\PH\Pq$, and $\cPqt\PH\PW$ cross sections are computed at NLO in QCD without including higher-order EW corrections~\cite{deFlorian:2016spz,Demartin:2015uha,Demartin:2016axk}. The \hmm partial width is computed with \textsc{hdecay}~\cite{Djouadi:1997yw,Spira:1997dg} at NLO in QCD and EW theory.

The DY process, which is the main background in the $\Pg\Pg\PH$ and VBF categories, is simulated at NLO in QCD using the \MGvATNLO generator with up to two partons in the final state at the ME level. The corresponding cross section is calculated with \FEWZ~v3.1b2~\cite{Li:2012wna} at NNLO in QCD and NLO accuracy in EW theory. The EW production of a $\PZ$ boson in association with two jets (\Zjjew) is an important background in the VBF category. This process is simulated at LO using the \MGvATNLO~v2.6.5 generator. The $\PW\PZ$, ${\Pq\cPaq\to\PZ\PZ}$, and $\PW\PW$ processes, which constitute the main backgrounds in the $\PV\PH$ category, are simulated at NLO in QCD using either the \POWHEG or \MGvATNLO generators. Their production cross sections are corrected with the NNLO/NLO $K$ factors taken from Refs.~\cite{Grazzini:2017ckn},~\cite{Grazzini:2015hta}, and~\cite{Gehrmann:2014fva}. The gluon-initiated loop-induced ZZ process (${\Pg\Pg \to \PZ\PZ}$) is simulated with the \MCFM~v7.0 generator~\cite{Campbell:2011bn} at LO and the corresponding production cross section is corrected to match higher-order QCD predictions, following the strategy detailed in Ref.~\cite{Sirunyan:2017exp}. Minor contributions from triboson processes ($\PW\PW\PW$, $\PW\PW\PZ$, $\PW\PZ\PZ$, and $\PZ\PZ\PZ$) are also taken into account and are simulated at NLO in QCD using the \MGvATNLO generator. The main backgrounds in the $\ttbar\PH$ category involve the production of top quarks. The $\ttbar$ background is simulated with NLO precision in QCD using the \POWHEG generator, and its cross section is obtained from the \textsc{top++}~v2.0~\cite{Czakon:2011xx} prediction that includes NNLO corrections in QCD and resummation of NNLL soft gluon terms. The single top quark processes are simulated at NLO in QCD via either \POWHEG or \MGvATNLO and their cross sections are computed, at the same order of precision, using \textsc{hathor}~\cite{Kant:2014oha}. Finally, contributions from the $\ttbar\PZ$, $\ttbar\PW$, $\ttbar\PW\PW$, ${\cPqt\cPaqt\cPqt\cPaqt}$, and ${\cPqt\PZ\cPq}$ processes are also considered and are simulated using the \MGvATNLO generator at NLO precision in QCD. For the simulated samples corresponding to the 2016 (2017--2018) data-taking periods, the NNPDF~v3.0~(v3.1) NLO (NNLO) parton distribution functions (PDFs) are used~\cite{Ball:2014uwa,Ball:2017nwa}. For processes simulated at NLO (LO) in QCD with the \MGvATNLO generator, events from the ME characterized by different parton multiplicities are merged via the FxFx (MLM) prescription~\cite{Alwall:2007fs,Frederix:2012ps}.

The simulated events at the ME level for both signal and background processes, except for \Zjjew production, are interfaced with \PYTHIA~v8.2.2 or higher~\cite{Sjostrand:2014zea} to simulate the shower and hadronization of partons in the initial and final states, along with the underlying event description. The CUETP8M1 tune~\cite{Khachatryan:2015pea} is used for simulated samples corresponding to the 2016 data-taking period, while the CP5 tune~\cite{Sirunyan:2019dfx} is used for the 2017 and 2018 simulated data. Simulated VBF signal events are interfaced with \PYTHIA but, rather than the standard \pt-ordered parton shower, the dipole shower is chosen to model the ISR and FSR~\cite{Cabouat:2017rzi}. The dipole shower correctly takes into account the structure of the colour flow between incoming and outgoing quark lines, and its predictions are found to be in good agreement with NNLO QCD calculations, as reported in Ref.~\cite{Jager:2020hkz}. In contrast, the parton shower (PS), hadronization, and simulation of the underlying event for the \Zjjew process are performed with the \HERWIGpp (2016 simulation) and \HERWIGSeven (2017 and 2018) programs~\cite{Bellm:2015jjp}, as they have shown to better match the observed data compared to the \pt-ordered \PYTHIA predictions in the description of the additional hadronic activity in the rapidity range between the two leading jets~\cite{Sirunyan:2017jej}. The EE5C~\cite{Khachatryan:2015pea} and CH3 tunes~\cite{CMS-PAS-GEN-19-001} are used in the \HERWIGpp and \HERWIGSeven simulated samples, respectively.

\section{Event selection}\label{sec:event_selection}

The analysis is performed using ${\sqrt{s}=13\TeV}$ $\Pp\Pp$ collision data collected by the {CMS} experiment from 2016 to 2018, corresponding to an integrated luminosity of 137\fbinv. Signal events considered in this analysis are expected to contain two prompt isolated muons, regardless of the targeted Higgs boson production mode. Events are initially selected by the L1 trigger, requiring at least one muon candidate reconstructed in the muon chambers with ${\pt > 22\GeV}$. Events of interest are selected by the HLT using single muon triggers that have a \pt threshold of 27~(24)\GeV for data recorded in 2017 (2016, 2018). 

After passing the trigger selections, each event is required to contain at least two oppositely charged muons with ${\pt > 20 \GeV}$, ${\abs{\eta} < 2.4}$, and passing certain selection requirements on the number of hits in the silicon tracker and in the muon systems, as well as on the quality of the fitted muon track~\cite{Sirunyan:2018fpa}. Each muon is also required to be isolated in order to reject events with nonprompt or misidentified muon candidates. The muon isolation variable, as defined in Section~\ref{sec:event_reconstruction}, is required to be less than 25\% of the muon \pt.  Muons from the Higgs boson decay satisfy these identification and isolation requirements with an average selection efficiency of about 95\%. In addition, at least one of the two muons is required to have ${\pt > 29~(26)\GeV}$ for data collected in 2017 (2016, 2018), ensuring nearly 100\% trigger efficiency.

The sensitivity of this analysis depends primarily on the resolution of the $m_{\mu\mu}$ peak in the signal events. This resolution depends on the precision with which the muon \pt is measured, which worsens with increasing muon $\abs{\eta}$. The relative \pt resolution of muons with ${\pt > 20\GeV}$ passing through the barrel region of the detector (${\abs{\eta} < 0.9}$) ranges from 1.5 to 2\%, whereas the \pt resolution of muons passing through the endcaps of the muon system (${\abs{\eta} > 1.2}$) ranges from 2 to 4\%. The muon momentum scale and resolution are calibrated in bins of \pt and $\eta$ using the decay products of known dilepton resonances, following the method described in Ref.~\cite{Bodek:2012id}. In signal events, the Higgs boson decays into a muon pair at the interaction point. Therefore, the precision of the muon \pt measurement can be improved by including the interaction point as an additional constraint in the muon track fit. This is implemented via an analytical correction to the muon \pt proportional to the product of the muon $\pt^{2}$, its charge, and the minimum distance in the transverse plane between the muon track and the beam position. The correction is derived in simulated \Zmm events and checked in both data and simulation to provide an equivalent result to refitting the muon track with the interaction point constraint. The resulting improvement in the expected $m_{\mu\mu}$ resolution in signal events ranges from 3 to 10\%, depending on muon $\pt$, $\eta$, and the data-taking period. 

In a nonnegligible fraction of signal events, a muon from the Higgs boson decay radiates a photon that carries away a significant fraction of the muon momentum. If not taken into account, this worsens the resolution of the dimuon invariant mass ($m_{\mu\mu}$) peak in signal events. Furthermore, if the FSR photon falls in the isolation cone of the corresponding muon candidate, it can significantly increase the value of the isolation sum, thereby creating an inefficiency in selecting signal events. Therefore, a procedure is implemented to identify and recover the contribution of FSR photons similar to that described in Ref.~\cite{Sirunyan:2017exp}. In order to preserve the overall signal acceptance of the dimuon selection described above, the FSR recovery is applied only to muons with ${\pt > 20\GeV}$ and ${\abs{\eta} < 2.4}$. Photons with ${\pt > 2 \GeV}$ and ${\abs{\eta}<2.5}$ that are not associated with reconstructed electrons are considered as FSR photon candidates if they lie inside a cone of ${R = 0.5}$ around a muon track. These candidates are then required to be loosely isolated and collinear with the muon such that ${(\Sigma_{i} \pt^{i}(\Delta R(\gamma,i)<0.3))/\pt(\gamma) < 1.8}$ and ${\Delta R(\mu,\gamma)/\pt^{2}(\gamma) < 0.012}$, where $\pt(\gamma)$ is the \pt of the FSR photon candidate and the index $i$ refers to the PF candidates other than the muon within a cone of ${R=0.3}$ around the photon. In order to suppress possible contaminations from ${\PH\to \PZ(\mu\mu)\gamma}$ decays, the ratio between the \pt of the FSR photon and that of the associated muon is required to be smaller than 0.4. In the case of multiple FSR candidates associated with a muon, the candidate with the smallest value of ${\Delta R(\mu, \gamma)/\pt^{2}(\gamma)}$ is chosen. The momentum of the photon is added to that of the muon and its contribution to the muon isolation sum is ignored. The FSR recovery increases the signal efficiency by about 2\% and improves the $m_{\mu\mu}$ resolution by about 3\%.

In order to maximize the analysis sensitivity, event candidates selected with the requirements described above are separated into independent and nonoverlapping classes based on the features of the final state expected from each production mode. Events with $\cPqb$-tagged jets are assigned to the $\ttbar\PH$ production category, which is further split into the hadronic and leptonic subclasses by the presence of additional charged leptons ($\mu$ or $\Pe$) in the final state. Dimuon events with one (two) additional charged lepton(s) and no $\cPqb$-tagged jets are assigned to the $\PW\PH$ ($\PZ\PH$) category. Events with neither additional charged leptons nor $\cPqb$-tagged jets belong to the VBF category if a pair of jets is present with large $\mjj$ and $\detajj$. The remaining untagged events, which constitute about 96\% of the total sample of dimuon candidate events, belong to the $\Pg\Pg\PH$-enriched category. In each production category, multivariate techniques are used to enhance the discrimination between the expected signal and background contributions by further dividing events into several subcategories with different signal-to-background ratios. The measured \hmm signal is then extracted via a simultaneous maximum-likelihood fit across all event categories to observables chosen for each category to maximize the overall measurement precision. In the following Sections, each production category is presented in order of decreasing sensitivity.

\section{The VBF production category}\label{sec:vbf_category}

A dimuon event passing the baseline selection detailed in Section~\ref{sec:event_selection} is considered in the VBF production category if it contains two or more jets, with the \pt of the leading jet ($\pt(\mathrm{j}_1)$) larger than 35\GeV, the \pt of the second-highest \pt jet ($\pt(\mathrm{j}_2)$) greater than 25\GeV, and the $\abs{\eta}$ of both jets less than 4.7. Jets overlapping with either of the two selected muons are discarded. In addition, the two highest \pt jets in the event are required to have ${\mjj > 400\GeV}$ and ${\abs{\detajj} > 2.5}$. An event is rejected from the VBF category if it contains one (two) jet(s) inside the silicon tracker fiducial volume (${\abs{\eta} < 2.5}$) with ${\pt > 25\GeV}$ and identified as a $\cPqb$ quark jet by the medium (loose) WP of the DeepCSV $\cPqb$-tagging algorithm. These requirements suppress the $\ttbar$ and single top quark backgrounds and ensure mutual exclusivity between the VBF and $\ttbar\PH$ categories. Moreover, events containing an additional muon (electron) with ${\pt > 20\GeV}$ and ${\abs{\eta} < 2.4~(2.5)}$ passing the selection criteria described in Section~\ref{sec:vh_category} are discarded. This requirement ensures no overlap between the analyses targeting VBF and V\PH production. Selected events are further grouped into two independent classes. Events in which the two muons form an invariant mass between 115 and 135\GeV belong to the signal region (VBF-SR), which is enriched in signal-like events. Events with ${110 < m_{\mu\mu} < 115\GeV}$ or ${135 < m_{\mu\mu} < 150\GeV}$ belong to the mass sideband region (VBF-SB), which is used as a control region to estimate the background. The VBF-SR is defined to be 20\GeV wide in order to be sensitive to Higgs boson mass hypotheses in the range of 120--130\GeV. A summary of the selection criteria used to define the VBF-SB and VBF-SR regions is reported in Table~\ref{tab:vbf_preselection}.

\begin{table*}[!htb]
  \centering
  \topcaption{Summary of the kinematic selections used to define the VBF-SB and VBF-SR regions.}
  \cmsTable{
    \begin{tabular}{lcc}
      \hline
      Observable & VBF-SB  & VBF-SR \\
      \hline
      Number of loose (medium) $\cPqb$-tagged jets & \multicolumn{2}{c}{$\leq$1~(0)} \\
      Number of selected muons & \multicolumn{2}{c}{$=$2} \\
      Number of selected electrons & \multicolumn{2}{c}{$=$0} \\
      Jet multiplicity ($\pt > 25 \GeV$, $\abs{\eta} < 4.7$) & \multicolumn{2}{c}{$\geq$2} \\
      Leading jet \pt   & \multicolumn{2}{c}{$\geq$35\GeV} \\ 
      Dijet mass (\mjj) & \multicolumn{2}{c}{$\geq$400\GeV} \\
      Pseudorapidity separation ($\abs{\detajj}$) & \multicolumn{2}{c}{$\geq$2.5} \\
      Dimuon invariant mass & $110 < m_{\mu\mu} < 115\GeV$ & $115 < m_{\mu\mu} < 135\GeV$ \\
       & or $135 < m_{\mu\mu} < 150\GeV$ & \\
      \hline
    \end{tabular}
  }
  \label{tab:vbf_preselection}
\end{table*}

A deep neural network (DNN) multivariate discriminant is trained to distinguish the expected signal from background events using kinematic input variables that characterize the signal and the main background processes in the VBF-SR. The DNN is implemented using \textsc{keras}~\cite{chollet2015keras} with \textsc{tensorflow}~\cite{tensorflow2015-whitepaper} as backend. The DNN inputs include six variables associated with the production and decay of the dimuon system, namely the $m_{\mu\mu}$,  the per-event uncertainty in the measured dimuon mass $\sigma(m_{\mu\mu})$, the dimuon transverse momentum ($\pt^{\mu\mu}$), the dimuon rapidity ($y_{\mu\mu}$), and the azimuthal angle ($\phi_{\mathrm{CS}}$) and the cosine of the polar angle ($\cos\theta_{\mathrm{CS}}$) computed in the dimuon Collins--Soper rest frame~\cite{PhysRevD.16.2219}. The DNN also takes as input a set of variables describing the properties of the dijet system, namely the full momentum vector of the two highest \pt jets in the event ($\pt(\mathrm{j}_1)$, $\pt(\mathrm{j}_2)$, $\eta(\mathrm{j}_1)$, $\eta(\mathrm{j}_2)$, $\phi(\mathrm{j}_1)$, and $\phi(\mathrm{j}_2)$), \mjj, and \detajj. In addition, observables sensitive to angular and \pt correlations between muons and jets are also included, namely the minimum $\Delta \eta$ between the dimuon system and each of the two leading jets, the Zeppenfeld variable ($z^{*}$)~\cite{Schissler:2013nga} constructed from $y_{\mu\mu}$ and the rapidities of the two jets as
\begin{linenomath}
\begin{equation}\label{eqn:Zeppenfeld}
  z^{*} = \frac{y_{\mu\mu}-(y_{\mathrm{j}_1}+y_{\mathrm{j}_2})/2}{\abs{y_{\mathrm{j}_1}-y_{\mathrm{j}_2}}},
\end{equation}
\end{linenomath}
and the \pt-balance ratio 
\begin{linenomath}
\begin{equation}\label{eqn:Rpt}
  R(\pt) = \frac{\abs{\vec{\pt}^{\mu\mu}+\vec{\pt}^{\mathrm{jj}}}}{\pt^{\mu\mu}+\pt(\mathrm{j}_1)+\pt(\mathrm{j}_2)}.
\end{equation}
\end{linenomath}
The VBF signal events are expected to have suppressed hadronic activity in the rapidity region between the two leading jets. This feature is exploited by considering ``soft track-jets'' in the event that are defined by clustering, via the anti-\kt algorithm with a distance parameter of 0.4, charged particles from the primary interaction vertex, excluding the two identified muons and those associated with the two VBF jets. The use of soft track-jet observables is a robust and validated method to reconstruct the hadronization products of partons with energy as low as a few GeV~\cite{CMS-PAS-JME-10-006}. The number of soft track-jets in an event with ${\pt > 5\GeV}$, as well as the scalar \pt sum of all track-jets with ${\pt > 2\GeV}$, are used as additional input variables. Finally, since jets in signal events are expected to originate from quarks, whereas in the DY process they can also be initiated by gluons, the quark-gluon likelihood~\cite{CMS:2017wyc} of the two leading jets is also used as input to the DNN.

The DNN is trained using simulated events from signal (VBF) and background (DY, \Zjjew, $\ttbar$, and diboson) processes selected in the VBF-SR. Signal events generated with ${\mh=125\GeV}$ are used in the DNN training. The last hidden layers of four intermediate networks are combined to form a single binary classifier: two networks exploit the full set of variables described above in order to optimize the separation between the VBF signal and the \Zjjew or DY background, while the other two optimize the separation between the VBF signal and the total expected background. The first of the two networks discriminating against the total background uses all the inputs except for $m_{\mu\mu}$, while the second uses only the dimuon mass and its resolution. Every network contains three or four hidden layers, each with a few tens of nodes. All trainings are performed using a four-fold strategy~\cite{Refaeilzadeh2009}, where 50\% of the events are used for training, 25\% for validation, and 25\% for testing. The validation sample is used to optimize the DNN hyper-parameters, while the test sample is used to evaluate the DNN performance and for the expected distributions in the signal extraction fit. The selected training epoch maximizes the expected significance, determined using the Asimov data set~\cite{Cowan:2010js}, defined as the minimum between the significances computed from the training and validation samples.

Events belonging to the VBF-SR are divided into nonoverlapping bins based on the DNN value, independently for each data-taking period. These bins are defined to achieve optimal sensitivity, while minimizing the total number of bins. From this optimization procedure, thirteen bins are obtained in each data-taking period characterized by different bin boundaries. Given the negligible correlation between the $m_{\mu\mu}$ and other input variables, the $m_{\mu\mu}$ variable can be marginalized from the DNN by replacing the $m_{\mu\mu}$ with a fixed value of 125\GeV during the DNN evaluation. The resulting DNN score is not significantly correlated with the $m_{\mu\mu}$. This mass-decorrelated DNN is used for events in the VBF-SB region and captures the main features of the DNN distribution in the VBF-SR. The signal is extracted from a binned maximum-likelihood fit to the output of the DNN discriminator performed simultaneously over the VBF-SR and VBF-SB regions. Because of significant variations in the detector response to forward jets during different data-taking periods, the fit is performed separately for data collected in 2016, 2017, and 2018. The contributions of the various background processes are estimated from simulation, following the same strategy employed in the measurement of the \Zjjew cross section with 13\TeV data~\cite{Sirunyan:2017jej}. This simulation-based strategy yields, in the VBF category, an improvement in sensitivity of about 20\% compared to an alternative strategy in which the background determination is entirely based on data. In this alternative analysis, a multivariate classifier is used to divide events into subcategories with different signal purity, and the signal is extracted by fitting the $m_{\mu\mu}$ distribution in each subcategory to parametric functions as in Ref.~\cite{Sirunyan:2018hbu}. In such data-driven analyses, the precision of the background estimate strictly depends on the number of observed events in the mass sidebands, thereby limiting the performance in the high purity subcategories that contain a small number of events. In contrast, the approach presented here relies on the precision with which the simulation is able to predict the different background components. The uncertainty in this prediction is validated and constrained using the signal-depleted sideband regions.

Theoretical uncertainties affect both the expected rate and the shape of signal and background histograms (templates) used in the fit. The Higgs boson production cross section for the various modes, and their corresponding uncertainties, are taken from Ref.~\cite{deFlorian:2016spz}. These include uncertainties in the choice of the PDF, as well as the QCD renormalization ($\mu_{\mathrm{R}}$) and factorization ($\mu_{\mathrm{F}}$) scales. The uncertainty in the prediction of \brhmm is also considered. For the VBF process, uncertainties in the modelling of the $\pt(\PH)$, $\pt(\PH\mathrm{jj})$, jet multiplicity, and \mjj distributions are considered. Their total uncertainty on the VBF signal prediction is about 2--4\%. Similarly, for the $\Pg\Pg\PH$ process, seven independent additional sources are included to account for the uncertainty in the modelling of the $\pt(\PH)$ distribution, the number of jets in the event, and its contamination in the VBF selected region, as described in Ref.~\cite{deFlorian:2016spz}. The magnitude of these uncertainties for $\Pg\Pg\PH$ events in the VBF category varies from about 15 to 25\%. The theoretical uncertainties described so far affect also the signal prediction in the $\Pg\Pg\PH$, $\ttbar\PH$, and $\PV\PH$ production categories reported in the next Sections. For each background process, template variations are built by changing the values of $\mu_{\mathrm{R}}$ and $\mu_{\mathrm{F}}$ by factors of 2 and 0.5 from the default values used in the ME calculation, excluding the combinations for which $\mu_{\mathrm{R}}/\mu_{\mathrm{F}} = 0.25$ or 4, as well as by comparing the nominal distributions with those obtained using the alternative PDFs of the NNPDF set. These theoretical uncertainties are correlated across years and regions (VBF-SR and VBF-SB) but are uncorrelated between processes. The shape uncertainty arising from the PS model is assessed by varying several parameters that control the properties of the ISR and FSR jets produced by \PYTHIA. The \Zjjew and VBF signal simulations are very sensitive to the PS model, as shown in Refs.~\cite{Sirunyan:2017jej,Jager:2020hkz}. A conservative PS uncertainty is assigned to the \Zjjew background and VBF signal, defined as the full symmetrized difference between \PYTHIA (dipole shower) and \HERWIG (angular-ordered shower) predictions in each DNN bin, which is larger than that obtained by varying the PS ISR and FSR parameters.

Several sources of experimental uncertainty are taken into account for both signal and background processes. These include the uncertainty in the measurement of the integrated luminosity, in the modelling of the pileup conditions during data taking, in the measurement of the muon selection and trigger efficiencies, in the muon momentum scale and resolution, in the efficiency of vetoing $\cPqb$ quark jets, and in the jet energy scale and resolution. If not explicitly mentioned, experimental uncertainties are considered correlated across event categories and data-taking periods. Most of the sources of uncertainty affecting the jet energy scale are correlated across processes and years, while those affecting the jet energy resolution are only correlated across processes but not across years. The uncertainty in the measurement of the integrated luminosity is partially correlated across years. The integrated luminosities of the 2016, 2017, and 2018 data-taking periods are individually known with uncertainties in the 2.3--2.5\% range~\cite{CMS-PAS-LUM-17-001,CMS-PAS-LUM-17-004,CMS-PAS-LUM-18-002}, while the total integrated luminosity has an uncertainty of 1.8\%. The improvement in precision reflects the (uncorrelated) time evolution of some systematic effects. During the 2016 and 2017 data-taking periods, a gradual shift in the timing of the inputs of the ECAL L1 trigger in the forward endcap region (${\abs{\eta} > 2.4}$) led to a specific inefficiency. A correction for this effect was determined using an unbiased data sample and is found to be relevant in events with high-\pt jets with ${2.4 < \abs{\eta} < 3.0}$. This correction is about 2~(3)\% at ${\mjj=400\GeV}$ in the 2016 (2017) data-taking period and it increases to about 6~(9)\% for ${\mjj > 2\TeV}$. A systematic uncertainty corresponding to 20\% of this correction is considered. Lastly, a significant fraction (about 30--35\%) of the DY background populating bins with low DNN score is comprised of events in which either the leading or subleading jet are in the forward region of the detector (${\abs{\eta}>3.0}$) and are not matched with a jet at the generator level. These jets originate either from the soft emissions produced by the PS or from pileup interactions. The normalization of this term is left floating in the fit and is directly constrained by the observed data events with low DNN score belonging to the VBF-SR and VBF-SB regions. Because of significant variations in the detector response in the forward region over time, these normalization parameters are considered uncorrelated across years. The normalization of the remaining DY component with at least two matched jets is taken from the simulation and constrained, as for the other background processes, within the systematic uncertainties described above.

The uncertainty arising from the limited size of simulated samples is also taken into account by allowing each bin of the total background template to vary within the corresponding statistical uncertainty using the Barlow--Beeston lite technique~\cite{BARLOW1993219,Conway:2011in}. These uncertainties are uncorrelated across the bins of the DNN templates used in the fit. Systematic uncertainties are modelled in the fit as nuisance parameters with log-normal or Gaussian external constraints.

\begin{figure*}[!]
  \centering
  \includegraphics[width=0.45\textwidth]{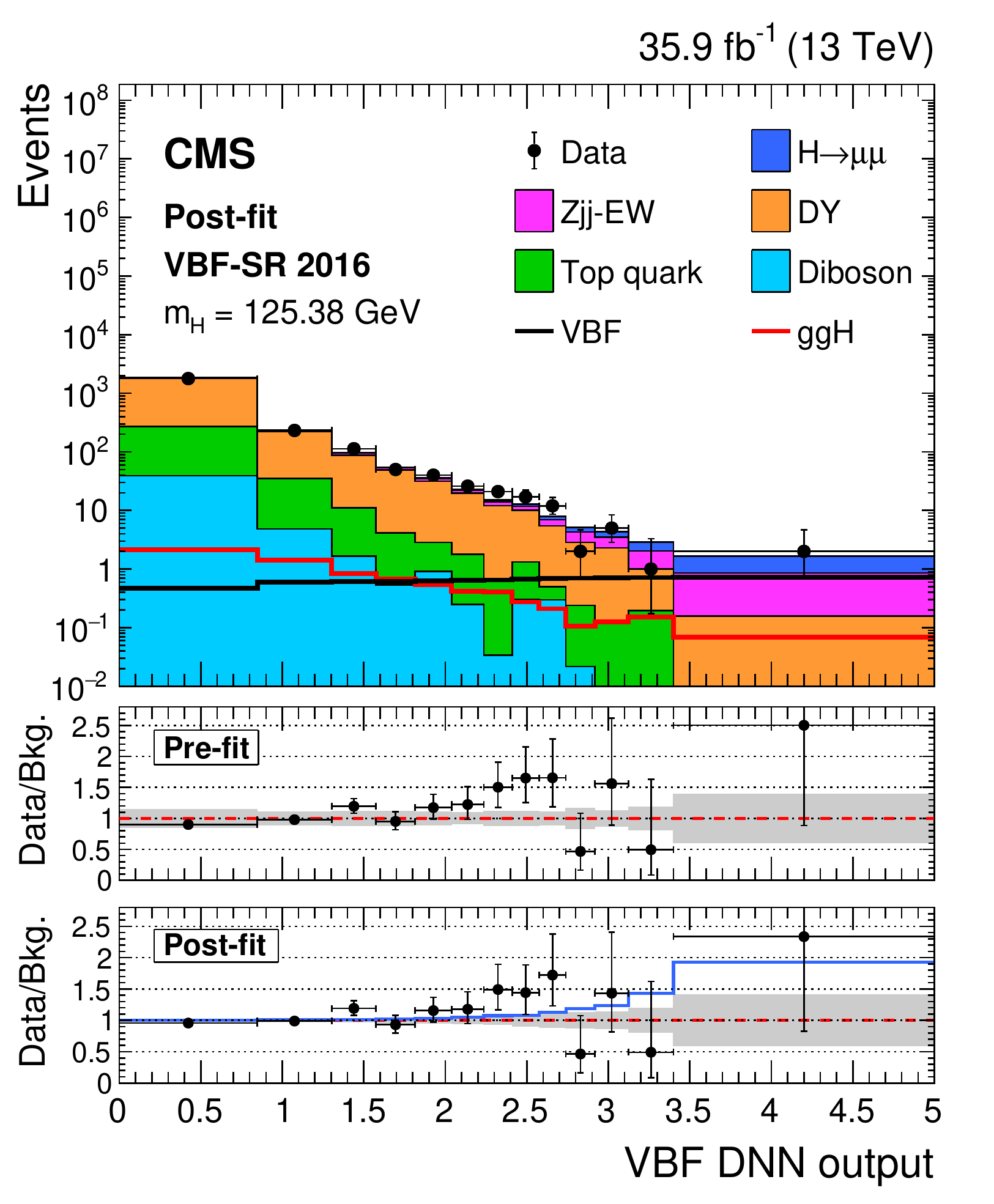}
  \includegraphics[width=0.45\textwidth]{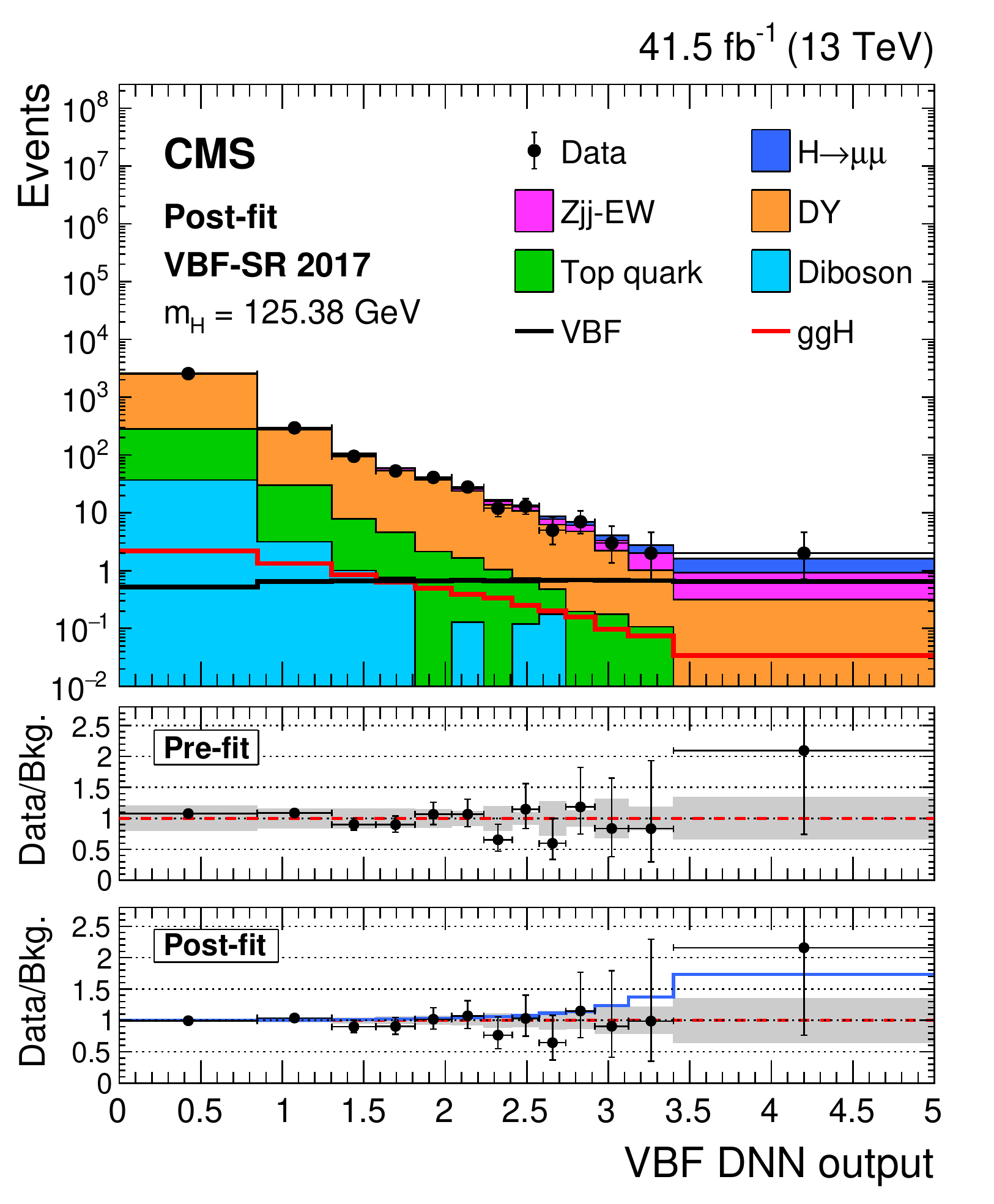}\\  
  \includegraphics[width=0.45\textwidth]{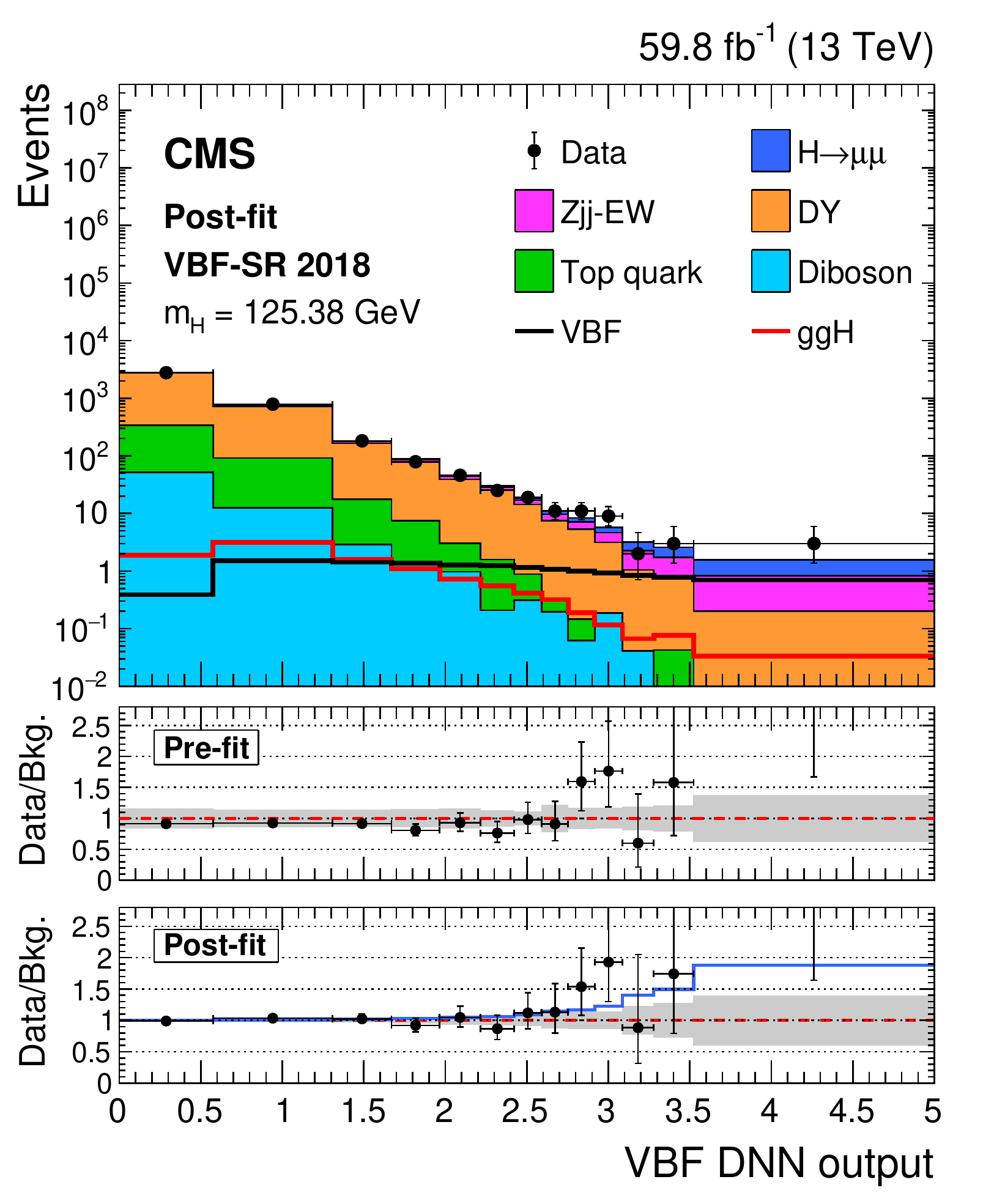}
  \caption{The observed DNN output distribution in the VBF-SR region for data collected in 2016 (first row, \cmsLeft), 2017 (first row, \cmsRight), and 2018 (second row) compared to the post-fit background estimate for the contributing SM processes. The post-fit distributions for the Higgs boson signal produced via $\Pg\Pg\PH$ (solid red) and VBF (solid black) modes with ${\mh = 125.38\GeV}$ are overlaid. The predicted backgrounds are obtained from a {S+B} fit performed across analysis regions and years. In the middle panel, the ratio between data and the pre-fit background prediction is shown. The grey band indicates the total pre-fit uncertainty obtained from the systematic sources previously described. The lower panel shows the ratio between data and the post-fit background prediction from the {S+B} fit. The grey band indicates the total background uncertainty after performing the fit. The blue histogram (upper panel) and solid line (lower panel) indicate the total signal extracted from the fit with ${\mh = 125.38\GeV}$.}
  \label{fig:dnn_vbf_sr}
\end{figure*}

\begin{figure*}[!]
  \centering
  \includegraphics[width=0.45\textwidth]{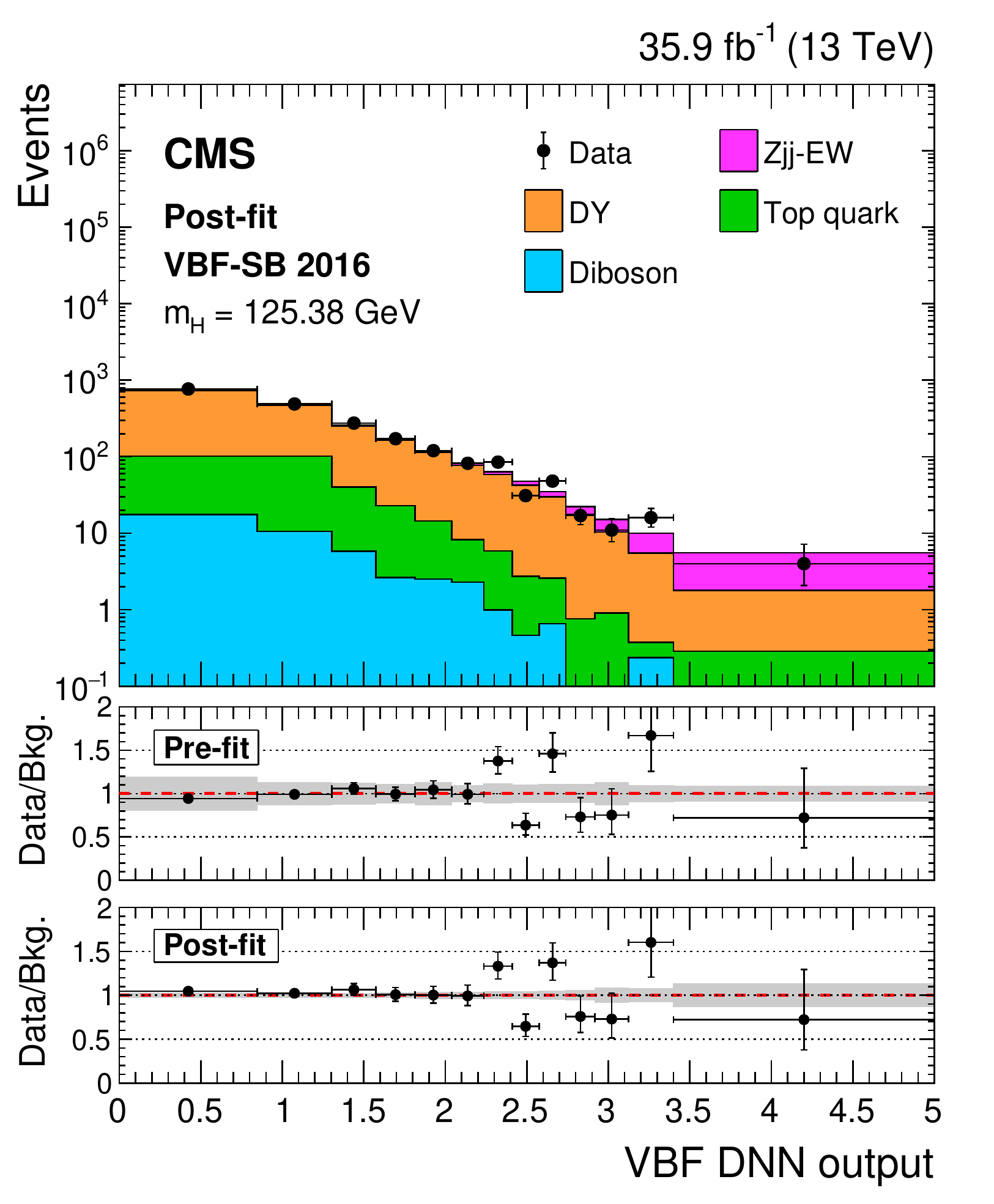}
  \includegraphics[width=0.45\textwidth]{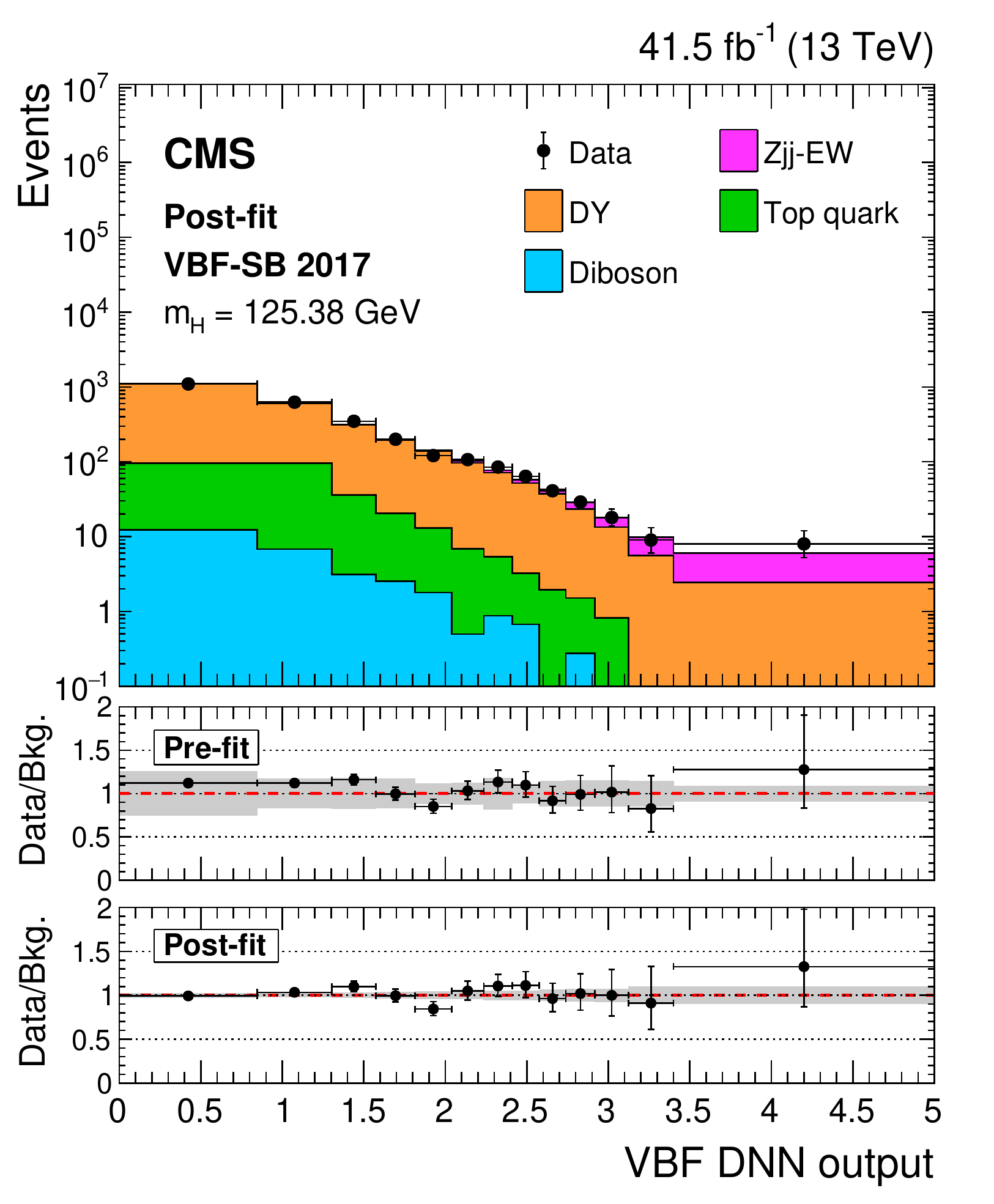}\\
  \includegraphics[width=0.45\textwidth]{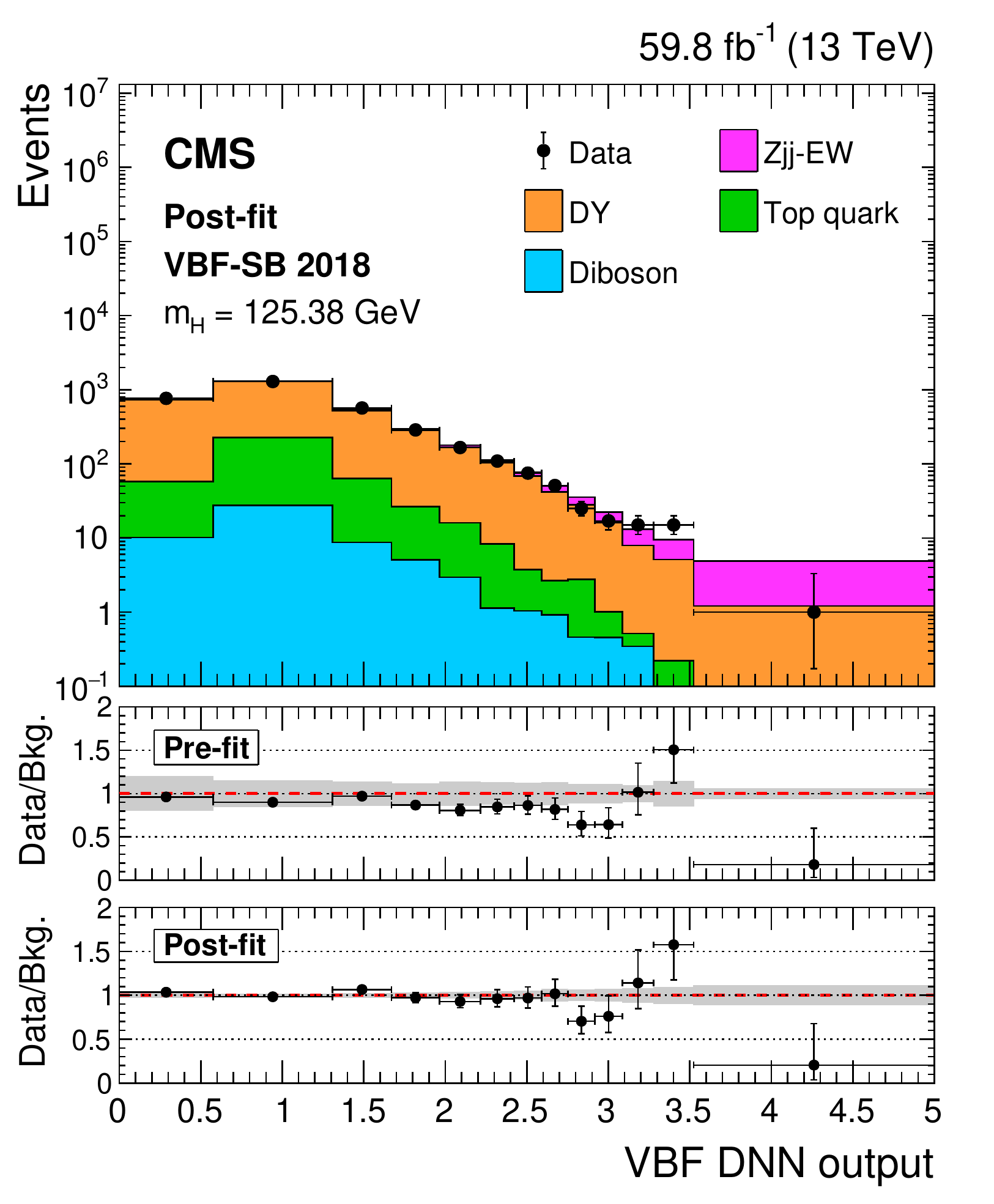}
  \caption{The observed DNN output distribution for data collected in 2016 (first row, \cmsLeft), 2017 (first row, \cmsRight), and 2018 (second row) in the VBF-SB region compared to the post-fit background estimate from SM processes. The predicted backgrounds are obtained from a {S+B} fit performed across analysis regions and years. The description of the three panels is the same as in Fig.~\ref{fig:dnn_vbf_sr}.}
  \label{fig:dnn_vbf_sb}
\end{figure*}

\begin{figure*}[!]
  \centering
  \includegraphics[width=0.45\textwidth]{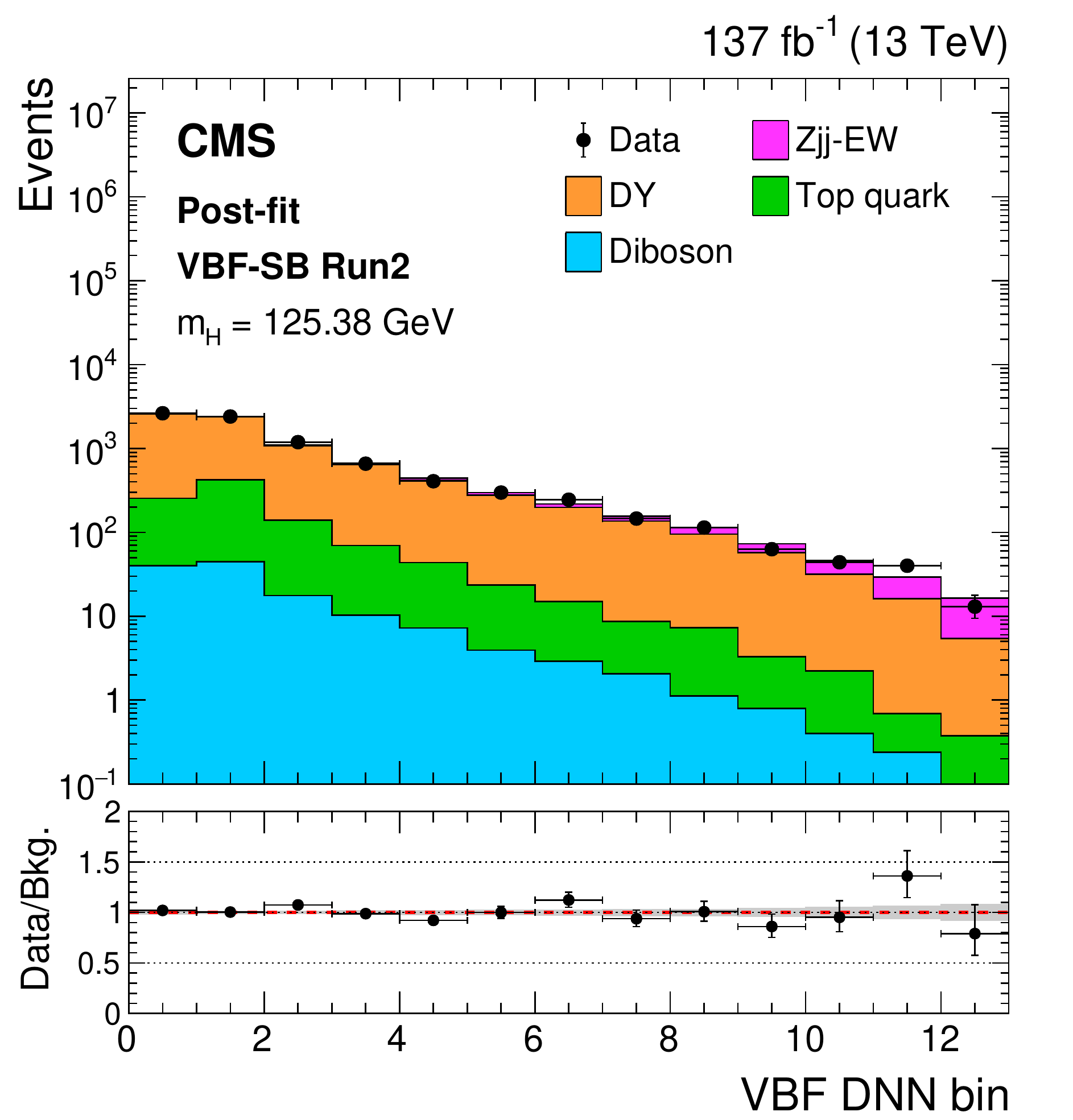}
  \includegraphics[width=0.45\textwidth]{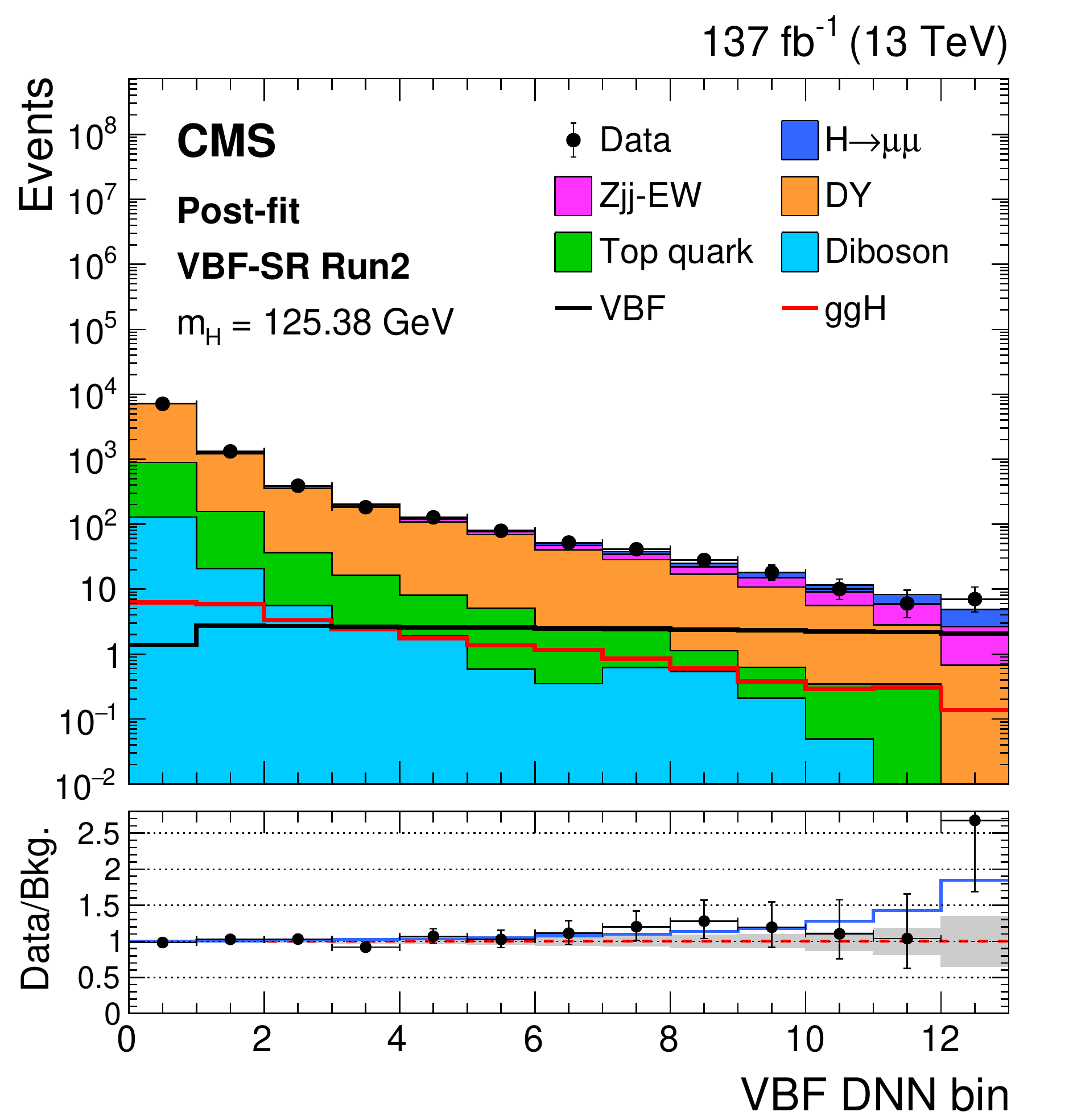}
  \caption{The observed DNN output distribution in the VBF-SB (\cmsLeft) and VBF-SR (\cmsRight) regions for the combination of 2016, 2017, and 2018 data, compared to the post-fit prediction from SM processes. The post-fit distributions for the Higgs boson signal produced via $\Pg\Pg\PH$ (solid red) and VBF (solid black) modes with ${\mh = 125.38\GeV}$ are overlaid. The lower panel shows the ratio between data and the post-fit background prediction from the {S+B} fit. The best fit \hmm signal contribution for ${\mh = 125.38\GeV}$ is indicated by the blue histogram (upper panel) and solid line (lower panel), while the grey band indicates the total background uncertainty.}
  \label{fig:dnn_vbf_run2}
\end{figure*}

Figure~\ref{fig:dnn_vbf_sr} shows the observed and predicted distributions of the DNN discriminant in the VBF-SR. The background prediction is obtained from a simultaneous signal-plus-background ({S+B}) fit performed across the VBF-SR and VBF-SB regions, as well as data-taking periods. The post-fit distributions for the Higgs boson signal produced via $\Pg\Pg\PH$ (solid red) and VBF (solid black) production with ${\mh = 125.38\GeV}$ are overlaid. The blue histogram indicates, instead, the total signal extracted from the fit. Similarly, Fig.~\ref{fig:dnn_vbf_sb} shows the distributions of the DNN discriminant in the VBF-SB, obtained after performing the same {S+B} fit. Figure~\ref{fig:dnn_vbf_run2} shows the observed and predicted DNN output distributions in the VBF-SB (\cmsLeft) and VBF-SR (\cmsRight) regions for the combination of 2016, 2017, and 2018 data. Since the bin boundaries are optimized separately per data-taking period, the distributions are combined by summing the corresponding observed and predicted number of events in each individual bin. The lower panel shows the ratio between the data and the post-fit background prediction, with the best fit signal contribution indicated by the blue line in the VBF-SR. Finally, Table~\ref{tab:vbf_dnn_yields} reports, for each bin or group of bins of the DNN output in the VBF-SR, the expected number of VBF and $\Pg\Pg\PH$ signal events (S), the observed number of events in data, the total background prediction (B) and its uncertainty ($\Delta\mathrm{B}$), and the {S/(S+B)} and ${\mathrm{S}/\sqrt{\smash[b]{\mathrm{{B}}}}}$ ratios obtained by summing the post-fit estimates from each of the three data-taking periods.

\begin{table*}[!htb]
 \centering
 \topcaption{Event yields in each bin or in group of bins defined along the DNN output in the VBF-SR for various processes. The expected signal contribution for $\mh=125.38\GeV$ (S), produced via VBF and $\Pg\Pg\PH$ modes and assuming SM cross sections and \brhmm, is shown. The background yields (B) and the corresponding uncertainties ($\Delta\mathrm{B}$) are obtained after performing a combined {S+B} fit across the VBF-SR and VBF-SB regions and each data-taking period. The observed event yields, {S/(S+B)} ratios and ${\mathrm{S}/\sqrt{\smash[b]{\mathrm{{B}}}}}$ ratios are also reported.}
  \begin{tabular}{l c c c c c c c}
      \hline 
      DNN bin & Total signal & VBF (\%) & $\Pg\Pg\PH$ (\%) & Bkg. $\pm~\Delta\mathrm{B}$ & Data & {S/(S+B)} (\%) & $\mathrm{S}/\sqrt{\smash[b]{\mathrm{B}}}$ \\
      \hline
      1--3 & 19.5 & 30 & 70 & 8890 $\pm$ 67 & 8815  & 0.22  & 0.21 \\
      4--6 & 11.6 & 57 & 43 & 394  $\pm$ 8  & 388   & 2.86  & 0.58 \\
      7--9 & 8.43 & 73 & 27 & 103  $\pm$ 4  & 121   & 7.56  & 0.83 \\
      10   & 2.30 & 85 & 15 & 15.1 $\pm$ 1.4 & 18   & 13.2 & 0.59 \\
      11   & 2.15 & 88 & 12 & 9.1  $\pm$ 1.2 & 10   & 19.1 & 0.71 \\
      12   & 2.10 & 87 & 13 & 5.8  $\pm$ 1.1 & 6    & 26.6 & 0.87 \\
      13   & 1.87 & 94 & 6  & 2.6  $\pm$ 0.9 & 7    & 41.8 & 1.16 \\
      \hline
    \end{tabular}
  \label{tab:vbf_dnn_yields}
\end{table*}

\section{The \texorpdfstring{$\Pg\Pg\PH$}{ggH} production category}\label{sec:ggh_category}

An event is considered in the $\Pg\Pg\PH$ category if it contains exactly two muons passing the baseline selection requirements detailed in Section~\ref{sec:event_selection}. Events with additional muons or electrons are rejected to avoid overlap with the $\PV\PH$ category. Any jets considered in the event must be spatially separated (${\Delta R > 0.4}$) from either of the two muons. In order to ensure mutual exclusivity with the VBF category, events containing two or more jets with ${\pt > 25\GeV}$ are only considered if the leading jet has ${\pt < 35\GeV}$, the invariant mass of the two highest \pt jets is smaller than 400\GeV, or the ${\abs{\detajj} < 2.5}$. Lastly, events containing at least two jets with ${\pt > 25 \GeV}$ and ${\abs{\eta} < 2.5}$ passing the loose WP of the DeepCSV $\cPqb$-tagging algorithm, or at least one jet passing the medium WP, are rejected, ensuring no overlap between the $\Pg\Pg\PH$ and $\ttbar\PH$ categories. A summary of the selection criteria used to define the $\Pg\Pg\PH$ category is reported in Table~\ref{tab:ggh_preselection}.

\begin{table*}[!htb]
  \centering
  \topcaption{Summary of the kinematic selections used to define the $\Pg\Pg\PH$ production category.}
  \cmsTable{
    \begin{tabular}{lc}
      \hline
      Observable & Selection \\
      \hline
      Number of loose (medium) $\cPqb$-tagged jets & $\leq$1~(0) \\
      Number of selected muons & $=$2 \\
      Number of selected electrons & $=$0 \\
      VBF selection veto & if $\mathrm{N}_{\mathrm{jets}} \geq 2$  \\
                         & $\mjj < 400\GeV$ or $\abs{\detajj} < 2.5$ or $\pt(\mathrm{j}_1) < 35\GeV$\\

      \hline
    \end{tabular}
  }
  \label{tab:ggh_preselection}
\end{table*}

A multivariate discriminant based on boosted decision trees (BDTs) is employed to discriminate between signal and background events. To account for the evolution in the detector response during data-taking periods, the BDT discriminant is trained separately for the 2016, 2017, and 2018 simulated samples using the \textsc{tmva} package~\cite{Hocker:2007ht}, resulting in three independent BDT outputs. The input variables are chosen such that the BDT discriminants are effectively uncorrelated with $m_{\mu\mu}$. This is required by the chosen analysis strategy, in which events are first divided into independent subcategories based on the BDT output, then a potential signal is extracted from each subcategory by searching for a narrow peak over a smoothly falling background in the $m_{\mu\mu}$ distribution. In this category, given the prior knowledge of the expected DY background shape and the large number of data events in the mass sideband around the peak that can be used to constrain the background, this strategy provides a robust background estimate from data while maximizing the analysis sensitivity.

The BDT discriminants include input variables that describe the production and decay of the dimuon system, namely $\pt^{\mu\mu}$, $y_{\mu\mu}$, $\phi_{\mathrm{CS}}$, and $\cos \theta_{\mathrm{CS}}$. In addition, the $\eta$ of each of the two muons and the ratio of each muon's \pt to $m_{\mu\mu}$ are also included. In order to increase the signal-to-background separation for events in which the gg\PH signal is produced in association with jets, the BDT discriminants also take into account the \pt and $\eta$ of the leading jet in the event with ${\pt > 25\GeV}$ and the absolute distance in $\eta$ and $\phi$ between the jet and the muon pair. For events with two or more jets with ${\pt > 25\GeV}$ in the final state, additional inputs are included: the \mjj, \detajj, and \dphijj of the two highest \pt jets. The \mjj, as well as the other dijet variables, is sensitive to the residual contribution from VBF and $\PV\PH$ modes, in which the vector boson decays hadronically. Furthermore, the Zeppenfeld variable defined in Eq.~(\ref{eqn:Zeppenfeld}) and the angular separation ($\Delta\eta$, $\Delta\phi$) between the dimuon system and each of the two leading jets are also included, which target residual VBF signal events in the $\Pg\Pg\PH$ selected region. Lastly, the total number of jets in the event with ${\pt > 25\GeV}$ and ${\abs{\eta} < 4.7}$ is also used as input to the BDT.

The signal simulation considered in the training of the multivariate discriminators includes the $\Pg\Pg\PH$, VBF, $\PV\PH$, and $\ttbar\PH$ processes. The $\Pg\Pg\PH$ sample used in the training is generated via \POWHEG since it provides positively weighted events at NLO in QCD. In later stages of the analysis, the prediction from \MGvATNLO is used instead since it provides a more accurate description of gluon fusion events accompanied by more than one jet, as detailed in Section~\ref{sec:event_simulation}. The background simulation consists of DY, $\ttbar$, single top quark, diboson, and \Zjjew processes. Only events with $m_{\mu\mu}$ in the range 115--135\GeV are included in the training. Signal and background events both contain two prompt muons in the final state, and the corresponding dimuon mass resolution ($\sigma_{\mu\mu}/m_{\mu\mu}$) does not discriminate between them. For this reason, $\sigma_{\mu\mu}/m_{\mu\mu}$ is not added as an input to the BDT. Instead, signal events in the BDT training are assigned a weight inversely proportional to the expected mass resolution, derived from the uncertainties in the \pt measurements of the individual muon tracks. This weighting improves the average signal $\sigma_{\mu\mu}/m_{\mu\mu}$ in the high-score BDT region by assigning increased importance to the high-resolution signal events. Apart from $m_{\mu\mu}$, the $\pt^{\mu\mu}$ is one of the most discriminating observables in the $\Pg\Pg\PH$ category. Discrepancies between data and simulation in the $\pt^{\mu\mu}$ spectrum for the DY background, similar to those reported in Ref.~\cite{Sirunyan:2019bzr}, are also observed in this analysis. In order to correctly model the $\pt^{\mu\mu}$ spectrum of the DY background during the training of the BDT discriminants, corrections are derived for each data-taking period by reweighting the $\pt^{\mu\mu}$ distribution of the DY simulation to reproduce the observation in data for dimuon events with ${70 < m_{\mu\mu} < 110\GeV}$. These corrections are obtained separately for events containing zero, one, and two or more jets with ${\pt > 25\GeV}$ and ${\abs{\eta}<4.7}$. 

\begin{figure*}[!htb]
  \centering
  \includegraphics[width=0.45\textwidth]{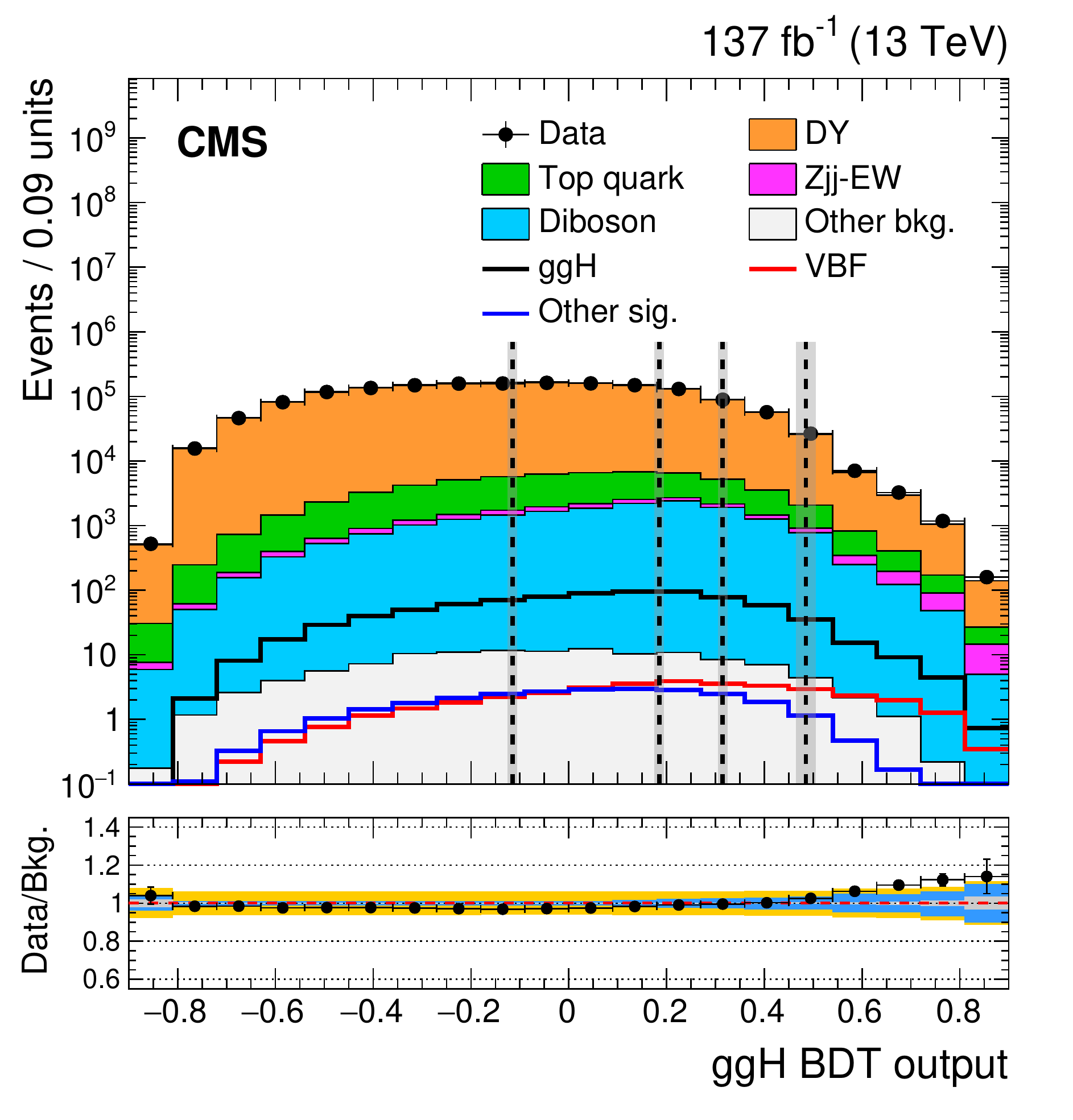}
  \includegraphics[width=0.45\textwidth]{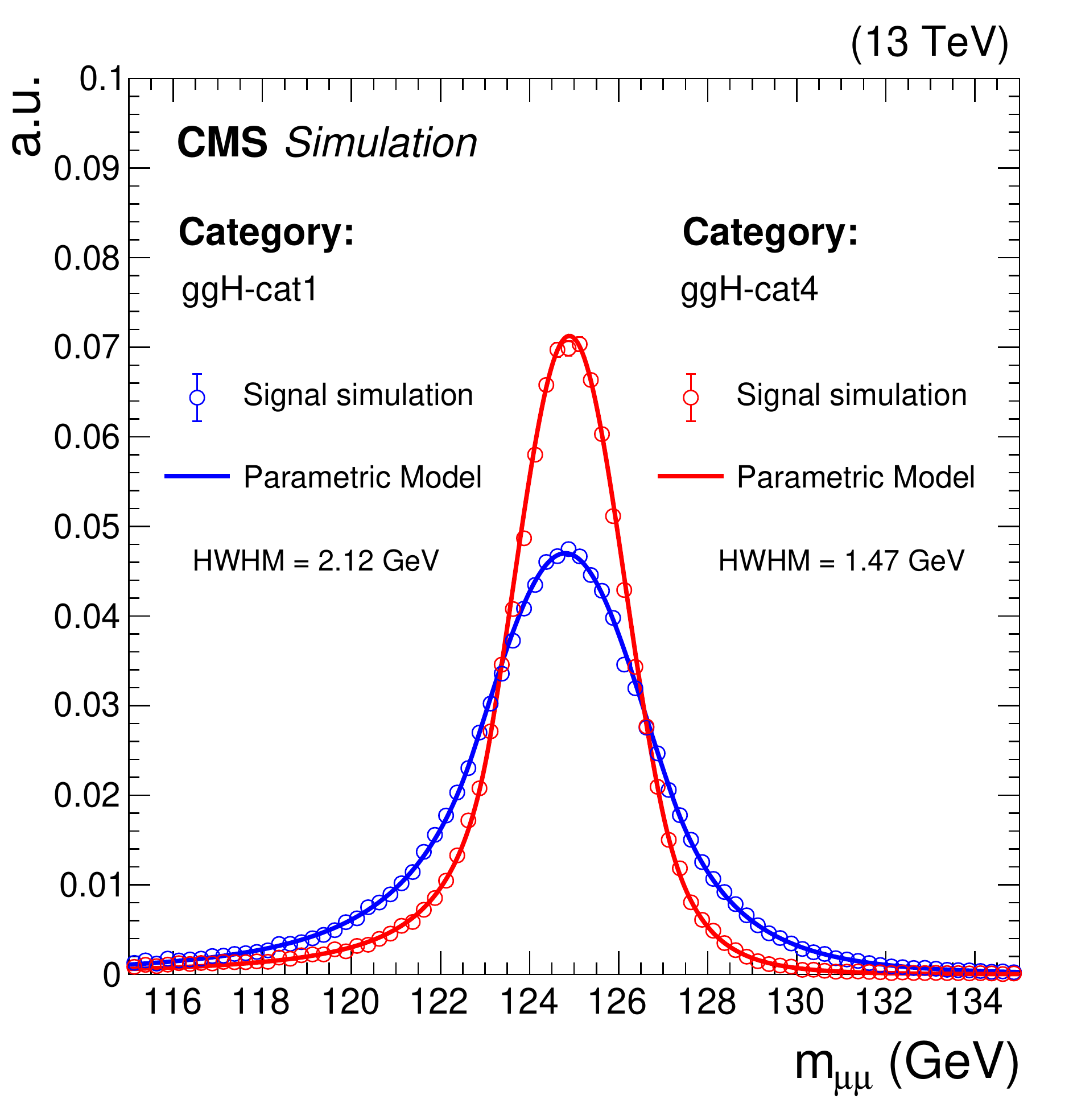}
  \caption{Left: the observed BDT output distribution compared to the prediction from the simulation of various SM background processes. Dimuon events passing the event selection requirements of the $\Pg\Pg\PH$ category, with $m_{\mu\mu}$ between 110--150\GeV, are considered. The expected distributions for $\Pg\Pg\PH$, VBF, and other signal processes are overlaid. The grey vertical bands indicate the range between the minimum and maximum BDT output values used to define the boundaries for the optimized event categories for different data-taking periods. In the lower panel, the ratio between data and the expected background is shown. The grey band indicates the uncertainty due to the limited size of the simulated samples. The azure band corresponds to the sum in quadrature between the statistical and experimental systematic uncertainties, while the orange band additionally includes the theoretical uncertainties affecting the background prediction. Right: the signal shape model for the simulated $\hmm$ sample with ${\mh=125\GeV}$ in the best (red) and the worst (blue) resolution categories.}
     \label{fig:ggH_bdt_output}
\end{figure*}

Figure~\ref{fig:ggH_bdt_output} (\cmsLeft) shows the BDT score distribution, comparing data to the prediction from simulation in events with ${110<m_{\mu\mu}<150\GeV}$, where the outputs of the individual BDTs obtained in each year are combined into a single distribution. The distributions for various signal processes ($\Pg\Pg\PH$, VBF, and $\PV\PH$+$\ttbar\PH$) are also shown. Five event subcategories are defined based on the output of these BDT discriminants. The subcategory boundaries are determined via an iterative process that aims to maximize the expected sensitivity of this analysis to \hmm decays of the SM Higgs boson. The expected sensitivity is estimated from {S+B} fits to the $m_{\mu\mu}$ distribution in simulated events with ${110<m_{\mu\mu}<150\GeV}$. In these fits, the Higgs boson signal is modelled using a parametric shape, the double-sided Crystal Ball function (DCB)~\cite{Oreglia:1980cs}
\begin{linenomath}
\begin{equation}\label{eq:dCBFunction}
\begin{aligned}
  \mathrm{DCB}(m_{\mu\mu})=
  \begin{cases}
    \re^{-(m_{\mu\mu}-\hat{m})^{2}/2\sigma^{2}}, & -\alpha_{\mathrm{L}} < \frac{m_{\mu\mu}-\hat{m}}{\sigma} < \alpha_{\mathrm{R}}\\
    \Big(\frac{n_{\mathrm{L}}}{\abs{\alpha_{\mathrm{L}}}}\Big)^{n_{\mathrm{L}}} \re^{-\alpha_{\mathrm{L}}^{2}/2} \Big(\frac{n_{\mathrm{L}}}{\abs{\alpha_{\mathrm{L}}}}-\abs{\alpha_{\mathrm{L}}}-\frac{m_{\mu\mu}-\hat{m}}{\sigma}\Big)^{-n_{\mathrm{L}}}, & \frac{m_{\mu\mu}-\hat{m}}{\sigma} \leq -\alpha_{\mathrm{L}}\\
    \Big(\frac{n_{\mathrm{R}}}{\abs{\alpha_{\mathrm{R}}}}\Big)^{n_{\mathrm{R}}} \re^{-\alpha_{\mathrm{R}}^{2}/2} \Big(\frac{n_{\mathrm{R}}}{\abs{\alpha_{\mathrm{R}}}}-\abs{\alpha_{\mathrm{R}}}+\frac{m_{\mu\mu}-\hat{m}}{\sigma}\Big)^{-n_{\mathrm{R}}}, & \frac{m_{\mu\mu}-\hat{m}}{\sigma} \geq \alpha_{\mathrm{R}} \\
  \end{cases}.
\end{aligned}
\end{equation}
\end{linenomath}
The core of the DCB function consists of a Gaussian distribution of mean $\hat{m}$ and standard deviation $\sigma$, while the tails on either side are modelled by a power-law function with parameters $\alpha_{\mathrm{L}}$ and $n_{\mathrm{L}}$ (low-mass tail), and $\alpha_{\mathrm{R}}$ and $n_{\mathrm{R}}$ (high-mass tail). The total expected background is modelled with a modified form of the Breit--Wigner function (mBW)~\cite{Sirunyan:2018hbu},
\begin{linenomath}
  \begin{equation}\label{eqn:BWZRedux}
    \mathrm{mBW}(m_{\mu\mu}; m_{\PZ}, \Gamma_{\PZ}, a_{1}, a_{2}, a_{3}) = \frac{\re^{a_{2} m_{\mu\mu} + a_{3} m_{\mu\mu}^{2}}}{(m_{\mu\mu}-m_{\mathrm{\PZ}})^{a_{1}}+(\Gamma_{\mathrm{\PZ}}/2)^{a_{1}}},
  \end{equation}
\end{linenomath}
 where the parameters $m_{\PZ}$ and $\Gamma_{\PZ}$ are fixed to the measured $\PZ$ boson mass of 91.19\GeV and width 2.49\GeV~\cite{pdg2020}, and the parameters $a_1$, $a_2$, and $a_3$ are free to float. A first boundary is selected by optimizing the total expected significance against all possible boundaries defined in quantiles of signal efficiency. This strategy accounts for the slight differences in the BDT shapes among data-taking periods for both signal and background processes. This process is repeated recursively to define additional subcategory boundaries until the further gain in the expected significance is less than 1\%. The optimized event categories are labelled as ``${\Pg\Pg\PH\textrm{-cat1}}$'', ``${\Pg\Pg\PH\textrm{-cat2}}$'', ``${\Pg\Pg\PH\textrm{-cat3}}$'', ``${\Pg\Pg\PH\textrm{-cat4}}$'', and ``${\Pg\Pg\PH\textrm{-cat5}}$'' corresponding to signal efficiency quantiles of 0--30, 30--60, 60--80, 80--95, and $>$95\%, respectively. The grey vertical bands in Figure~\ref{fig:ggH_bdt_output} (\cmsLeft) indicate the small range of variation, among the data-taking years, of the BDT boundaries for the optimized event categories described above.

A simultaneous binned maximum-likelihood fit to the observed $m_{\mu\mu}$ distributions is performed over the mass range 110--150\GeV to extract the \hmm signal. A bin size of 50\MeV is chosen for the $m_{\mu\mu}$ distributions, which is about one order of magnitude smaller than the expected resolution of the signal peak. In each event category, simulated signal distributions from the different production modes ($\Pg\Pg\PH$, VBF, $\PW\PH$, $\PZ\PH$, and $\ttbar\PH$) are modelled independently with DCB functions, and the best fit values of the DCB tail parameters are treated as constants in the final fit to the data. The $\hat{m}$ and $\sigma$ parameters of the DCB function represent the peak position and resolution of the Higgs boson resonance, respectively. These are the only signal shape parameters allowed to vary in the fit. Their predicted values from simulation are constrained by Gaussian priors with widths corresponding to the muon momentum scale (up to 0.2\%) and resolution uncertainties (up to 10\%) in each event category. Figure~\ref{fig:ggH_bdt_output} (\cmsRight) shows the total signal model for ${\mh=125\GeV}$ obtained by summing the contributions from the different production modes in the best and the worst resolution subcategories of the $\Pg\Pg\PH$ category, ${\Pg\Pg\PH\textrm{-cat4}}$ and ${\Pg\Pg\PH\textrm{-cat1}}$, where HWHM represents the half-width at half maximum of the signal peak. The category with the highest signal purity ($\Pg\Pg\PH$-cat5) uses particular kinematic features ($\pt^{\mu\mu}$, $\Delta \eta$ and $\Delta \phi$ between the dimuon system and jets) to isolate the signal, while $\Pg\Pg\PH$-cat4 relies more heavily on the $m_{\mu\mu}$ resolution itself. Therefore, the mass resolution for signal events in $\Pg\Pg\PH$-cat4 is expected to be about 2\% better than in $\Pg\Pg\PH$-cat5.

The theoretical and experimental sources of systematic uncertainties affecting the expected signal rate in each event category are similar to those described in the VBF analysis. Experimental uncertainties in the measurement of the muon selection efficiencies (0.5--1\% per event category), jet energy scale (1--4\% per event category) and resolution (1--6\% per event category), the modelling of the pileup conditions (0.3--0.8\% per event category), the integrated luminosity, and the efficiency for vetoing $\cPqb$ quark jets (0.1--0.5\% per event category) are considered. Theoretical uncertainties in the prediction of the Higgs boson production cross section, decay rate, and acceptance are also included, corresponding to a total uncertainty in the $\Pg\Pg\PH$ yield ranging from 6--12\% depending on the event category. Rate uncertainties are modelled in the signal extraction as nuisance parameters acting on the relative signal yield with log-normal constraints.

The background contribution in each subcategory is modelled with parametric functions. No prior knowledge of the shape parameters of these functions or the yield of the total background is assumed. These parameters are therefore constrained directly by the observed data in the {S+B} fit. Since the background composition expected from simulation is very similar across subcategories and largely dominated by the DY process, the background shape in $m_{\mu\mu}$ is similar in all event categories. There are, however, variations in the overall slope of the $m_{\mu\mu}$ spectrum across the BDT score categories. The function describing the background in each event category is therefore defined as the product of a ``core'' shape that is common among all event categories, with parameters correlated across categories, and a Chebyshev polynomial term (shape modifier) specific to each event category that modulates the core shape. This background modelling approach is referred to as the ``core-pdf method''. The core background shape is obtained from an envelope of three distinct functions: the mBW defined in Eq.~(\ref{eqn:BWZRedux}), a sum of two exponentials, and the product of a nonanalytical shape derived from the \FEWZ~v3.1 generator~\cite{Li:2012wna} and a third-order Bernstein polynomial. Each of these functions contains three freely floating shape parameters. The nonanalytical shape derived from the \FEWZ generator is obtained by simulating DY events at NNLO precision in QCD and NLO accuracy in EW theory and interpolating the resulting $m_{\mu\mu}$ distribution using a spline function~\cite{Bourilkov:2016qum,Bourilkov:2016oet}. In a given subcategory, each of the three core functions is modulated by either a third- (${\Pg\Pg\PH\textrm{-cat1}}$ and ${\Pg\Pg\PH\textrm{-cat2}}$) or a second-order polynomial, with parameters uncorrelated across event categories. A discrete profiling method~\cite{Dauncey:2014xga} is employed, which treats the choice of the core function used to model the background as a discrete nuisance parameter in the signal extraction. 

The following strategy is adopted to estimate the uncertainty in the measured signal due to the choice of parametric function for the background model. In each event category, background-only fits to the data are performed using different types of functions: the mBW, a sum of two exponentials, a sum of two power-law functions, a Bernstein polynomial, the product between the nonanalytical shape described above and a Bernstein polynomial, the product between the ``BWZ'' function, defined as
\begin{linenomath}
\begin{equation}\label{eqn:BWZ}
  \mathrm{BWZ}(m_{\mu\mu};a,m_{\PZ},\Gamma_{\PZ}) = \frac{\Gamma_{\PZ}\re^{am_{\mu\mu}}}{(m_{\mu\mu}-m_{\PZ})^{2}+(\Gamma_{\PZ}/2)^{2}},
\end{equation}
and a Bernstein polynomial, and the ``$\mathrm{BWZ}\gamma$'' function~\cite{Khachatryan:2014aep}
\begin{equation}\label{eqn:BWZGamma}
  \mathrm{BWZ}\gamma(m_{\mu\mu};a,f,m_{\PZ},\Gamma_{\PZ}) = f\,\mathrm{BWZ}(m_{\mu\mu};a,m_{\PZ},\Gamma_{\PZ})+(1-f)\,\frac{\re^{am_{\mu\mu}}}{m_{\mu\mu}^2}.
\end{equation}
\end{linenomath}
The $\mathrm{BWZ}\gamma$ function is the sum of a Breit--Wigner function and a $1/m_{\mu\mu}^2$ term, which are used to model the $\PZ$ boson and the photon contributions to the $m_{\mu\mu}$ spectrum in DY events, respectively. Both terms are multiplied by an exponential function to approximate the effect of the PDFs. The BWZ function is a Breit--Wigner distribution with an exponential tail. For the functions including Bernstein polynomials, a Fisher test~\cite{10.2307/2340521} is used to determine the maximum degree of the polynomials to be considered in the fit. The chosen functional forms fit the data with a $\chi^{2}$ probability larger than 5\% in all event categories.

Pseudodata sets are generated across all event categories from the post-fit background shapes obtained for each type of function in each subcategory, taking into account the uncertainties in the fit parameters as well as their correlations, and injecting a given number of signal events. Signal-plus-background fits are performed on the pseudodata sets using the core-pdf method. The median difference between the measured and injected signal yields, relative to the post-fit uncertainty in the signal yields, gives an estimate of the bias due to the choice of the background model. The bias measured in each BDT category, as well as from pseudodata sets in which the signal is injected simultaneously in all event categories, is smaller than 20\% of the post-fit uncertainty on the signal yield. Including these observed deviations as spurious signals leads to a change in the overall uncertainty in the measured signal rate of less than 1\% and is therefore neglected. The core-pdf method employed in this analysis yields an improvement in sensitivity of about 10\% with respect to the background functions used in the previous result~\cite{Sirunyan:2018hbu}. It also ensures a negligible bias in the measured signal with significantly fewer total degrees of freedom in the signal extraction fit.

\begin{figure*}[!]
  \centering
  \includegraphics[width=0.42\textwidth]{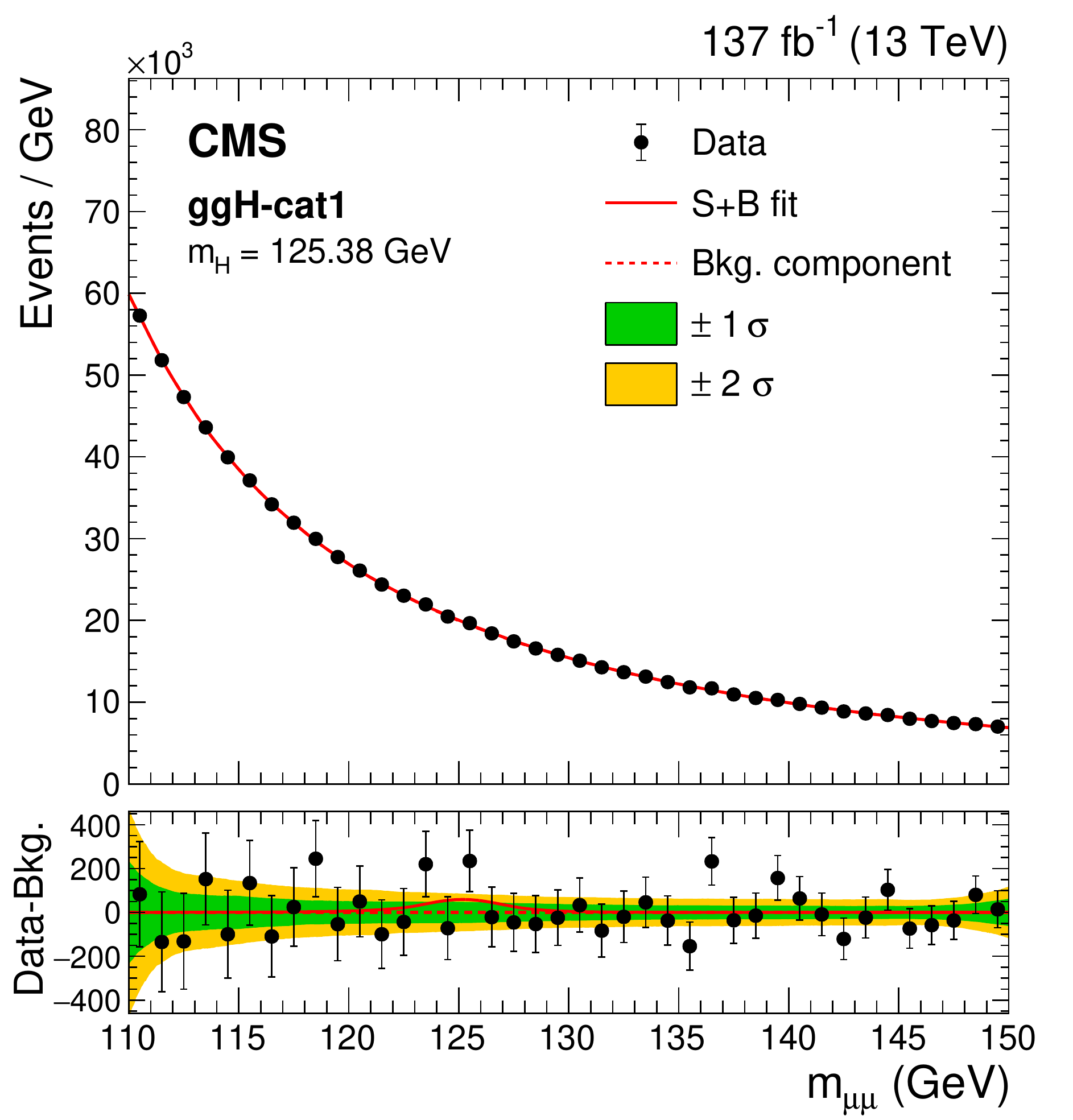}
  \includegraphics[width=0.42\textwidth]{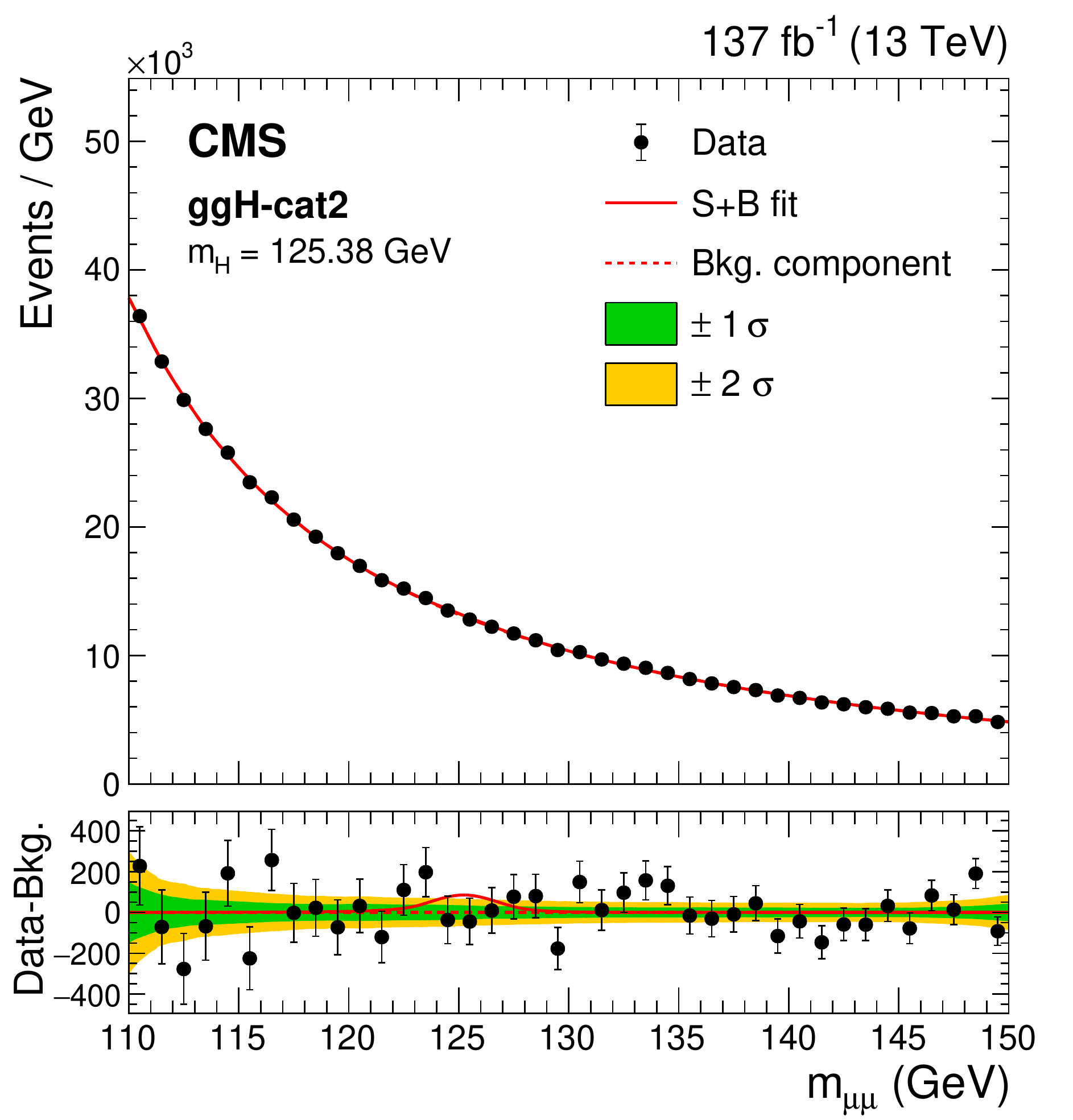}\\
  \includegraphics[width=0.42\textwidth]{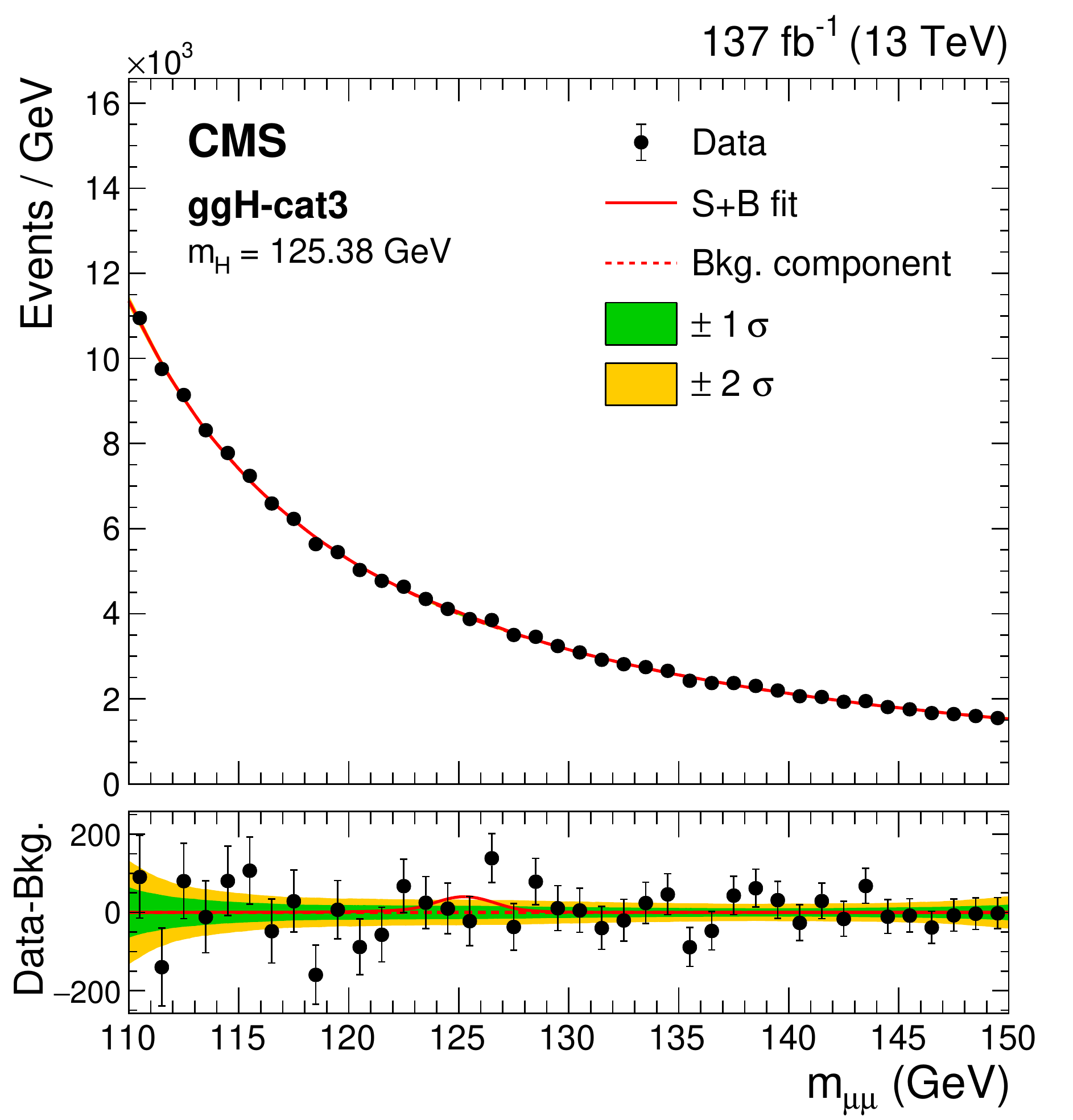}
  \includegraphics[width=0.42\textwidth]{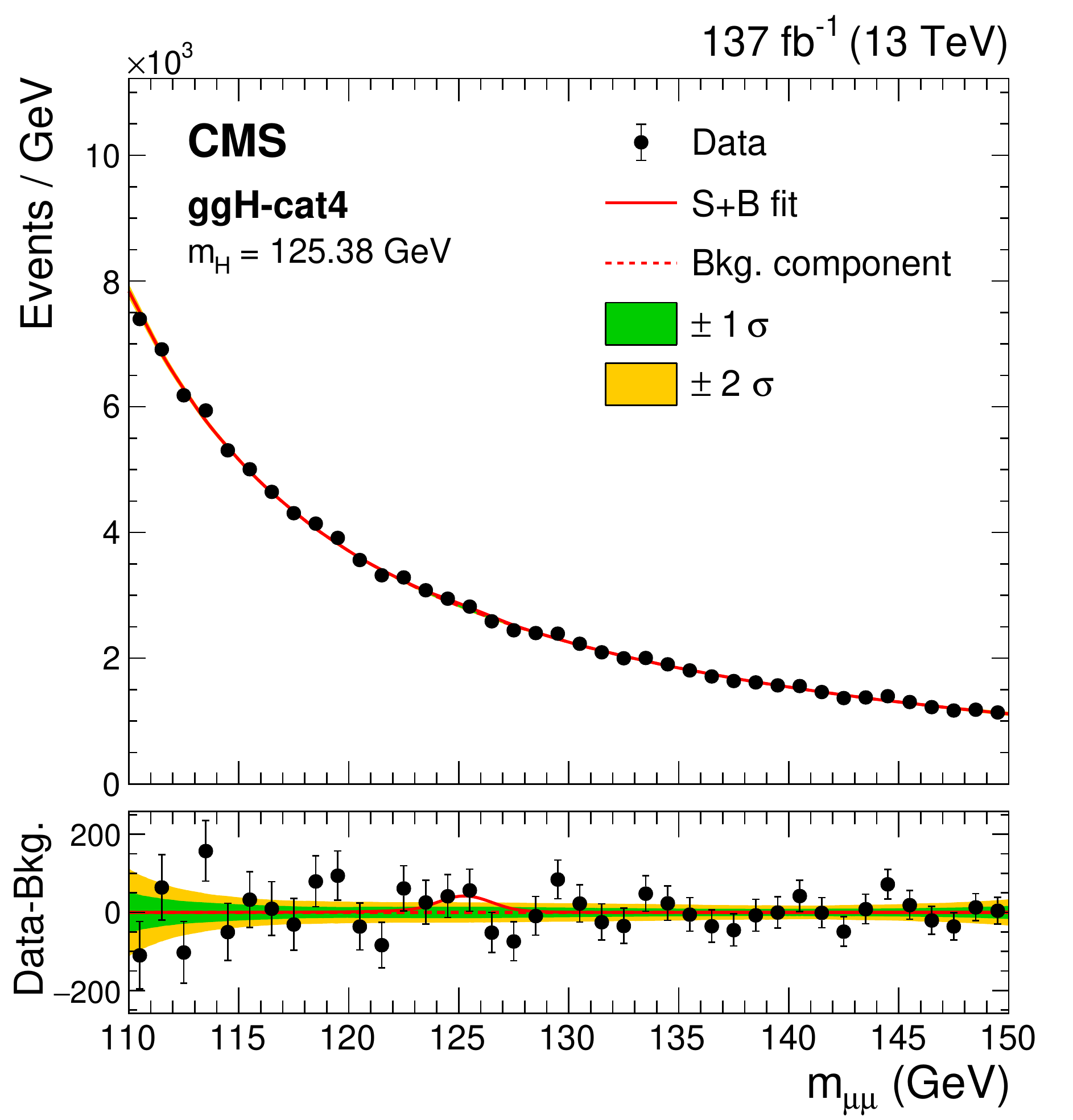}\\
  \includegraphics[width=0.42\textwidth]{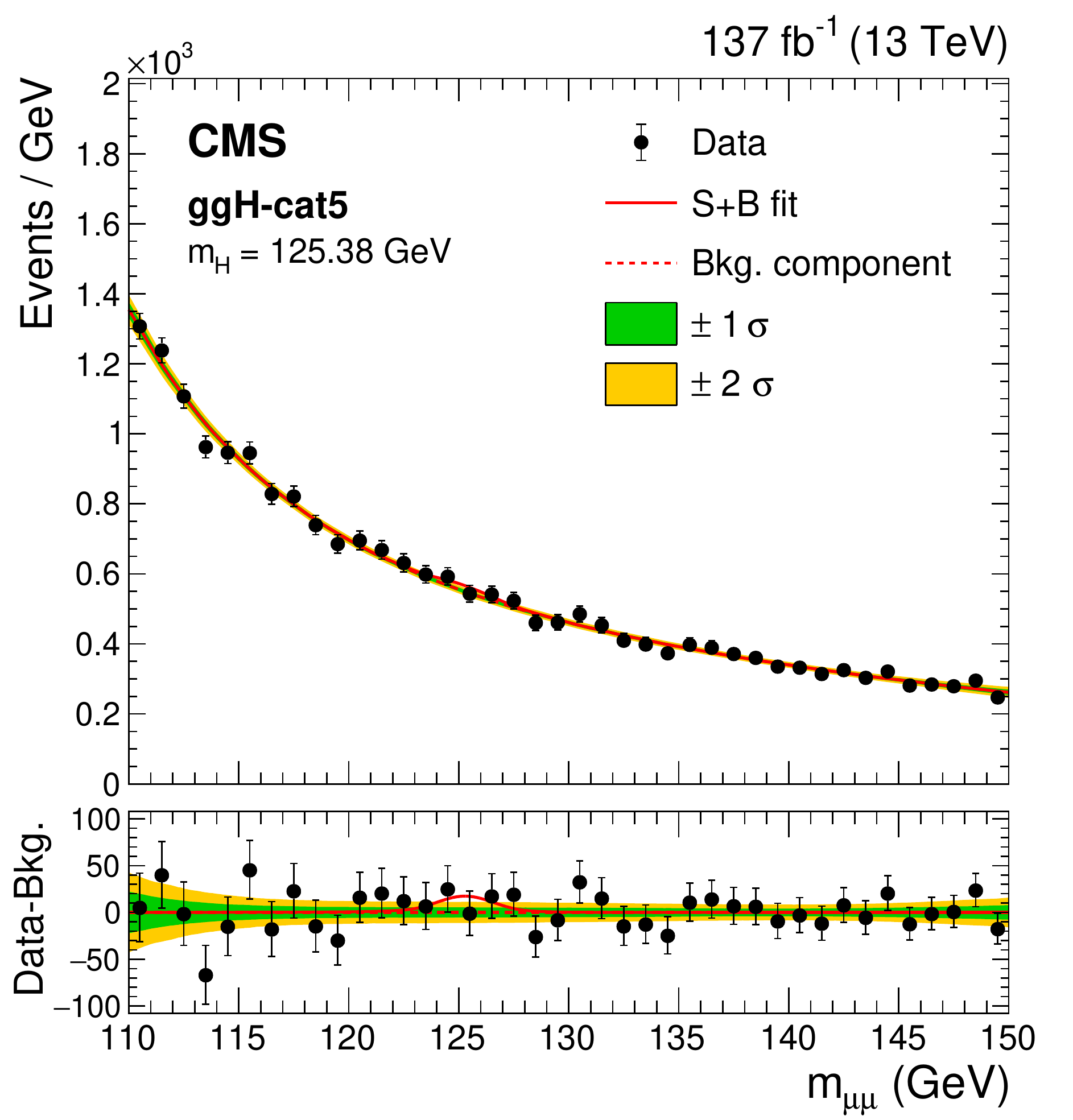}
  \caption{Comparison between the data and the total background extracted from a {S+B} fit performed across the various $\Pg\Pg\PH$ subcategories. The one (green) and two (yellow) standard deviation bands include the uncertainties in the background component of the fit. The lower panel shows the residuals after background subtraction and the red line indicates the signal with ${\mh=125.38\GeV}$ extracted from the fit.}
  \label{fig:ggh_fit_mass}
\end{figure*}

Figure~\ref{fig:ggh_fit_mass} shows the $m_{\mu\mu}$ distributions in each of the $\Pg\Pg\PH$ subcategories, in which the signal is extracted by performing a binned maximum-likelihood fit using a DCB function to model the signal contribution, while the background is estimated with the core-pdf method. Table~\ref{tab:ggh_categories} reports the total number of expected signal events (S), the signal composition in each $\Pg\Pg\PH$ category, and the HWHM of the expected signal shape. In addition, the estimated number of background events (B), the observation in data, the {S/(S+B)}, and the ${\mathrm{S}/\sqrt{\smash[b]{\mathrm{B}}}}$ ratios computed within the HWHM range around the signal peak are listed.

\begin{table*}[!htb]
  \centering
  \topcaption{The total expected number of signal events with ${\mh=125.38\GeV}$ (S), the ratio of the expected contributions from different production modes to the total signal yield (``Other'' represents the sum of $\PV\PH$, $\ttbar\PH$, and $\bbbar\PH$ contributions), the HWHM of the signal peak, the estimated number of background events (B) and the observation in data within ${\pm\,\mathrm{HWHM}}$, and the {S/(S+B)} and the ${\mathrm{S}/\sqrt{\smash[b]{\mathrm{B}}}}$ ratios within  ${\pm\,\mathrm{HWHM}}$, for each of the optimized $\Pg\Pg\PH$ event categories.}
  \cmsTable{
    \begin{tabular}{l c c c c c c c c c c}
      \hline
      Event           & Total  & $\Pg\Pg\PH$  & VBF      & Other   & HWHM        & Bkg.   & Data    & {S/(S+B)} (\%)  & ${\mathrm{S}/\sqrt{\smash[b]{\mathrm{B}}}}$  \\
      category        & signal & (\%)         & (\%)     & (\%)    & ({\GeVns})  & @HWHM  & @HWHM   & @HWHM           & @HWHM   \\
      \hline
      gg\PH-cat1      & 268    & 93.7         & 2.9      & 3.4     & 2.12        & 86\,360   & 86\,632   & 0.20   & 0.60  \\
      gg\PH-cat2      & 312    & 93.5         & 3.4      & 3.1     & 1.75        & 46\,350   & 46\,393   & 0.46   & 0.98  \\
      gg\PH-cat3      & 131    & 93.2         & 4.0      & 2.8     & 1.60        & 12\,660   & 12\,738   & 0.70   & 0.80  \\
      gg\PH-cat4      & 126    & 91.5         & 5.5      & 3.0     & 1.47        & 8260      & 8377      & 1.03   & 0.96  \\
      gg\PH-cat5      & 53.8   & 83.5         & 14.3     & 2.2     & 1.50        & 1680      & 1711      & 2.16   & 0.91  \\
      \hline
    \end{tabular}
  }
  \label{tab:ggh_categories}
\end{table*}

\section{The \texorpdfstring{$\ttbar\PH$}{ttH} production category}\label{sec:tth_category}

The $\ttbar\PH$ process has the smallest cross section among the targeted Higgs boson production modes at the LHC. However, the presence of a top quark-antiquark pair in addition to the Higgs boson helps to reduce the background to a level that is comparable to the expected signal rate. The top quark decays predominantly into a $\cPqb$ quark and a $\PW$ boson~\cite{pdg2020}, therefore a sample of events enriched in $\ttbar\PH$ production is selected by requiring the presence of at least two jets passing the loose WP of the DeepCSV $\cPqb$-tagging algorithm, or at least one $\cPqb$-tagged jet passing the medium WP. This requirement suppresses background processes in which jets originate mainly from the hadronization of light-flavour quarks, such as DY and diboson production. This selection also ensures mutual exclusivity between the $\ttbar\PH$ category and the other production categories considered in this analysis.

In order to increase the signal selection efficiency in events with large hadronic activity, as expected for the $\ttbar\PH$ signal process, the isolation requirement on all muons described in Section~\ref{sec:event_selection} is relaxed to be less than 40\% of the muon \pt. In addition, the isolation cone size decreases dynamically with the muon \pt (${R = 0.2}$ for ${\pt < 50\GeV}$, ${R = 10/\pt}$ for ${50 < \pt < 200\GeV}$, and ${R = 0.05}$ for ${\pt>200\GeV}$), following the approach used in Ref.~\cite{Sirunyan:2018shy}. Electron candidates are required to have ${\pt > 20 \GeV}$, ${\abs{\eta} < 2.5}$, and to pass identification requirements imposed on the properties of the ECAL cluster associated with the electron track, as well as the consistency between the electron momentum measured by the inner silicon tracker and its ECAL energy deposit. Each electron is also required to be isolated following the same strategy as for muons, and the magnitude of the transverse and longitudinal impact parameters must be smaller than $0.05$ and $0.1$\unit{cm}, respectively. In order to suppress backgrounds containing nonprompt leptons produced in the decay of heavy quarks, muons and electrons are rejected when the jet with ${\pt > 15\GeV}$ that is nearest to the lepton in $\Delta R$ separation is $\cPqb$-tagged according to the DeepCSV medium WP. Furthermore, all muons and electrons in the $\ttbar\PH$ category are required to pass the medium WP of a multivariate lepton identification discriminant specifically designed to reject nonprompt leptons~\cite{Sirunyan:2018zgs}, resulting in a selection efficiency of about 95~(92)\% per prompt muon (electron).

The $\ttbar\PH$ signal events may contain additional charged leptons, depending on the decay of the top quarks. Events with one or two additional charged leptons in the final state are grouped in the $\ttbar\PH$ leptonic category. An event in the $\ttbar\PH$ leptonic category containing three~(four) charged leptons is further required to have the net sum of the lepton electric charges equal to one~(zero). In the case of events with more than one pair of oppositely charged muons with ${110 < m_{\mu\mu} < 150\GeV}$, the pair with the largest dimuon \pt is chosen as the Higgs boson candidate. The invariant mass of each pair of same-flavour, opposite-sign leptons is required to be greater than 12\GeV to suppress backgrounds arising from quarkonium decays. An event is vetoed if it contains a pair of oppositely charged electrons or muons with an invariant mass in the range 81--101\GeV, consistent with the decay of an on-shell $\PZ$ boson. In contrast, events with exactly two oppositely charged muons with ${110 < m_{\mu\mu} < 150\GeV}$, no identified electrons, and at least one combination of three jets in the final state with invariant mass ($m_{\mathrm{jjj}}$) between 100 and 300\GeV belong to the $\ttbar\PH$ hadronic category. Each jet must have ${\pt > 25\GeV}$ and ${\abs{\eta} < 4.7}$. A summary of the selection criteria used to define the $\ttbar\PH$ hadronic and leptonic categories is reported in Table~\ref{tab:tth_preselection}.

\begin{table*}[!htb]
  \centering
  \topcaption{Summary of the kinematic requirements used to define the $\ttbar\PH$ hadronic and leptonic production categories}
  \cmsTable{
    \begin{tabular}{lcc}
      \hline
      Observable & $\ttbar\PH$ hadronic & $\ttbar\PH$ leptonic\\
      \hline 
      Number of $\cPqb$ quark jets & \multicolumn{2}{c}{$>$0 medium or $>$1 loose $\cPqb$-tagged jets}\\
      Number of leptons ($\mathrm{N}(\ell=\mu,\Pe)$) & $=$2 & $=$3 or 4 \\
      Lepton charge     ($q(\ell)$) & $\sum q(\ell) =$0 & $\mathrm{N}(\ell)=3~(4) \to \sum q(\ell) = \pm1~(0)$\\
      Jet multiplicity  (${\pt>25\GeV}$, ${\abs{\eta}< 4.7}$) &  $\geq$3 & $\geq$2 \\      
      Leading jet \pt   &  $>$50\GeV & $>$35\GeV  \\
      $\PZ$ boson veto    & \NA & $\abs{m_{\ell\ell} - m_{\PZ}} > 10\GeV$ \\
      Low-mass resonance veto & \NA & $m_{\ell\ell} > 12\GeV$ \\
      Jet triplet mass  & $100 < m_{\mathrm{jjj}} < 300\GeV$ & \NA \\
      \hline
    \end{tabular}
  }
  \label{tab:tth_preselection}
\end{table*}

The dominant background in the $\ttbar\PH$ hadronic category comes from fully leptonic $\ttbar$ decays, while the main backgrounds in the $\ttbar\PH$ leptonic category are the $\ttbar\PZ$ and $\ttbar$ processes. In order to obtain an optimal discrimination between the $\ttbar\PH$ signal and the expected backgrounds, BDT-based multivariate discriminants are trained in both the hadronic and leptonic categories. The input variables are chosen to account for both the kinematic properties of the dimuon system and the properties of the top quark decay products, while ensuring that the BDT outputs remain uncorrelated with $m_{\mu\mu}$. A common set of observables is used as input to the two BDT discriminants. These include variables that characterize the production and decay of the Higgs boson candidate, namely the $\pt^{\mu\mu}$, $y_{\mu\mu}$, $\phi_{\mathrm{CS}}$, and $\cos\theta_{\mathrm{CS}}$. In addition, the $\eta$ of each of the two muons and the ratio of each muon's \pt to $m_{\mu\mu}$ are also considered. To account for the large hadronic activity in $\ttbar\PH$ signal events, the \pt and $\eta$ of the three leading jets, the maximum DeepCSV value of jets not overlapping with charged leptons ($\Delta R(\ell,j) > 0.4$), the number of jets, and the scalar (vectorial) \pt sum \HT ($\abs{\htvecmiss}$) of all identified leptons and jets (${\pt > 25\GeV,~\abs{\eta}<2.5}$) are included. The \ptmiss is also considered along with the $\Delta\zeta$ variable~\cite{Aad:2012cfr}, which is defined as the projection of the \ptvecmiss on the bisector of the dimuon system in the transverse plane. Signal events are weighted during the BDT training with the inverse of the per-event mass resolution, following the same approach used in the $\Pg\Pg\PH$ categories.

In the $\ttbar\PH$ leptonic category, several additional variables are used in the BDT discriminant that target the kinematic properties of a leptonic top quark decay. These include the azimuthal separation between the Higgs boson candidate and the highest \pt additional charged lepton ($\ell_{\cPqt}$), the invariant mass formed by $\ell_{\cPqt}$ and the jet with the highest DeepCSV score, the transverse mass formed by $\ell_{\cPqt}$ and \ptvecmiss in the event, and the flavour of $\ell_{\cPqt}$. In the $\ttbar\PH$ hadronic category, the resolved hadronic top tagger (RHTT), which combines a kinematic fit and a BDT-based multivariate discriminant, is used to identify top quark decays to three resolved jets following a similar approach to the one reported in Ref.~\cite{Sirunyan:2018dub}. The jet triplet with ${100 < m_{\mathrm{jjj}} < 300\GeV}$ and the highest RHTT score is selected as a hadronic top quark candidate. The corresponding RHTT score is used as input to the BDT discriminant. Furthermore, the \pt of the top quark candidate and the \pt balance of the top quark and the muon pair are also considered.

\begin{figure*}[!htb]
  \centering
  \includegraphics[width=0.45\textwidth]{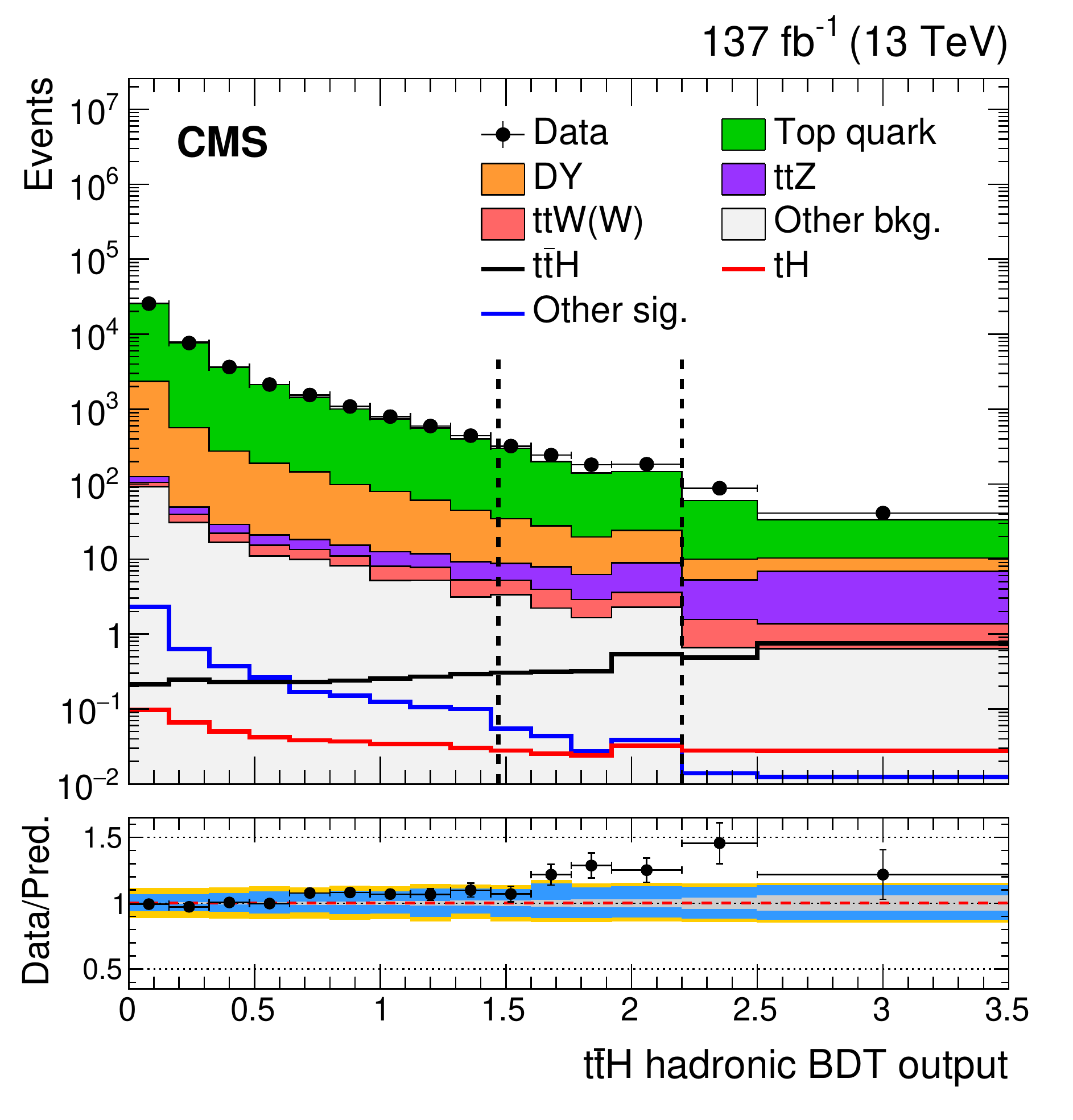}
  \includegraphics[width=0.45\textwidth]{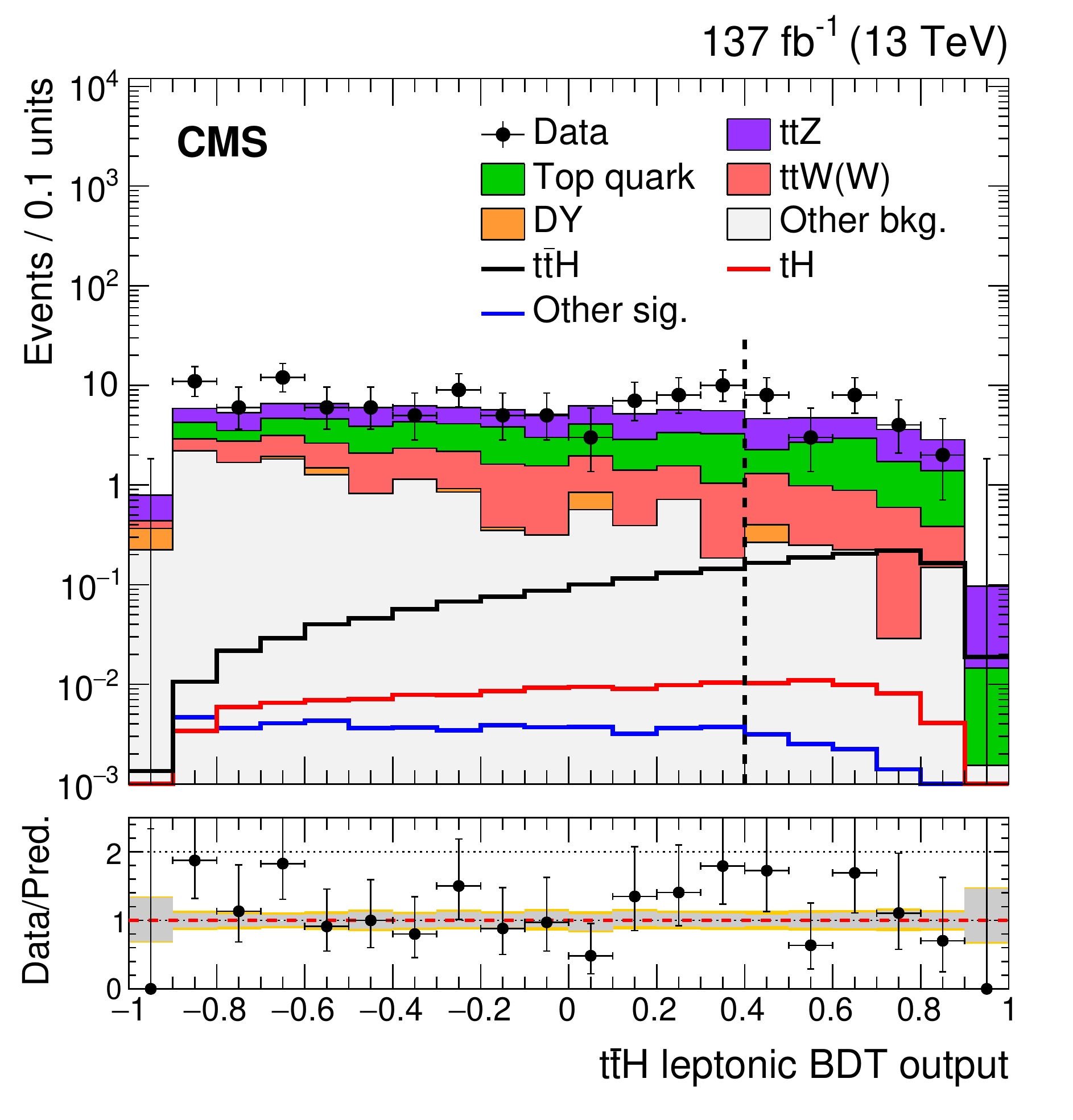}
  \caption{The observed BDT output distribution in the $\ttbar\PH$ hadronic (\cmsLeft) and leptonic (\cmsRight) categories compared to the prediction from the simulation of various SM background processes. Signal distributions expected from different production modes of the Higgs boson with ${\mh = 125\GeV}$ are overlaid. The dashed vertical lines indicate the boundaries of the optimized event categories. The description of the ratio panels is the same as in Fig.~\ref{fig:ggH_bdt_output}.}
  \label{fig:tth_bdt_output}
\end{figure*}

Figure~\ref{fig:tth_bdt_output} shows the output of the BDT discriminant in the $\ttbar\PH$ hadronic (\cmsLeft) and leptonic (\cmsRight) categories. The high BDT score region of the $\ttbar\PH$ hadronic category is enriched in events with large jet multiplicity, where the $\ttbar$ and DY background predictions rely on a significant number of jets from the PS and are known to not entirely reproduce the data~\cite{Sirunyan:2018ucr}. The signal prediction, however, relies largely on jets derived from the ME calculation. Since the background prediction is extracted from the data, the observed differences between data and background simulation do not affect the fit result. Based on the BDT output, events in the $\ttbar\PH$ leptonic category are further divided into two subcategories, labelled as ``${\ttbar\PH\textrm{lep-cat1}}$'' and ``${\ttbar\PH\textrm{lep-cat2}}$'', corresponding to signal efficiency quantiles of 0--52 and $>$52\%, respectively. Similarly, events in the $\ttbar\PH$ hadronic category are divided into three subcategories labelled ``${\ttbar\PH\textrm{had-cat1}}$'', ``${\ttbar\PH\textrm{had-cat2}}$'', and ``${\ttbar\PH\textrm{had-cat3}}$'', corresponding to signal efficiency quantiles of 0--70, 70--86, and $>$86\%, respectively. The BDT score boundaries of these event categories, indicated in Fig.~\ref{fig:tth_bdt_output} by black dashed vertical lines, are optimized following the same strategy adopted for events in the $\Pg\Pg\PH$ category. In the optimization, exponential functions are used to model the background in both the ttH hadronic and leptonic subcategories.

\begin{figure*}[!]
  \centering
  \includegraphics[width=0.42\textwidth]{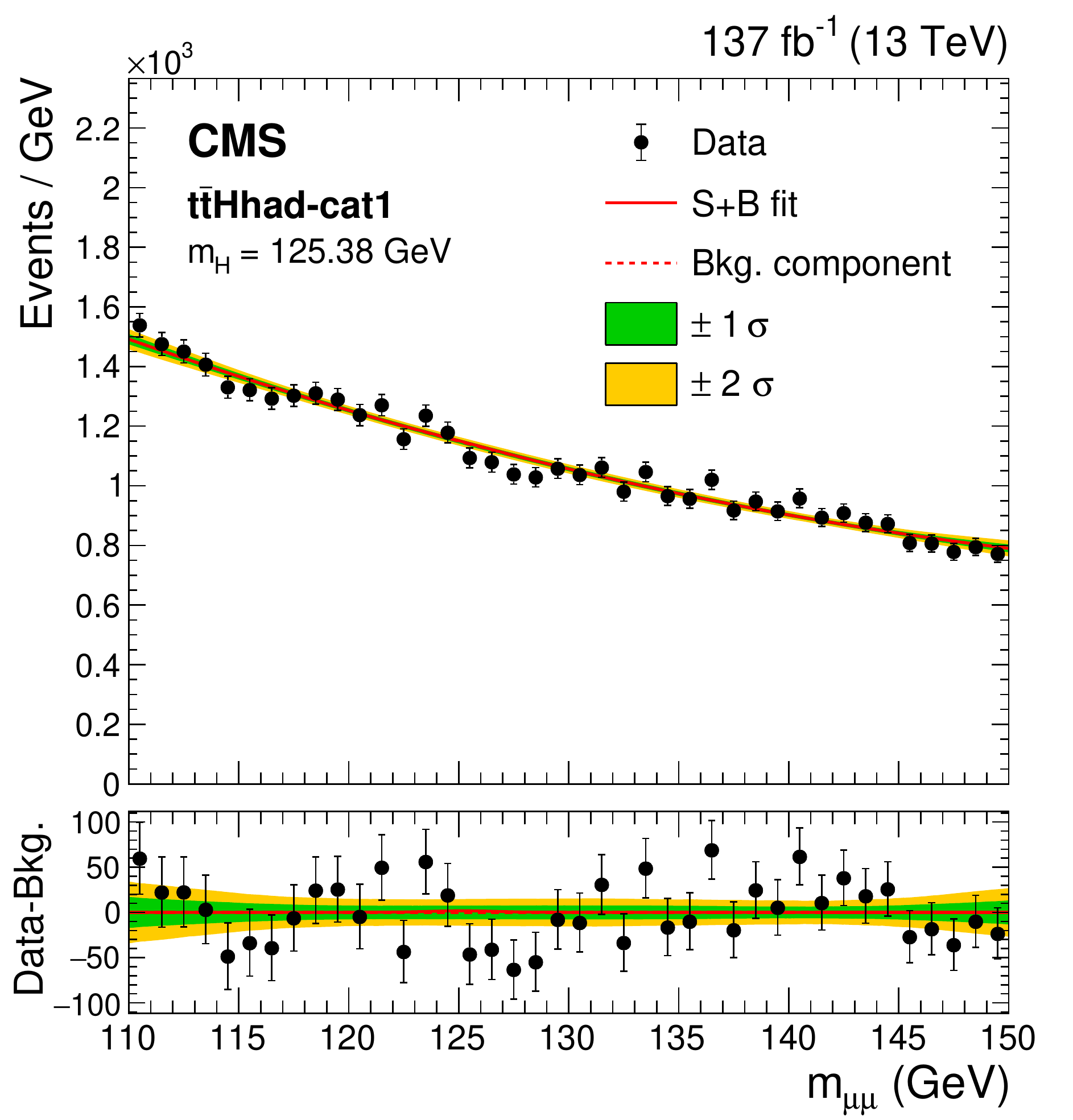}
  \includegraphics[width=0.42\textwidth]{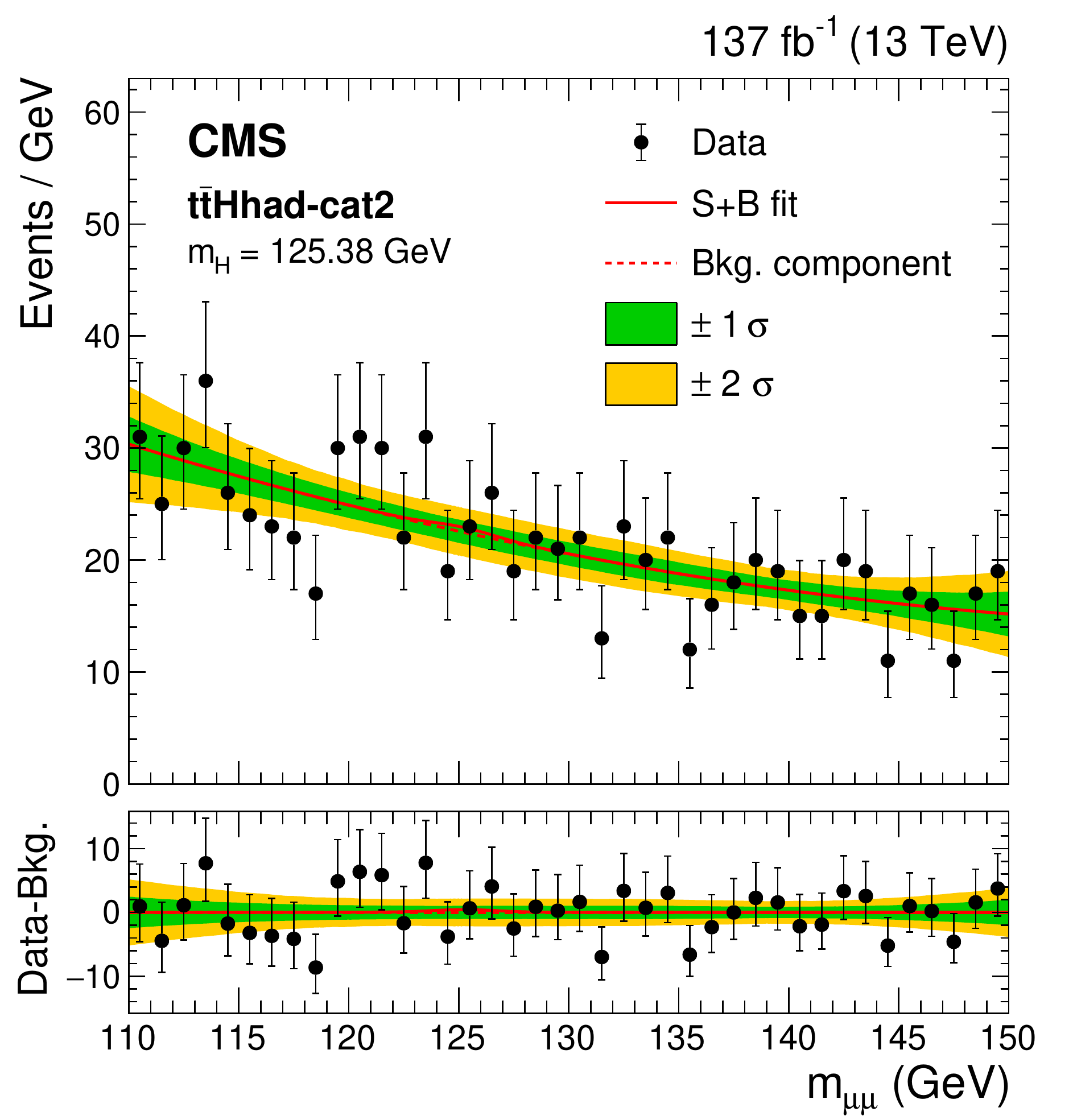}\\
  \includegraphics[width=0.42\textwidth]{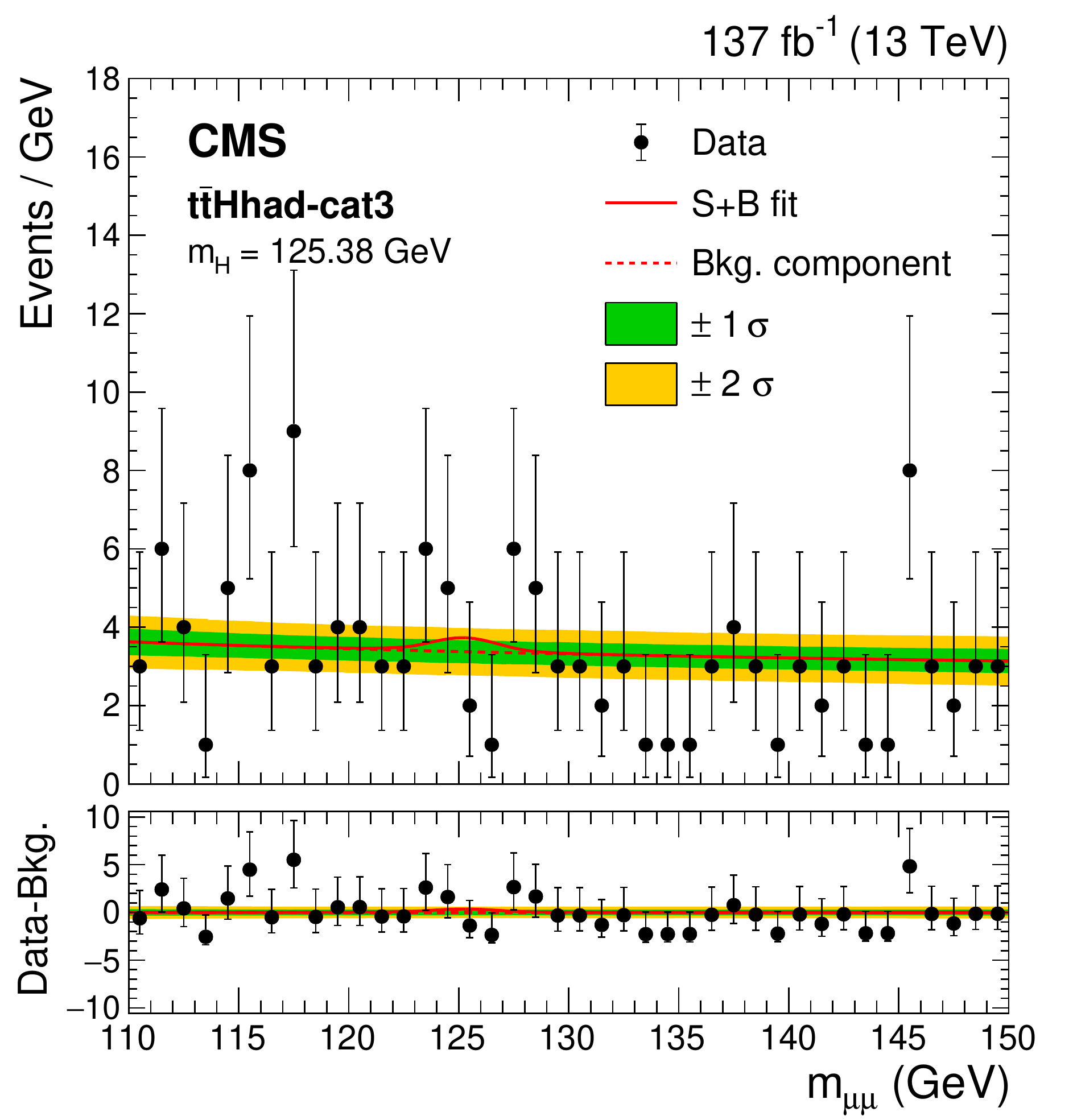}
  \includegraphics[width=0.42\textwidth]{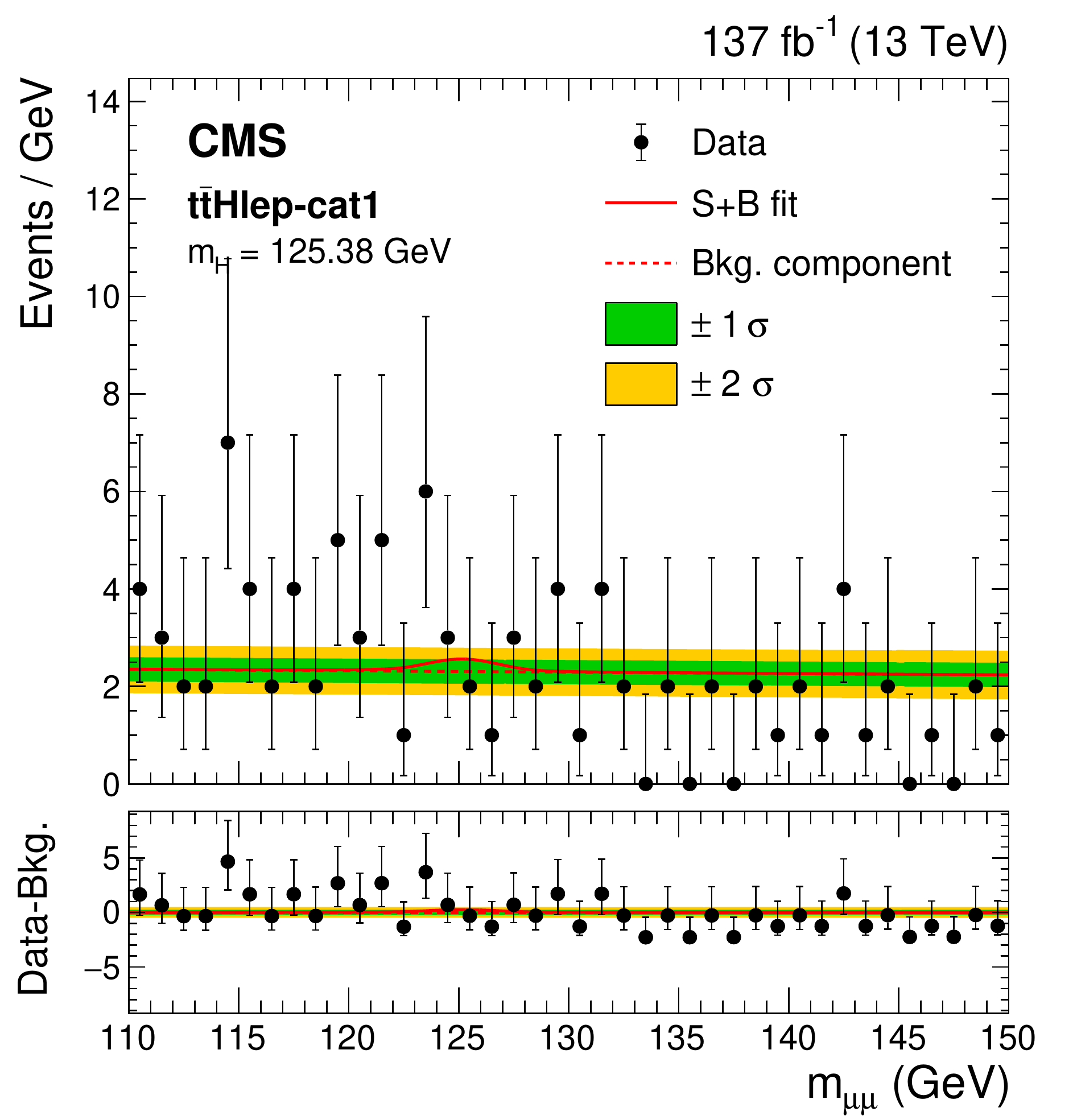}\\
  \includegraphics[width=0.42\textwidth]{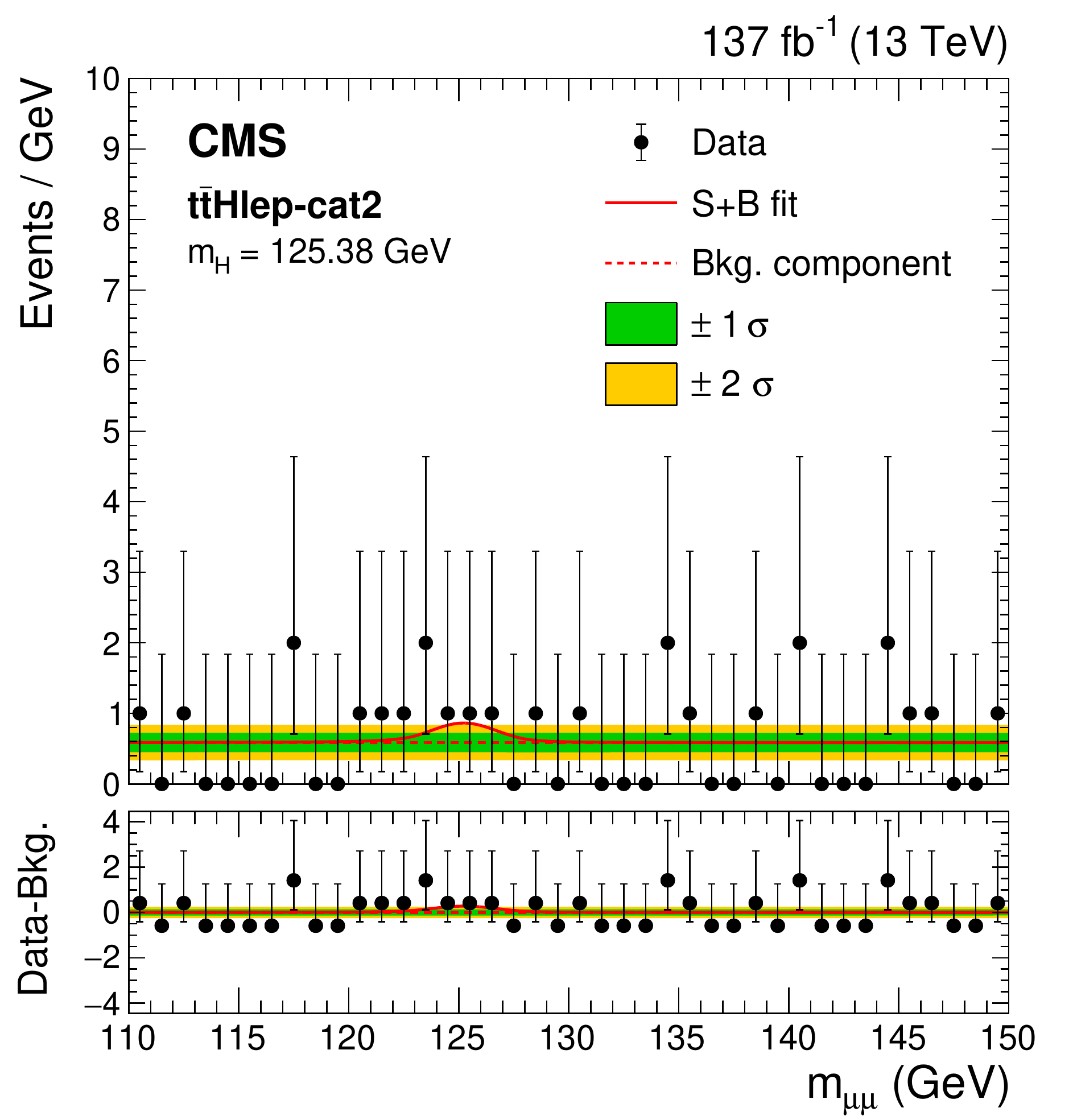}
  \caption{Comparison between the data and the total background extracted from a {S+B} fit performed across the various $\ttbar\PH$ hadronic and leptonic event subcategories. The one (green) and two (yellow) standard deviation bands include the uncertainties in the background component of the fit. The lower panel shows the residuals after the background subtraction, where the red line indicates the signal with ${\mh=125.38\GeV}$ extracted from the fit.}
  \label{fig:tth_fit_mass}
\end{figure*}

Figure~\ref{fig:tth_fit_mass} shows the $m_{\mu\mu}$ distributions in the $\ttbar\PH$ hadronic and leptonic event categories. The signal is extracted by performing a binned maximum-likelihood fit to these $m_{\mu\mu}$ distributions (bin size of 50\MeV), where signal is modelled using the DCB function and the background is modelled using a second-order Bernstein polynomial (Bern(2)) in ${\ttbar\PH\textrm{had-cat1}}$ and ${\ttbar\PH\textrm{had-cat2}}$, a sum of two exponentials (S-Exp) in ${\ttbar\PH\textrm{had-cat3}}$, and a single exponential (Exp) in the $\ttbar\PH$ leptonic event categories. Table~\ref{tab:tth_categories} reports the expected signal composition of each $\ttbar\PH$ subcategory, along with the HWHM of the expected signal shape. In addition, the estimated number of background events, the observation in data, and the {S/(S+B)} and $\mathrm{S}/\sqrt{\smash[b]{\mathrm{B}}}$ ratios within the HWHM of the signal shape are shown.

\begin{table*}[!htb]
  \centering
  \topcaption{The total expected number of signal events with ${\mh=125.38\GeV}$ (S), the ratio of the expected contributions from different production modes to the total signal yield (``Other'' represents the sum of $\cPqt\PH$, VBF, and $\bbbar\PH$ contributions), the HWHM of the signal peak, the functional form used for the background modelling, the estimated number of background events (B) and the observed number of events within ${\pm\,\mathrm{HWHM}}$, and the {S/(S+B)} and $\mathrm{S}/\sqrt{\smash[b]{\mathrm{B}}}$ ratios computed within the HWHM of the signal peak, for each of the optimized event categories defined along the $\ttbar\PH$ hadronic and leptonic BDT outputs.}
  \cmsTable{
    \begin{tabular}{l c c c c c c c c c c c}
      \hline
      Event       & Total  & $\ttbar\PH$ & $\Pg\Pg\PH$  & $\PV\PH$  & Other   & HWHM       & Bkg. fit & Bkg.   & Data  & {S/(S+B)} (\%) & $\mathrm{S}/\sqrt{\smash[b]{\mathrm{B}}}$ \\
      category    & signal & (\%)        & (\%)         & (\%)      & (\%)    & ({\GeVns}) & function & @HWHM  & @HWHM & @HWHM  & @HWHM\\
      \hline
      $\ttbar\PH$had-cat1  & 6.87  & 32.3  & 40.3   & 17.2   & 10.2  & 1.85   & Bern(2) & 4298   & 4251 & 1.07   & 0.07   \\ 
      $\ttbar\PH$had-cat2  & 1.62  & 84.3  & 3.8    & 5.6    & 6.2   & 1.81   & Bern(2) & 82.0   & 89   & 1.32   & 0.12   \\ 
      $\ttbar\PH$had-cat3  & 1.33  & 94.0  & 0.3    & 1.3    & 4.4   & 1.80   & S-Exp   & 12.3   & 12   & 6.87   & 0.26   \\
      [\cmsTabSkip]
      $\ttbar\PH$lep-cat1  & 1.06  & 85.8  & \NA    & 4.7    & 9.5   & 1.92   & Exp     & 9.00   & 13   & 7.09   & 0.22   \\ 
      $\ttbar\PH$lep-cat2  & 0.99  & 94.7  & \NA    & 1.0    & 4.3   & 1.75   & Exp     & 2.08   & 4    & 24.5   & 0.47   \\ 
      \hline
    \end{tabular}
  }
  \label{tab:tth_categories}
\end{table*}

The systematic uncertainties considered account for possible mismodelling of the signal shape and rate. Uncertainties in the calibration of the muon momentum scale and resolution are propagated to the shape of the signal $m_{\mu\mu}$ distribution, yielding variations of up to 0.1\% in the peak position and up to 10\% in width. Experimental uncertainties from the measurement of the electron and muon selection efficiencies (0.5--1.5\% per event category), muon momentum scale and resolution (0.1--0.8\% per event category), jet energy scale and resolution (2--6\% per event category), efficiency of identifying $\cPqb$ quark jets (1--3\% per event category), integrated luminosity, and modelling of the pileup conditions (0.2--1\% per event category) affect the predicted signal rate. Furthermore, theoretical uncertainties in the prediction of the Higgs boson production cross sections, decay rate, and acceptance are also included, as already described for the $\Pg\Pg\PH$, VBF, and $\PV\PH$ analyses. Rate uncertainties are included in the signal extraction as nuisance parameters acting on the relative signal yield with log-normal constraints.

In order to estimate the potential bias arising from the choice of the parametric function used to model the background, alternative functions able to fit the data with a $\chi^{2}$ $p$-value larger than 5\% are considered. These include Bernstein polynomials, sum of exponentials, and sum of power laws. In each event category, background-only fits to the data are performed with each function listed above. From each of these fits, pseudodata sets are generated taking into account the uncertainties in the fit parameters and their correlations, and injecting a certain number of signal events. A {S+B} fit is then performed on these pseudodata sets using, in each category, the parametric functions listed above. The corresponding bias is observed to be smaller than 20\% of the post-fit uncertainty on the signal yield and is therefore neglected in the signal extraction. The chosen functions maximize the expected sensitivity to the 125\GeV Higgs boson.

\section{The \texorpdfstring{$\PV\PH$}{VH} production category}\label{sec:vh_category}

Events considered in the $\PV\PH$ category contain at least two muons passing the selection requirements listed in Section~\ref{sec:event_selection}. In order to ensure no overlap with the $\ttbar\PH$ category, events containing at least two $\cPqb$-tagged jets with ${\pt>25\GeV}$ and ${\abs{\eta}<2.5}$ passing the loose WP of the DeepCSV $\cPqb$-tagging algorithm, or at least one jet passing the medium WP, are discarded. Events are also required to have at least one additional charged lepton (electron or muon), which is expected from the leptonic decay of the $\PW$ or $\PZ$ boson. The additional muons (electrons) must have ${\pt > 20 \GeV}$, ${\abs{\eta} < 2.4~(2.5)}$, and pass certain isolation and identification requirements with an average efficiency of 95~(90)\%. Furthermore, all muons and electrons in this category are required to pass the medium WP of a multivariate discriminant developed in Ref.~\cite{Sirunyan:2018zgs} to identify and suppress nonprompt leptons, with a selection efficiency of about 95~(92)\% per prompt muon (electron).

Events containing exactly one additional charged lepton belong to the $\PW\PH$ category, which targets signal events where the Higgs boson is produced in association with a leptonically decaying $\PW$ boson. If the additional lepton is a muon, the two pairs of oppositely charged muons are required to have ${m_{\mu\mu} > 12\GeV}$ to suppress background events from quarkonium decays. Moreover, neither of the two oppositely charged muon pairs can have an invariant mass consistent with $m_{\PZ}$ within 10\GeV. Finally, at least one of these two muon pairs must have $m_{\mu\mu}$ in the range 110--150\GeV. If both $m_{\mu\mu}$ pairs satisfy this criterion, the pair with the highest $\pt^{\mu\mu}$ is considered as the Higgs boson candidate. If the additional lepton is an electron, the only requirement imposed is that ${110 < m_{\mu\mu} < 150\GeV}$. 

The $\PZ\PH$ category targets signal events where the Higgs boson is produced in association with a $\PZ$ boson that decays to a pair of electrons or muons. Events in the $\PZ\PH$ category are therefore required to contain four charged leptons, with a combined lepton number and electric charge of zero. As in the $\PW\PH$ category, the invariant mass of each pair of same-flavour, opposite-sign leptons is required to be greater than 12\GeV. An event is rejected if it does not contain exactly one pair of same-flavour, opposite-sign leptons with invariant mass compatible with the $\PZ$ boson within 10~(20)\GeV for muon (electron) pairs. In addition, each event must contain one oppositely charged muon pair satisfying ${110 < m_{\mu\mu} < 150\GeV}$. For events with four muons, the muon pair with $m_{\mu\mu}$ closer to $m_{\PZ}$ is chosen as the $\PZ$ boson candidate, while the other muon pair is selected as the Higgs boson candidate. A summary of the selection criteria applied in the $\PW\PH$ and $\PZ\PH$ production categories is reported in Table~\ref{tab:vh_preselection}.

\begin{table*}[!htb]
  \centering
  \topcaption{Summary of the kinematic selection used to define the $\PW\PH$ and $\PZ\PH$ production categories.}
  \begin{tabular}{lcccc}
    \hline
    Observable & \multicolumn{2}{c}{$\PW\PH$ leptonic} & \multicolumn{2}{c}{$\PZ\PH$ leptonic}\\
    & $\mu\mu\mu$ & $\mu\mu\Pe$ & $4\mu$ & $2\mu2\Pe$\\
    \hline
    Number of loose (medium) $\cPqb$-tagged jets & $\leq$1~(0) &$\leq$1~(0) & $\leq$1~(0) & $\leq$1~(0) \\
    Number of selected muons & $=$3  & $=$2 & $=$4 & $=$2 \\
    Number of selected electrons & $=$0  & $=$1 & $=$0 & $=$2 \\
    Lepton charge ($q(\ell)$)    & \multicolumn{2}{c}{$\sum q(\ell) = \pm1$} & \multicolumn{2}{c}{$\sum q(\ell) = 0$} \\
    Low-mass resonance veto & \multicolumn{4}{c}{$m_{\ell\ell} > 12\GeV$}\\
    $\mathrm{N}(\mu^{+}\mu^{-})$ pairs with $110<m_{\mu\mu}<150\GeV$ & $\geq$1 & $=$1 & $\geq$1 & $=$1\\
    $\mathrm{N}(\mu^{+}\mu^{-})$ pairs with ${\abs{m_{\mu\mu}-m_{\PZ}} < 10\GeV}$ & $=$0 & $=$0 & $=$1 & $=$0\\
    $\mathrm{N}(\Pe^{+}\Pe^{-})$ pairs with ${\abs{m_{\Pe\Pe}-m_{\PZ}} < 20\GeV}$ & $=$0 & $=$0 & $=$0 & $=$1\\
    \hline
  \end{tabular}
  \label{tab:vh_preselection}
\end{table*}

Two BDT discriminants are trained to discriminate between signal and background events in the $\PW\PH$ and $\PZ\PH$ categories. The input variables are selected such that the BDT outputs are not significantly correlated with the $m_{\mu\mu}$ of the Higgs boson candidate. This is required by the chosen analysis strategy, which is analogous to that adopted for the signal extraction in the $\Pg\Pg\PH$ category. The impact of the $m_{\mu\mu}$ resolution, which evolves as a function of muon \pt and $\eta$, is taken into account during the BDT training by applying weights to the simulated signal events that are inversely proportional to the per-event mass resolution, estimated from the uncertainty in the measured $m_{\mu\mu}$ following the same strategy described in Section~\ref{sec:ggh_category} and~\ref{sec:tth_category}. 

The BDT discriminant used in the $\PW\PH$ category takes as inputs several variables that exploit the kinematic features of the three charged leptons in the event, as well as the \ptmiss. These variables include the full kinematic information, apart from the invariant mass, of the dimuon system corresponding to the Higgs boson candidate. In addition, the $\Delta\phi$ and $\Delta\eta$ separations between the additional lepton ($\ell_{\PW}$) and the Higgs boson candidate, between $\ell_{\PW}$ and both muons from the Higgs boson candidate, and between $\ell_{\PW}$ and \htvecmiss are considered. The \htvecmiss is defined as the negative vector \pt sum of all jets in the event with ${\pt > 30\GeV}$ and ${\abs{\eta} < 4.7}$. Finally, the transverse mass of the combined $\ell_{\PW}$ and \htvecmiss system, the flavour of $\ell_{\PW}$, and the \pt of $\ell_{\PW}$ are added as inputs to the BDT. The particular kinematic properties in signal events of the $\ell_{\PW}$ and \mht enable a large suppression of the residual DY background. The BDT discriminant trained in the $\PZ\PH$ category considers several input observables constructed from the lepton pair associated with the $\PZ$ boson decay ($\ell\ell_{\PZ}$) and the muon pair considered as the Higgs boson candidate ($\mu\mu_{\PH}$). These include the \pt and $\eta$ of both $\PZ$ and Higgs boson candidates, the $\Delta\phi$ ($\Delta R$) between the muons (charged leptons) of the $\mu\mu_{\PH}$ ($\ell\ell_{\PZ}$) system, $m_{\ell\ell_{\PZ}}$, $\Delta\eta(\mu\mu_{\PH},\ell\ell_{\PZ})$, and the cosine of the polar angle between the $\mu\mu_{\PH}$ and $\ell\ell_{\PZ}$ candidates. The flavour of the lepton pair associated with the $\PZ$ boson decay is also included as an input variable.

\begin{figure*}[!htb]
  \centering
  \includegraphics[width=0.45\textwidth]{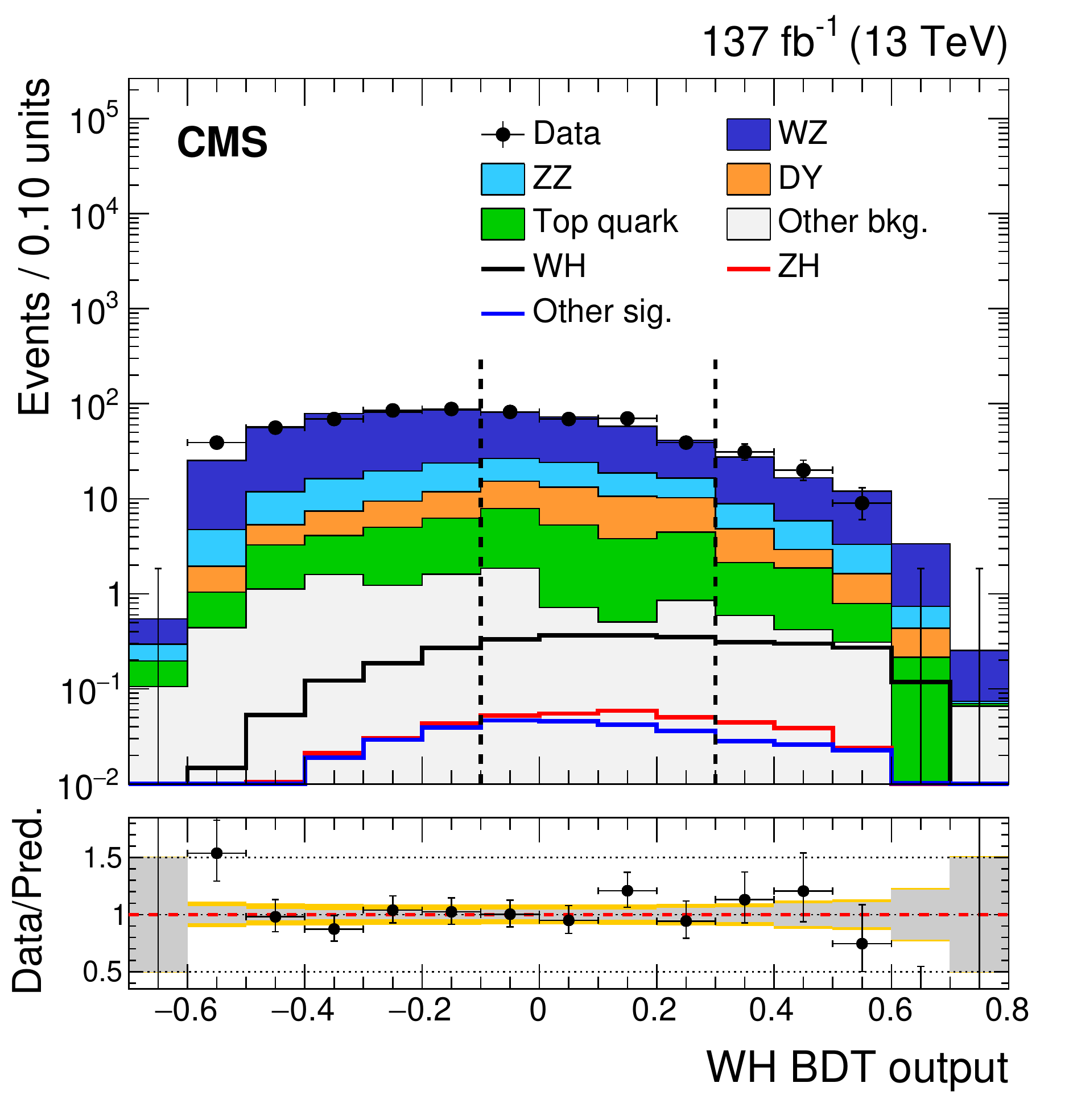}
  \includegraphics[width=0.45\textwidth]{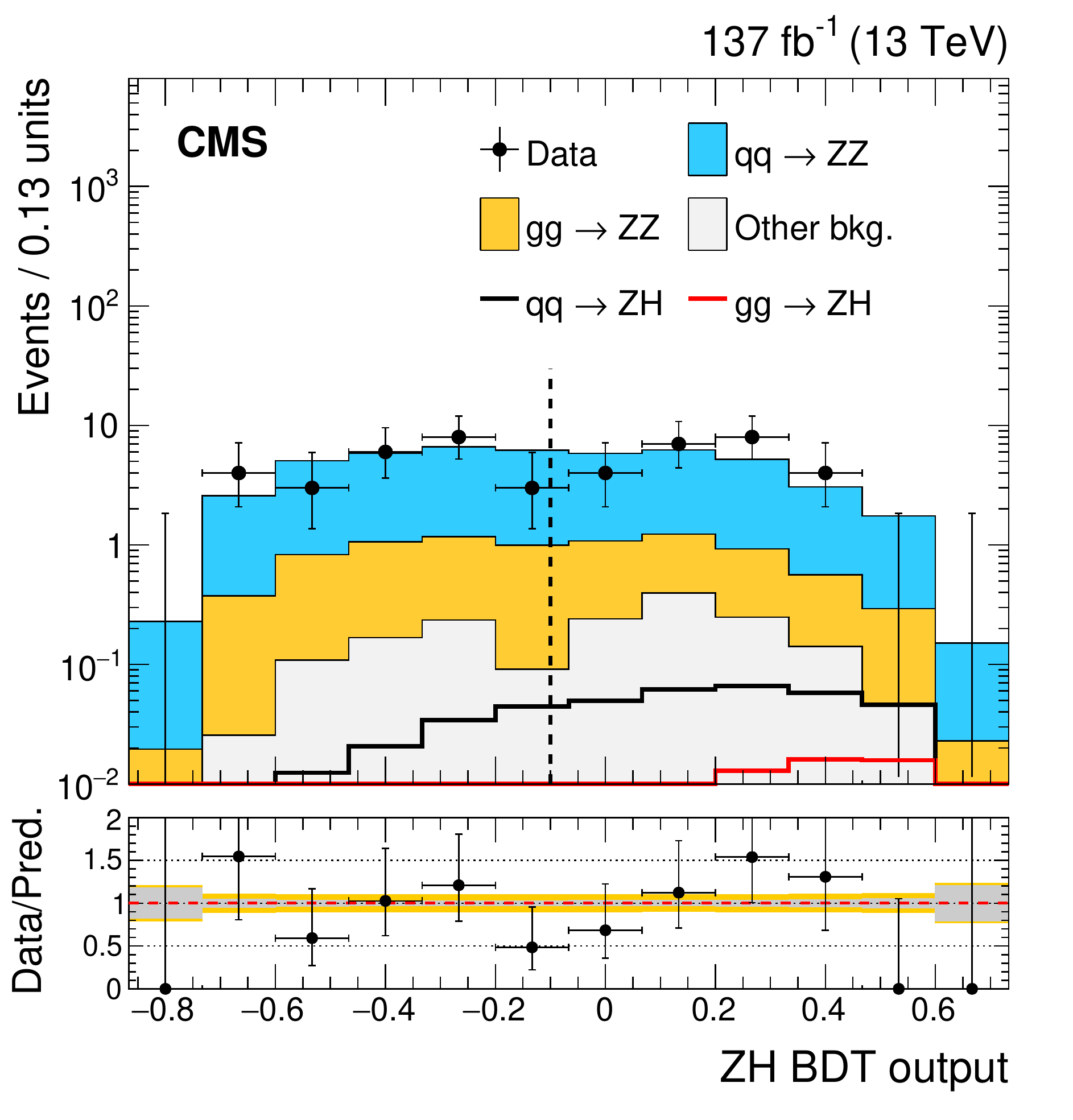}
  \caption{The observed BDT output distribution in the $\PW\PH$ (\cmsLeft) and $\PZ\PH$ (\cmsRight) categories compared to the prediction from the simulation of various SM background processes. Signal distributions expected from different production modes of the Higgs boson with ${\mh = 125\GeV}$ are overlaid. The description of the ratio panel is the same as in Fig.~\ref{fig:ggH_bdt_output}. The dashed vertical lines indicate the boundaries of the optimized event categories.}
  \label{fig:vh_bdt_output}
\end{figure*}

Figure~\ref{fig:vh_bdt_output} shows the output of the BDT classifiers in the $\PW\PH$ (\cmsLeft) and $\PZ\PH$ (\cmsRight) categories. Based on these outputs, events in the $\PW\PH$ category are further divided into three subcategories termed ``${\PW\PH\textrm{-cat1}}$'', ``${\PW\PH\textrm{-cat2}}$'', and ``${\PW\PH\textrm{-cat3}}$'' corresponding to signal efficiency quantiles of 0--22, 22--70, $>$70\%, respectively. Similarly, events in the $\PZ\PH$ category are divided into two subcategories, labelled ``${\PZ\PH\textrm{-cat1}}$'' and ``${\PZ\PH\textrm{-cat2}}$'' corresponding to signal efficiency quantiles of 0--52 and $>$52\%, respectively. The boundaries of these subcategories, defined in terms of the BDT discriminant and indicated in Fig.~\ref{fig:vh_bdt_output} by black dashed vertical lines, are chosen via the same optimization strategy adopted in the $\Pg\Pg\PH$ and $\ttbar\PH$ categories. In the $\PV\PH$ category, the BWZ function is used to estimate the total background instead of mBW.

\begin{figure*}[!]
  \centering
  \includegraphics[width=0.42\textwidth]{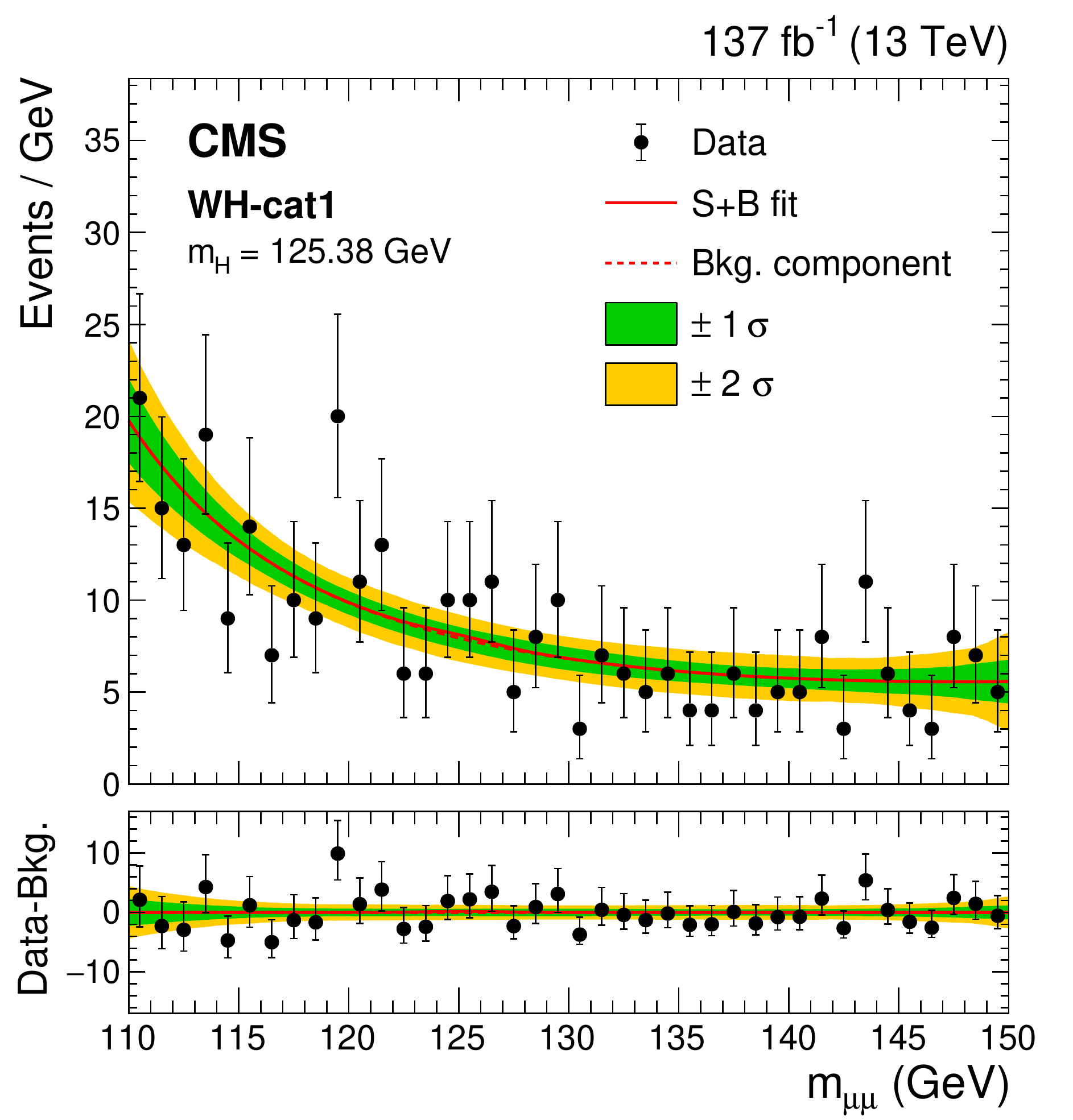}
  \includegraphics[width=0.42\textwidth]{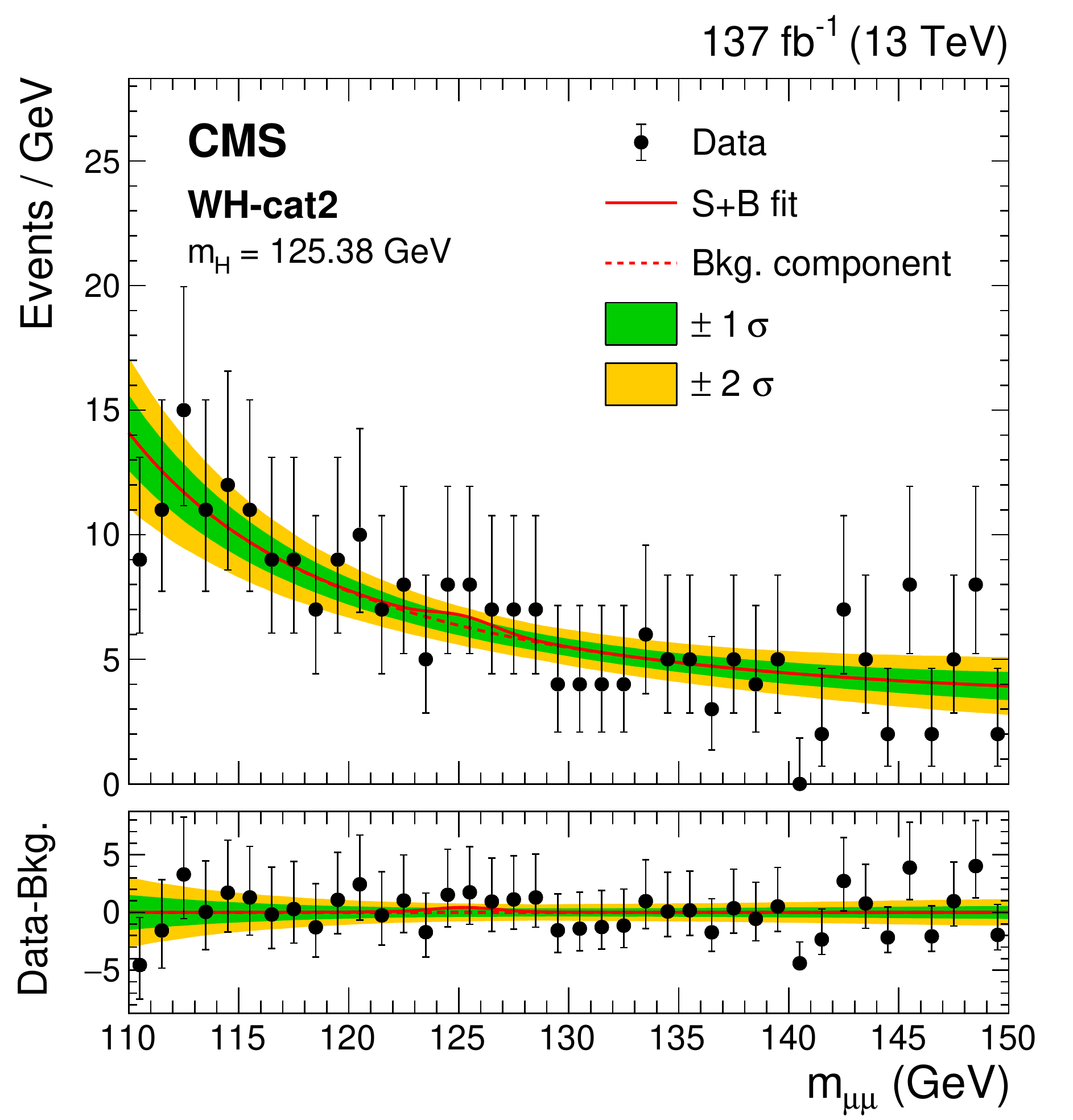}\\
  \includegraphics[width=0.42\textwidth]{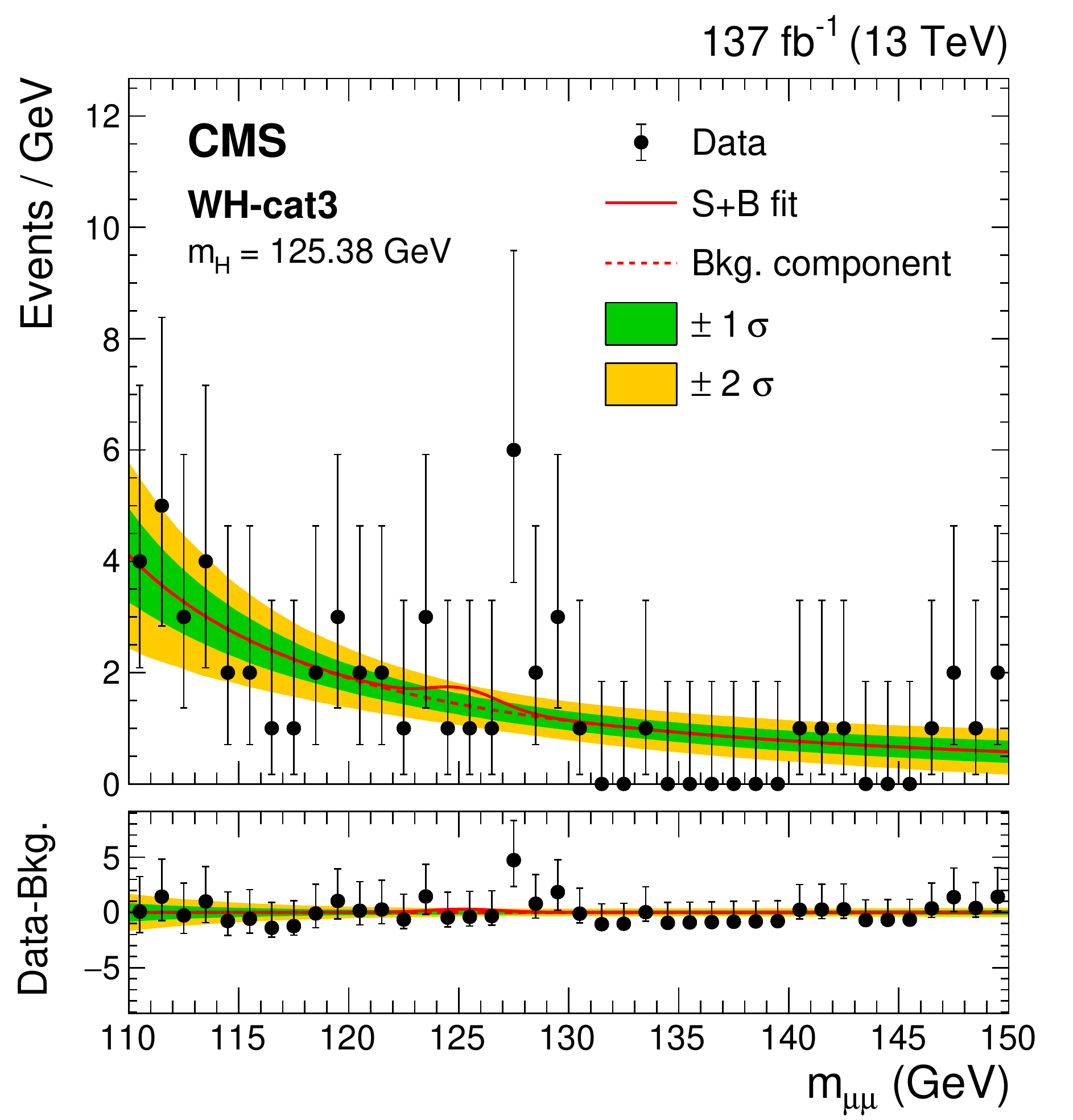}
  \includegraphics[width=0.42\textwidth]{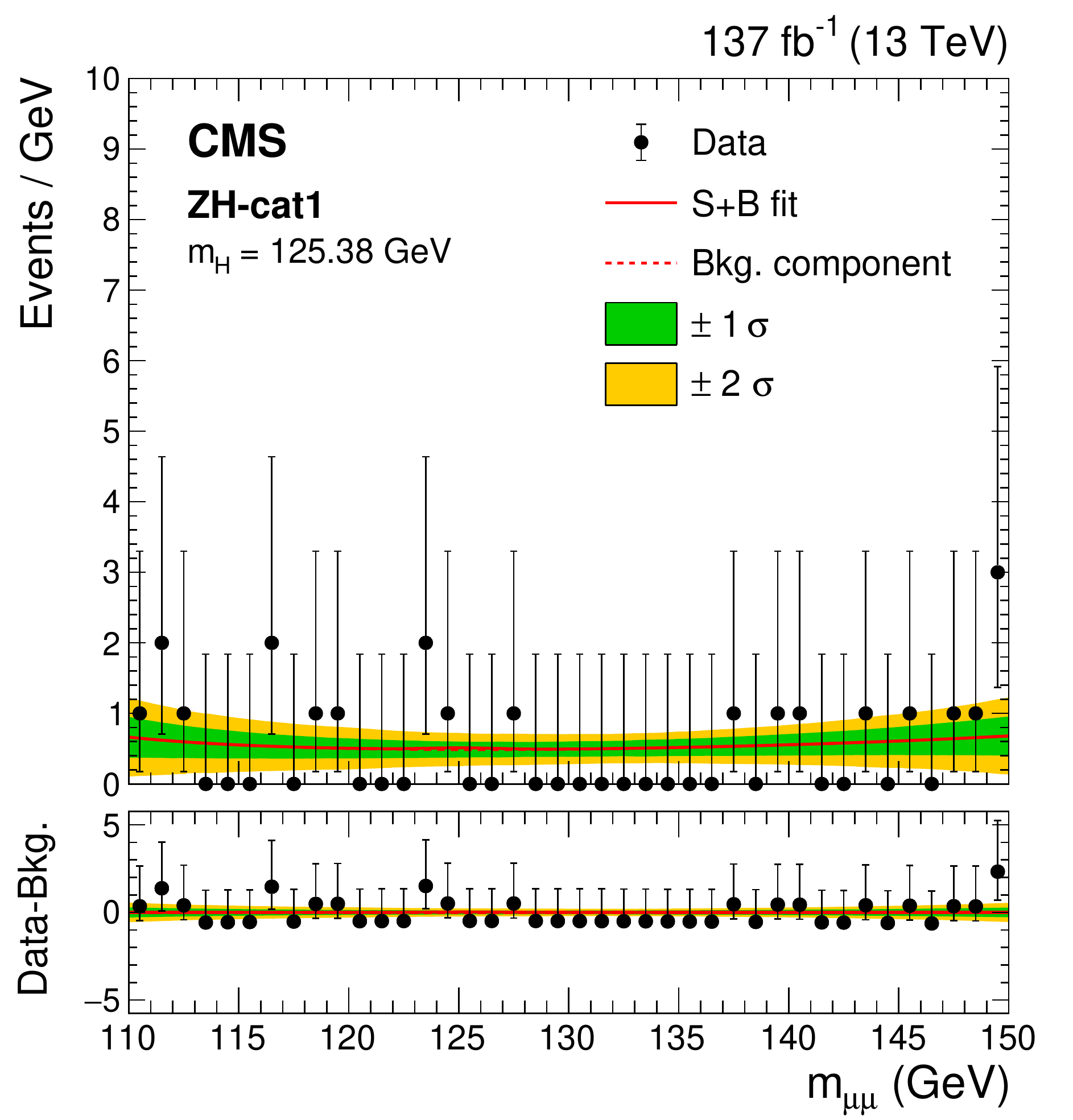}\\
  \includegraphics[width=0.42\textwidth]{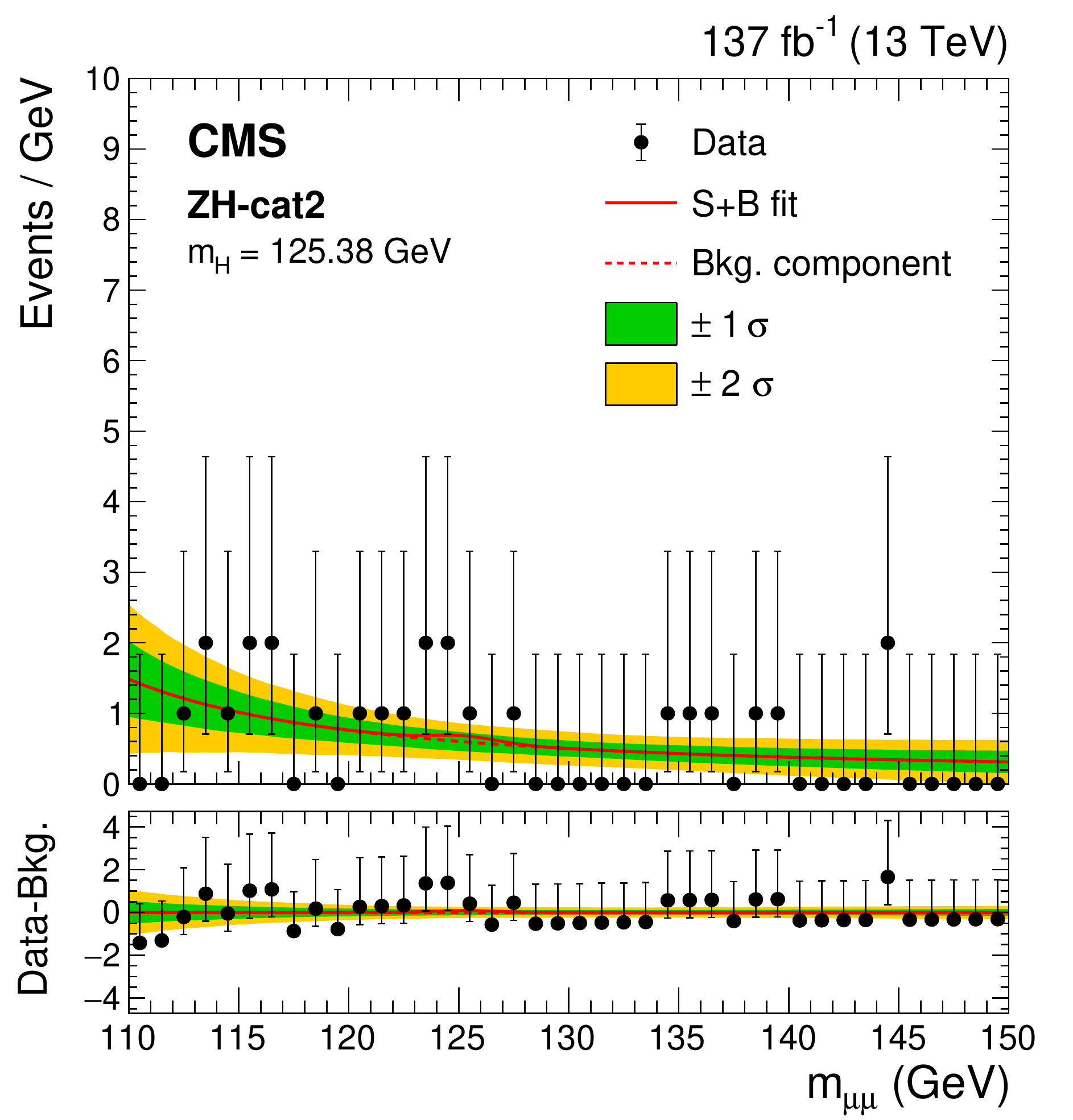}
  \caption{Comparison between the data and the total background extracted from a {S+B} fit performed across the various $\PW\PH$ and $\PZ\PH$ event subcategories. The one (green) and two (yellow) standard deviation bands include the uncertainties in the background component of the fit. The lower panel shows the residuals after the background subtraction, where the red line indicates the signal with ${\mh=125.38\GeV}$ extracted from the fit.}
  \label{fig:vh_fit_mass}
\end{figure*}

Figure~\ref{fig:vh_fit_mass} shows the $m_{\mu\mu}$ distributions in the $\PW\PH$ and $\PZ\PH$ event categories. The signal is extracted via a binned maximum-likelihood fit in each event category, where the signal is modelled with a DCB function and the background is modelled with the $\mathrm{BWZ}\gamma$ function in ${\PW\PH\textrm{-cat1}}$, as defined in Eq.~(\ref{eqn:BWZGamma}) and the BWZ function in the remaining subcategories, as defined in Eq.~(\ref{eqn:BWZ}). Table~\ref{tab:vh_categories} reports the signal composition in the $\PW\PH$ and $\PZ\PH$ subcategories, along with the HWHM of the expected signal shape. In addition, the estimated number of background events, the {S/(S+B)} and ${\mathrm{S}/\sqrt{\smash[b]{\mathrm{B}}}}$ ratios, and the observation in data within the HWHM of the signal peak are also listed.

\begin{table*}[!htb]
  \centering
  \topcaption{The total expected number of signal events with ${\mh=125.38\GeV}$ (S), the ratio of the expected contributions from different production modes to the total signal yield, the HWHM of the signal peak, the functional form used for the background modelling, the estimated number of background events (S) and the observed number of events within ${\pm\,\mathrm{HWHM}}$, and the {S/(S+B)} and the ${\mathrm{S}/\sqrt{\smash[b]{\mathrm{B}}}}$ ratios computed within the HWHM of the signal peak for each of the optimized event categories defined along the $\PW\PH$ and $\PZ\PH$ BDT outputs.}
  \cmsTable{
    \begin{tabular}{l c c c c c c c c c c c c}
      \hline
      Event     & Total  & $\PW\PH$ & $\cPq\cPq\PZ\PH$ & $\Pg\Pg\PZ\PH$ & $\ttbar\PH$+$\cPqt\PH$ & HWHM       & Bkg. fit & Bkg.  & Data  & {S/(S+B)} (\%)  & ${\mathrm{S}/\sqrt{\smash[b]{\mathrm{B}}}}$ \\
      category  & signal & (\%)     & (\%)             & (\%)           & (\%)                   & ({\GeVns}) & function & @HWHM & @HWHM & @HWHM  & @HWHM \\
      \hline
      $\PW\PH$-cat1    & 0.82  & 76.2  & 9.6   & 1.6  & 12.6   & 2.00  & $\mathrm{BWZ}\gamma$     & 32.0      & 34          & 1.54        & 0.09  \\
      $\PW\PH$-cat2    & 1.72  & 80.1  & 9.1   & 1.5  & 9.3    & 1.80  & BWZ          & 23.1      & 27        & 4.50        & 0.23  \\
      $\PW\PH$-cat3    & 1.14  & 85.7  & 6.7   & 1.8  & 4.8    & 1.90  & BWZ          & 5.48      & 4         & 12.6        & 0.35  \\
      [\cmsTabSkip]
      $\PZ\PH$-cat1    & 0.11  & \NA   & 82.8  & 17.2 & \NA    & 2.07  & BWZ          & 2.05      & 4         & 3.29        & 0.05  \\
      $\PZ\PH$-cat2    & 0.31  & \NA   & 79.6  & 20.4 & \NA    & 1.80  & BWZ          & 2.19      & 4         & 8.98        & 0.14  \\
      \hline
    \end{tabular}
  }
  \label{tab:vh_categories}
\end{table*}

The systematic uncertainties considered in this category account for possible mismodelling in the signal shape and rate. The shape of the reconstructed Higgs boson resonance, modelled using the DCB function defined in Eq.~(\ref{eq:dCBFunction}), is affected by the uncertainty in the muon momentum scale and resolution. Uncertainties in the calibration of these values are propagated to the shape of the $m_{\mu\mu}$ distribution, yielding variations of up to 0.2\% in the peak position and up to 10\% in the width. Experimental systematic uncertainties from the measurement of the electron and muon selection efficiencies (1--3\% per event category), jet energy scale and resolution (0.5--2\% per event category), the efficiency of vetoing $\cPqb$ quark jets (1--3\% per event category), the integrated luminosity, and the pileup model (0.5--2\% per event category) affect the predicted signal rate. Furthermore, theoretical uncertainties in the prediction of the Higgs boson production cross section, decay rate, and acceptance are also considered. Rate uncertainties are taken into account in the signal extraction as nuisance parameters acting on the relative signal yield with log-normal constraints.

The potential bias due to the choice of the parametric function used to model the background is estimated using the same procedure employed in the $\ttbar\PH$ analysis, detailed in Section~\ref{sec:tth_category}. The set of parametric functional forms considered in the bias studies includes BWZ, $\mathrm{BWZ}\gamma$, sum of exponentials, Bernstein polynomials, and sum of power laws. The chosen parametrization maximizes the expected sensitivity without introducing a significant bias in the measured signal yield. The corresponding bias is found to be smaller than 20\% and is therefore neglected in the signal extraction. The chosen functions maximize the expected sensitivity to the 125\GeV Higgs boson.

\section{Results}\label{sec:results}

A simultaneous fit is performed across all event categories, with a single overall signal strength modifier ($\mu$) free to float in the fit. The signal strength modifier is defined as the ratio between the observed Higgs boson rate in the \hmm decay channel and the SM expectation, ${\mu = (\sigma\brhmm)_{\mathrm{obs}}/(\sigma\brhmm)_{\mathrm{SM}}}$. The relative contributions from the different Higgs boson production modes are fixed to the SM prediction within uncertainties. Confidence intervals on the signal strength are estimated using a profile likelihood ratio test statistic~\cite{Cowan:2010js}, in which systematic uncertainties are modelled as nuisance parameters following a modified frequentist approach~\cite{CMS-NOTE-2011-005}. The profile likelihood ratio is defined as
\begin{linenomath}
\begin{equation*}
  \begin{aligned}
    q_{\mu}  &  = -2 \Delta \ln \mathcal{L} = -2 \ln \frac{\mathcal{L}(\mathrm{data}|\mu,\hat{\theta}_{\mu})}{\mathcal{L}(\mathrm{data}|\hat{\mu},\hat{\theta})},
  \end{aligned}
\end{equation*}
\end{linenomath}
where $\hat{\mu}$ represents the value of the signal strength that maximizes the likelihood $\mathcal{L}$ for the data, while $\hat{\theta}$ and $\hat{\theta}_{\mu}$ denote the best fit estimate for the nuisance parameters and the estimate for a given fixed value of $\mu$, respectively. Theoretical uncertainties affecting the signal prediction are correlated among all the event categories included in the fit. Similarly, experimental uncertainties in the measurement of the integrated luminosity in each year, jet energy scale and resolution, $\cPqb$ quark jet identification, modelling of the pileup conditions, and selection efficiencies of muons and electrons are also correlated across categories. Because of the different analysis strategy employed in the VBF category, the acceptance uncertainties from the muon energy scale and resolution are correlated only among the $\Pg\Pg\PH$, $\PW\PH$, $\PZ\PH$, and $\ttbar\PH$ categories. Furthermore, their effect on the position and width of the signal peak are assumed to be uncorrelated across event categories.

\begin{figure*}[htb!]
  \centering
  \includegraphics[width=0.45\textwidth]{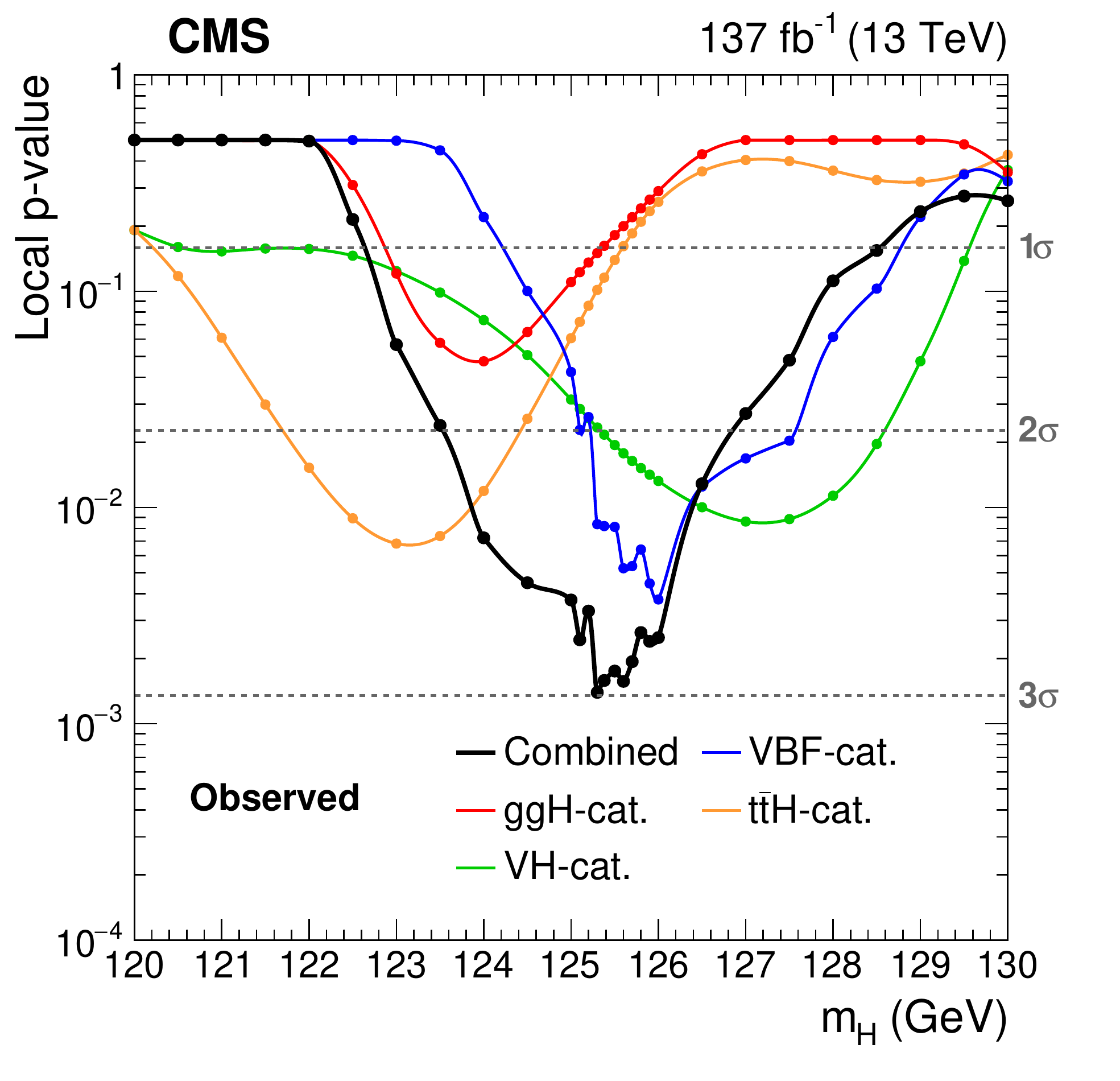}
  \includegraphics[width=0.45\textwidth]{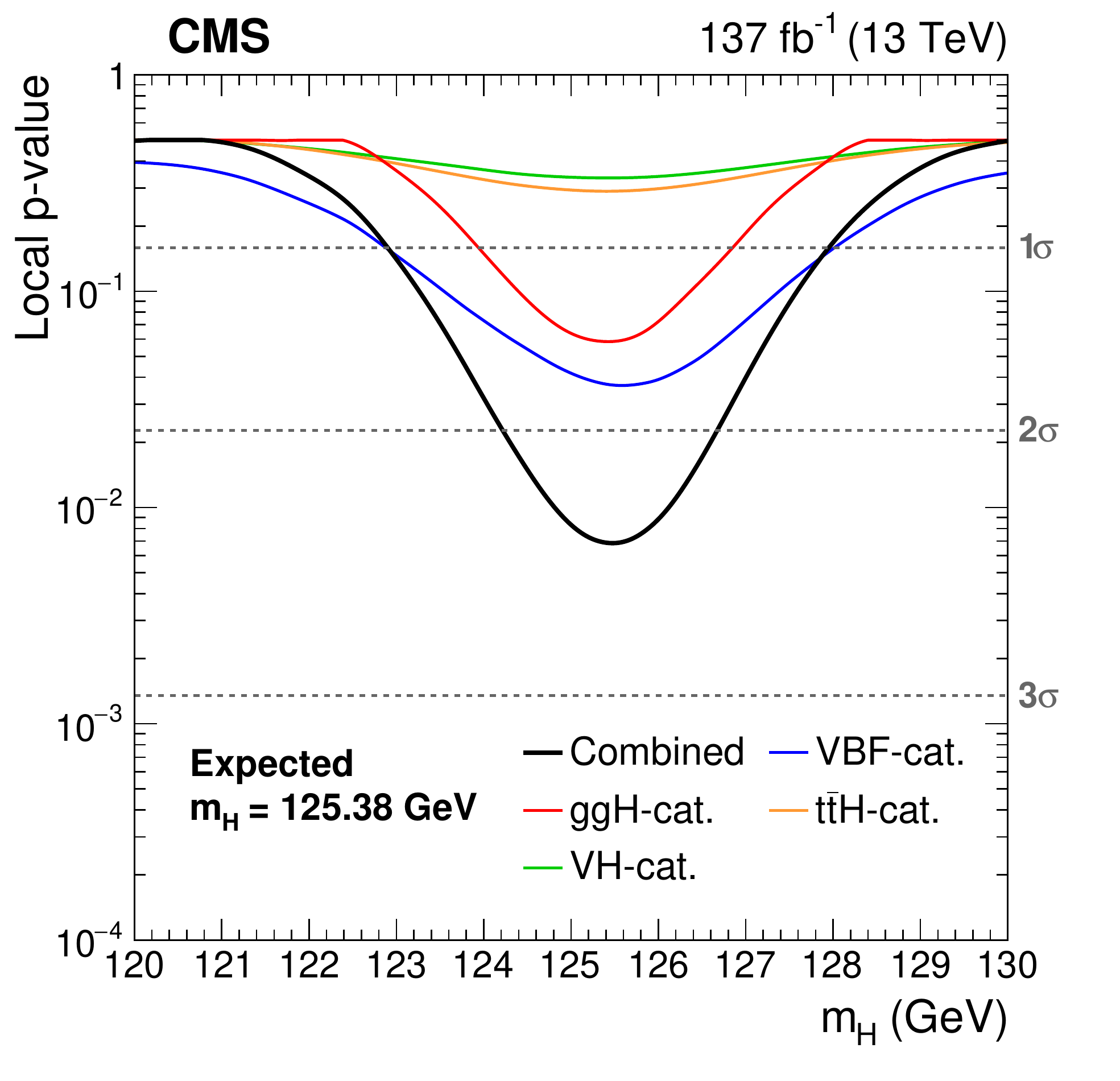}
  \caption{Left: observed local $p$-values as a function of $\mh$, extracted from the combined fit as well as from each individual production category, are shown. The solid markers indicate the mass points for which the observed $p$-values are computed. Right: the expected $p$-values are calculated using the background expectation obtained from the {S+B} fit and injecting a signal with ${\mh=125.38\GeV}$ and ${\mu=1}$.}
  \label{fig:pvalue_vs_mass}
\end{figure*}

The local $p$-value quantifies the probability for the background to produce a fluctuation larger than the apparent signal observed in the search region. Figure~\ref{fig:pvalue_vs_mass} (\cmsLeft) shows the observed local $p$-value for the combined fit, and for each individual production category, as a function of \mh in a 5\GeV window around the expected Higgs boson mass. The solid markers indicate the mass points for which the observed $p$-values are computed. Figure~\ref{fig:pvalue_vs_mass} (\cmsRight) shows the expected $p$-values computed for the combined fit, and for each production category, on an Asimov data set~\cite{Cowan:2010js} generated from the background expectation obtained from the {S+B} fit with a ${\mh=125.38\GeV}$ signal injected. The observed p-values as a function \mh are compatible, within the statistical variation, with the expectation for the Higgs boson with ${\mh = 125.38\GeV}$. In the $\Pg\Pg\PH$, $\PV\PH$, and $\ttbar\PH$ categories, in order to evaluate $p$-values for masses different from 125\GeV, signal models are derived using alternative \hmm signal samples generated with $\mh$ fixed to 120 and 130\GeV. Signal shape parameters and the expected rate for each production mode in each event category are then interpolated using a spline function within ${120<\mh<130\GeV}$, providing a signal model for any mass value in the ${\mh = 125\pm5\GeV}$ range. A different strategy is employed in the VBF category since $m_{\mu\mu}$ is a DNN input variable. As described in Section~\ref{sec:vbf_category}, the DNN output can be decorrelated from the $m_{\mu\mu}$ information by fixing its value to 125\GeV. Therefore, a potential signal with mass $m'$ different from 125\GeV can be extracted by fitting the data with an alternative set of signal and background templates, obtained by shifting the mass value used as input to the DNN evaluation by ${\Delta m = 125\GeV-m'}$ and adjusting the expected signal yields by the corresponding differences in the production cross section and decay rate. Variations in the acceptance per DNN bin as a function of $\Delta m$ are found to be negligible in the mass range of interest. This procedure is also applied to the data, yielding for each tested mass hypothesis a different observed DNN distribution to fit. Throughout the explored mass range, ${120<\mh<130\GeV}$, the VBF category has the highest expected sensitivity to \hmm decays, followed by the $\Pg\Pg\PH$, $\ttbar\PH$, and $\PV\PH$ categories, respectively. The observed (expected for ${\mu=1}$) significance at ${\mh=125.38\GeV}$ of the incompatibility with the background-only hypothesis is 3.0 (2.5) standard deviations. The 95\% \CL upper limit (UL) on the signal strength, computed with the asymptotic \CLs criterion~\cite{Junk:1999kv,Read:2002av,Cowan:2010js}, is also derived from the combined fit performed across all event categories. The observed (expected for ${\mu=0}$) UL on $\mu$ at 95\% \CL for ${\mh = 125.38\GeV}$ is 1.9 (0.8). Discrete fluctuations in the observed $p$-value for the VBF category and the combined fit arise from event migrations in data between neighbouring bins when reevaluating the VBF category DNN for different mass hypotheses, following the procedure described above.

The best fit signal strength for the Higgs boson with mass of 125.38\GeV, and the corresponding 68\% \CL interval, is ${\hat{\mu} = 1.19~^{+0.41}_{-0.40}\stat^{+0.17}_{-0.16}\syst}$. Assuming SM production cross sections for the various modes, the \hmm branching fraction is constrained at 95\% \CL to be within $0.8 \times 10^{-4} < \brhmm < 4.5 \times 10^{-4}$. The statistical component of the post-fit uncertainty is separated by performing a likelihood scan as a function of $\mu$ in which nuisance parameters associated with systematic uncertainties are fixed to their best fit values. The systematic uncertainty component is then taken as the difference in quadrature between the total and the statistical uncertainties. The individual contributions to the uncertainty in the measured signal strength from experimental uncertainties, the limited size of the simulated samples, and theoretical uncertainties are also evaluated following a similar procedure. The individual uncertainty components are summarized in Table~\ref{tab:combo_sys_impact}. The uncertainty in the measured signal rate is dominated by the limited number of events in data.

\begin{figure*}[htb!]
  \centering
  \includegraphics[width=0.45\textwidth]{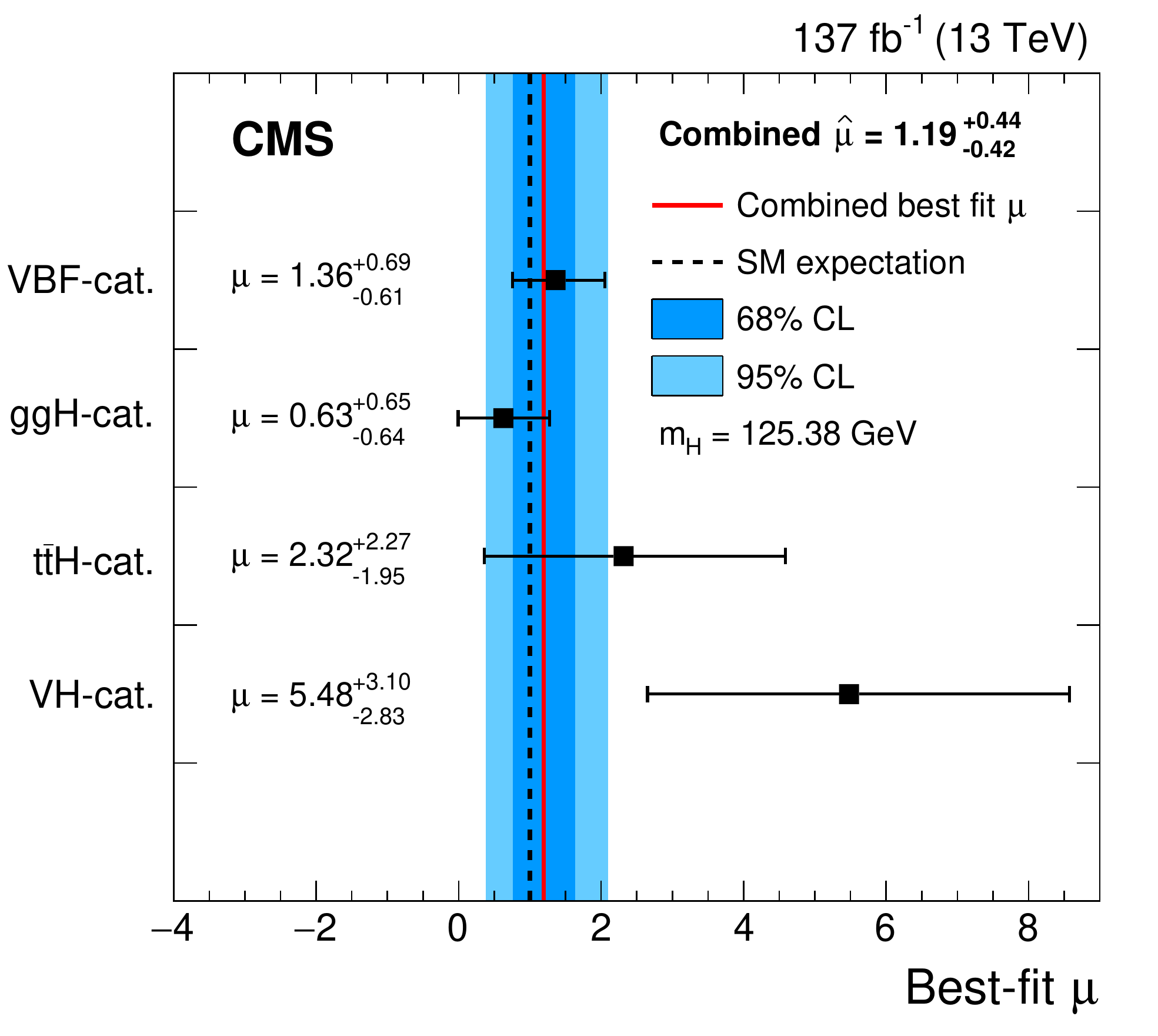}
  \includegraphics[width=0.45\textwidth]{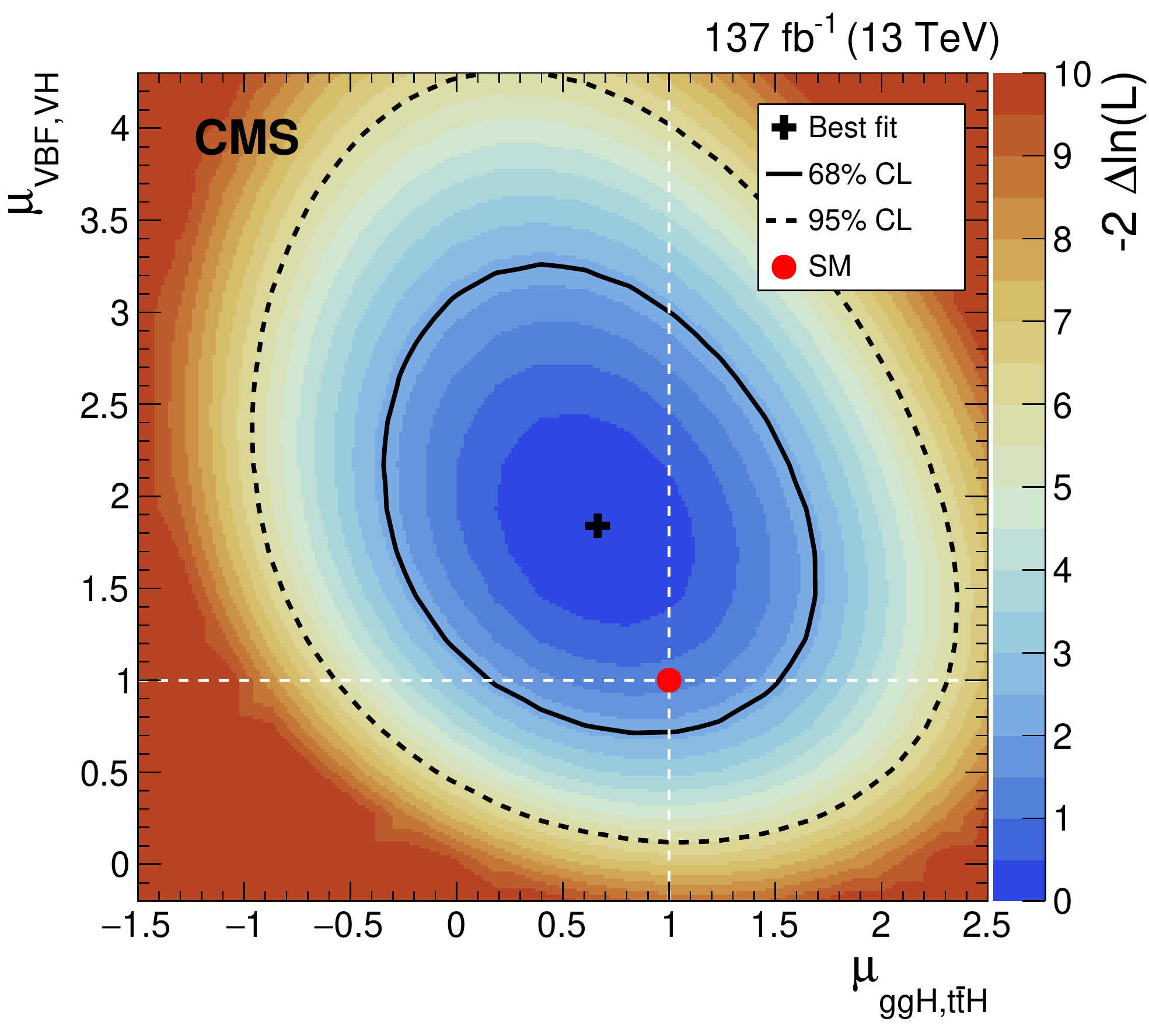}
  \caption{Left: signal strength modifiers measured for ${\mh=125.38\GeV}$ in each production category (black points) are compared to the result of the combined fit (solid red line) and the SM expectation (dashed grey line). Right: scan of the profiled likelihood ratio as a function of $\mu_{\Pg\Pg\PH,\ttbar\PH}$ and $\mu_{\mathrm{VBF},\PV\PH}$ with the corresponding $1\sigma$ and $2\sigma$ uncertainty contours. The black cross indicates the best fit values $(\hat{\mu}_{\Pg\Pg\PH,\ttbar\PH},\hat{\mu}_{\mathrm{VBF},\PV\PH})=(0.66,1.84)$, while the red circle represents the SM expectation.}
  \label{fig:combo_likelihood_scan}
\end{figure*}

Figure~\ref{fig:combo_likelihood_scan} (\cmsLeft) reports a summary of the best fit values for the signal strength and the corresponding 68\% \CL intervals obtained from a profile likelihood scan in each production category. The best fit signal strengths in each production category are consistent with the combined fit result as well as the SM expectation. A likelihood scan is performed in which the four main Higgs boson production mechanisms are associated to either fermion ($\Pg\Pg\PH$ and $\ttbar\PH$) or vector boson (VBF and V\PH) couplings. Two signal strength modifiers, denoted as $\mu_{\Pg\Pg\PH,\ttbar\PH}$ and $\mu_{\mathrm{VBF},\PV\PH}$, are varied independently as unconstrained parameters in the fit. Figure~\ref{fig:combo_likelihood_scan} (\cmsRight) shows the $1\sigma$ and $2\sigma$ contours, computed as variations around the minimum of ${-2 \, \Delta\ln(\mathcal{L})}$ for ${\mh = 125.38\GeV}$, for the signal strength modifiers $\mu_{\Pg\Pg\PH,\ttbar\PH}$ and $\mu_{\mathrm{VBF},\PV\PH}$. The best fit values for these parameters are ${\hat{\mu}_{\Pg\Pg\PH,\ttbar\PH}=0.66^{+0.67}_{-0.66}}$ and ${\hat{\mu}_{\mathrm{VBF},\PV\PH}=1.84^{+0.89}_{-0.77}}$, consistent with the SM expectation. 

\begin{table*}[!htb]
  \centering
  \topcaption{Major sources of uncertainty in the measurement of the signal strength $\mu$ and their impact. The total post-fit uncertainty on $\mu$ is divided into the statistical and systematic components. The systematic component is further separated into three parts depending on the origin of the different sources of uncertainty: experimental, theoretical, and size of the simulated samples. The uncertainty due to the limited statistics of the simulated samples only affects the VBF category results.}  
  \def\arraystretch{1.2}
  \begin{tabular}{l c c}
    \hline
    Uncertainty source & \multicolumn{2}{c}{$\Delta\mu$} \\
    \hline
    Post-fit uncertainty           & $+0.44$ & $-0.42$ \\
    [\cmsTabSkip]
    Statistical uncertainty        & $+0.41$ & $-0.40$ \\
    [\cmsTabSkip]
    Systematic uncertainty         & $+0.17$ & $-0.16$ \\
    [\cmsTabSkip]
    Experimental uncertainty       & $+0.12$ & $-0.11$ \\ 
    Theoretical uncertainty        & $+0.10$ & $-0.11$ \\
    Size of simulated samples      & $+0.07$ & $-0.06$ \\
    \hline
  \end{tabular}
  \label{tab:combo_sys_impact}
\end{table*}

\begin{figure*}[htb!]
  \centering
  \includegraphics[width=0.45\textwidth]{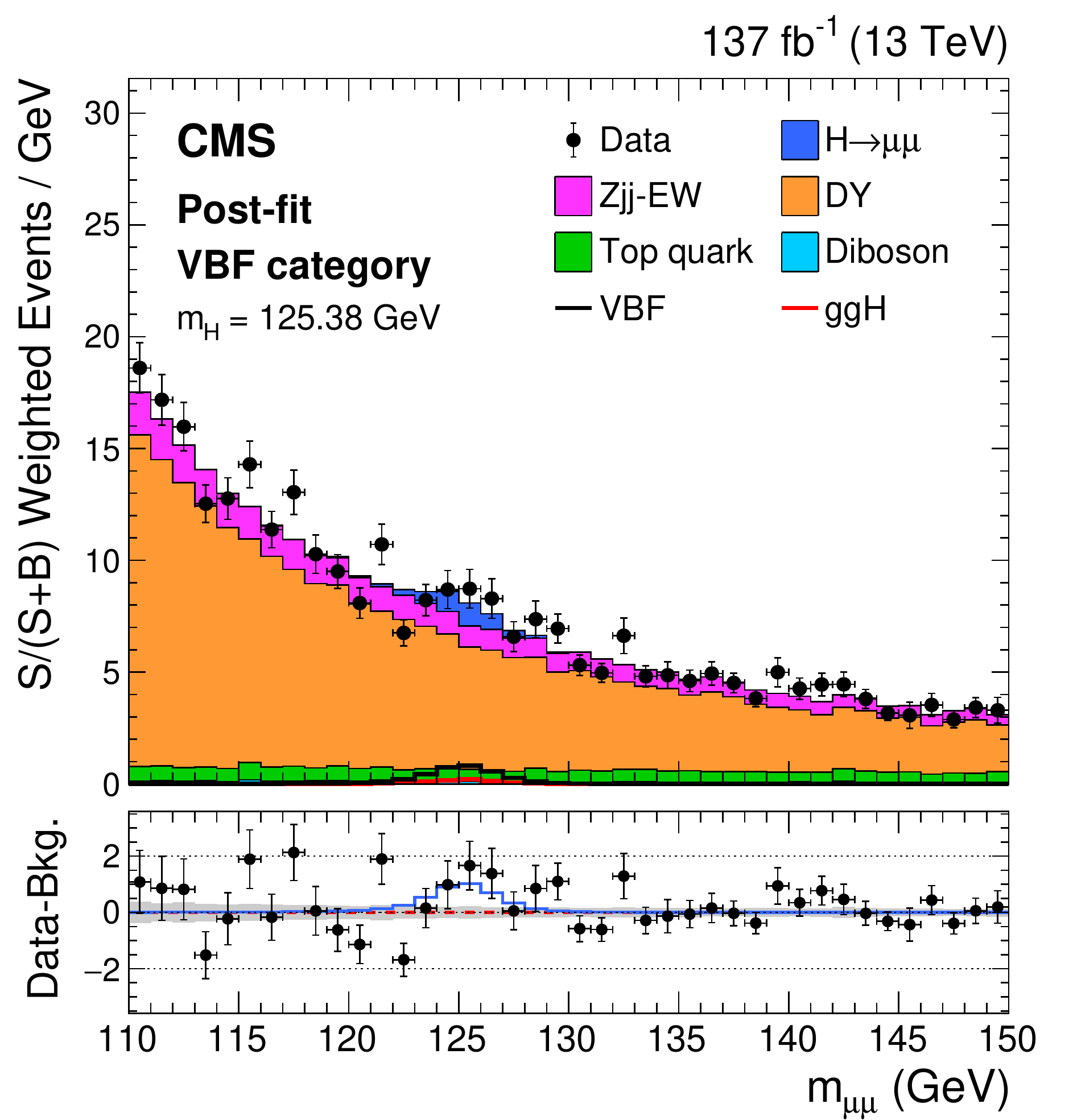}
  \includegraphics[width=0.45\textwidth]{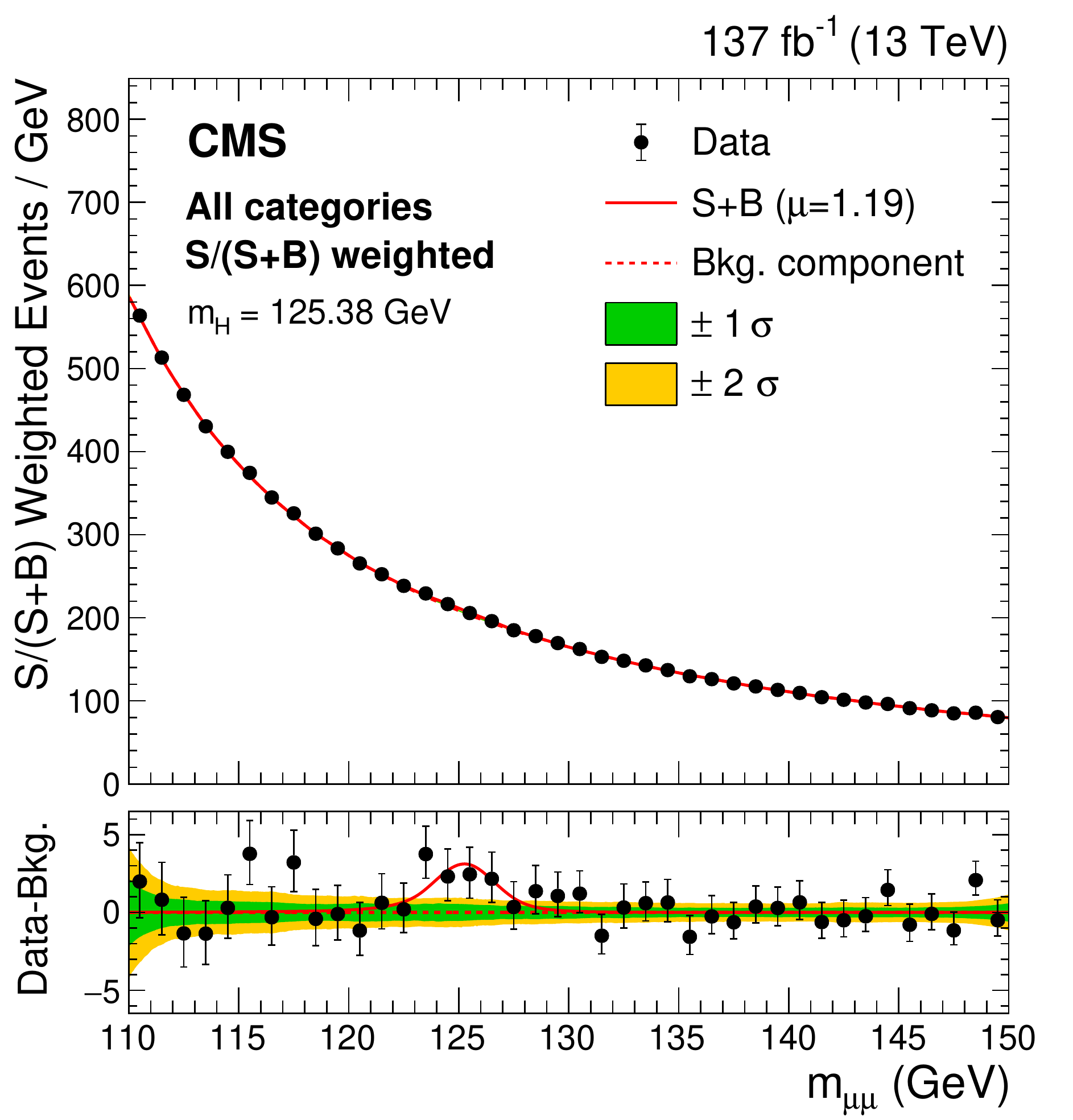}
  \caption{Left: the $m_{\mu\mu}$ distribution for the weighted combination of VBF-SB and VBF-SR events. Each event is weighted proportionally to the {S/(S+B)} ratio, calculated as a function of the mass-decorrelated DNN output. The lower panel shows the residuals after subtracting the background prediction from the {S+B} fit. The best fit \hmm signal contribution is indicated by the blue line and histogram, while the grey band indicates the total background uncertainty from the background-only fit. Right: the $m_{\mu\mu}$ distribution for the weighted combination of all event categories. The lower panel shows the residuals after background subtraction, with the best fit SM \hmm signal contribution for ${\mh = 125.38\GeV}$ indicated by the red line.}
 \label{fig:fit_summary_plot}    
\end{figure*}

An unbiased mass distribution representative of the fit result in the VBF category is obtained by weighting both simulated and data events from the VBF-SR and VBF-SB regions by the {S/(S+B)} ratio. The {S/(S+B)} weights are computed as a function of the mass-decorrelated DNN output, defined in Section~\ref{sec:vbf_category}, for events within ${m_{\mu\mu} = 125.38\GeV {\pm\,\mathrm{HWHM}}}$ and using the same bin boundaries as displayed in Fig.~\ref{fig:dnn_vbf_sr}. The HWHM of the signal peak in the VBF category is about 2\GeV. The best fit estimates for the nuisance parameters and signal strength are propagated to the $m_{\mu\mu}$ distribution. This distribution is not used for any of the measurements presented in this paper, but only to visualize the fit result. Figure~\ref{fig:fit_summary_plot} (\cmsLeft) shows the observed and predicted weighted $m_{\mu\mu}$ distributions for events in the VBF-SB and VBF-SR regions, combining 2016, 2017, and 2018 data. The lower panel shows the residuals between the data and the post-fit background prediction, along with the post-fit uncertainty obtained from the background-only fit. The best fit signal contribution with ${\mh = 125.38\GeV}$ is indicated by the blue line. 
An excess is observed in the weighted data distribution that is consistent with the expected resonant mass distribution for the signal with $\mh$ near 125\GeV and compatible with the excess observed at high DNN score in Fig.~\ref{fig:dnn_vbf_run2}. The signal and background distributions are then interpolated with a spline function in order to obtain a continuous spectrum that can be summed with the parametric fit results in the $\Pg\Pg\PH$, $\PW\PH$, $\PZ\PH$, and $\ttbar\PH$ categories. Figure~\ref{fig:fit_summary_plot} (\cmsRight) shows the $m_{\mu\mu}$ distribution for the weighted combination of all event categories. The $\Pg\Pg\PH$, $\PV\PH$, and $\ttbar\PH$ categories are weighted proportionally to the corresponding {S/(S+B)} ratio, where S and B are the number of expected signal and background events with mass within ${\pm\,\mathrm{HWHM}}$ of the expected signal peak with ${\mh = 125.38\GeV}$. The weighted data in the upper panel are dominated by the $\Pg\Pg\PH$ event categories with many data events but relatively small {S/(S+B)}. The lower panel shows the residuals after background subtraction, with the best fit SM signal contribution with ${\mh = 125.38\GeV}$ indicated by the red line. An excess of events over the background-only expectation is observed near $m_{\mu\mu} = 125\GeV$.

\begin{figure*}[htb!]
  \centering
  \includegraphics[width=0.60\textwidth]{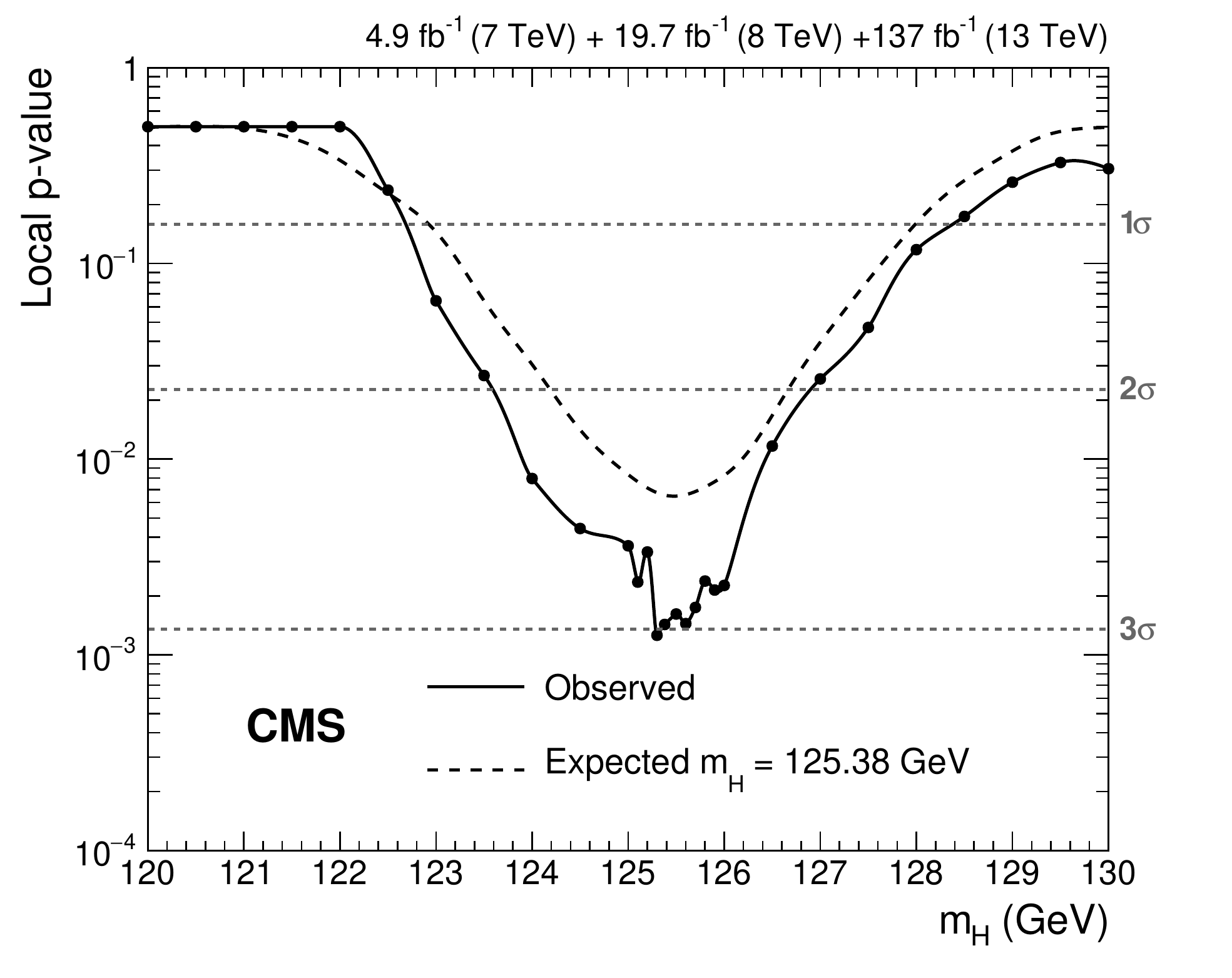}
  \caption{Observed (solid black) and expected (dashed black) local $p$-values as a function of $\mh$, extracted from the combined fit performed on data recorded at ${\sqrt{s}=7}$, 8, and 13\TeV, are shown. The expected $p$-values are calculated using the background expectation obtained from the {S+B} fit and injecting a signal with ${\mh=125.38\GeV}$ and ${\mu=1}$.}
  \label{fig:pval_comb_run1_run2}
\end{figure*}

The result is combined with that obtained from data recorded at centre-of-mass energies of 7 and 8\TeV. The 7+8\TeV search described in Ref.~\cite{Khachatryan:2014aep} has been updated using for the Higgs boson production cross sections and branching fractions the values reported in Ref.~\cite{deFlorian:2016spz}. Systematic uncertainties in the inclusive signal production cross sections and \brhmm are correlated across the 7, 8, and 13\TeV analyses. Experimental uncertainties affecting the measured properties of the various physics objects (muons, electrons, jets, and \cPqb quark jets), the measurement of the integrated luminosity, and the modelling of the pileup conditions are assumed to be uncorrelated between the 7+8 and 13\TeV analyses. Table~\ref{tab:combo_signif_table} reports the observed and expected significances over the background-only expectation at ${\mh=125.38\GeV}$ and the 95\% \CL ULs on $\mu$ in each production category, as well as for the 13\TeV and the 7+8+13\TeV combined fits. The combination improves, relative to the 13\TeV-only result, both the expected and the observed significance at $\mh = 125.38\GeV$ by about 1\%. Figure~\ref{fig:pval_comb_run1_run2} shows the observed (solid black) and the expected (dashed black) local $p$-values derived from the 7+8+13\TeV combined fit as a function of \mh in a 5\GeV window around the expected Higgs boson mass. The expected $p$-value is computed on an Asimov data set generated from the background expectation obtained from the {S+B} fit with a ${\mh=125.38\GeV}$ signal injected. As in Fig.~\ref{fig:pvalue_vs_mass}, the solid markers indicate the mass points for which the observed $p$-values are computed. The best fit signal strength, and the corresponding  68\% \CL interval, obtained from the 7+8+13\TeV combination for the Higgs boson with mass of 125.38\GeV is ${1.19^{+0.40}_{-0.39}\stat^{+0.15}_{-0.14}\syst}$.

\begin{table*}[!htb]
  \centering
  \topcaption{Observed and expected significances for the incompatibility with the background-only hypothesis for ${\mh = 125.38\GeV}$ and the corresponding 95\% \CL upper limits on $\mu$ (in the absence of \hmm decays) for each production category, as well as for the 13\TeV and the 7+8+13\TeV combined fits.}
  \begin{tabular}{l c c}
    \hline
    Production category & Observed (expected) signif. & Observed (expected) UL on $\mu$\\
    \hline
    VBF             & 2.40 (1.77) & 2.57 (1.22) \\
    $\Pg\Pg\PH$     & 0.99 (1.56) & 1.77 (1.28) \\
    $\ttbar\PH$     & 1.20 (0.54) & 6.48 (4.20) \\
    $\PV\PH$ & 2.02 (0.42) & 10.8 (5.13)\\
    [\cmsTabSkip]
    Combined $\sqrt{s} = 13\TeV$ & 2.95 (2.46) & 1.94 (0.82) \\
    [\cmsTabSkip]     
    Combined $\sqrt{s} = 7,~8,~13\TeV$ & 2.98 (2.48) & 1.93 (0.81) \\
    \hline
  \end{tabular}
  \label{tab:combo_signif_table}
\end{table*}

\begin{figure*}[htb!]
  \centering
  \includegraphics[width=0.45\textwidth]{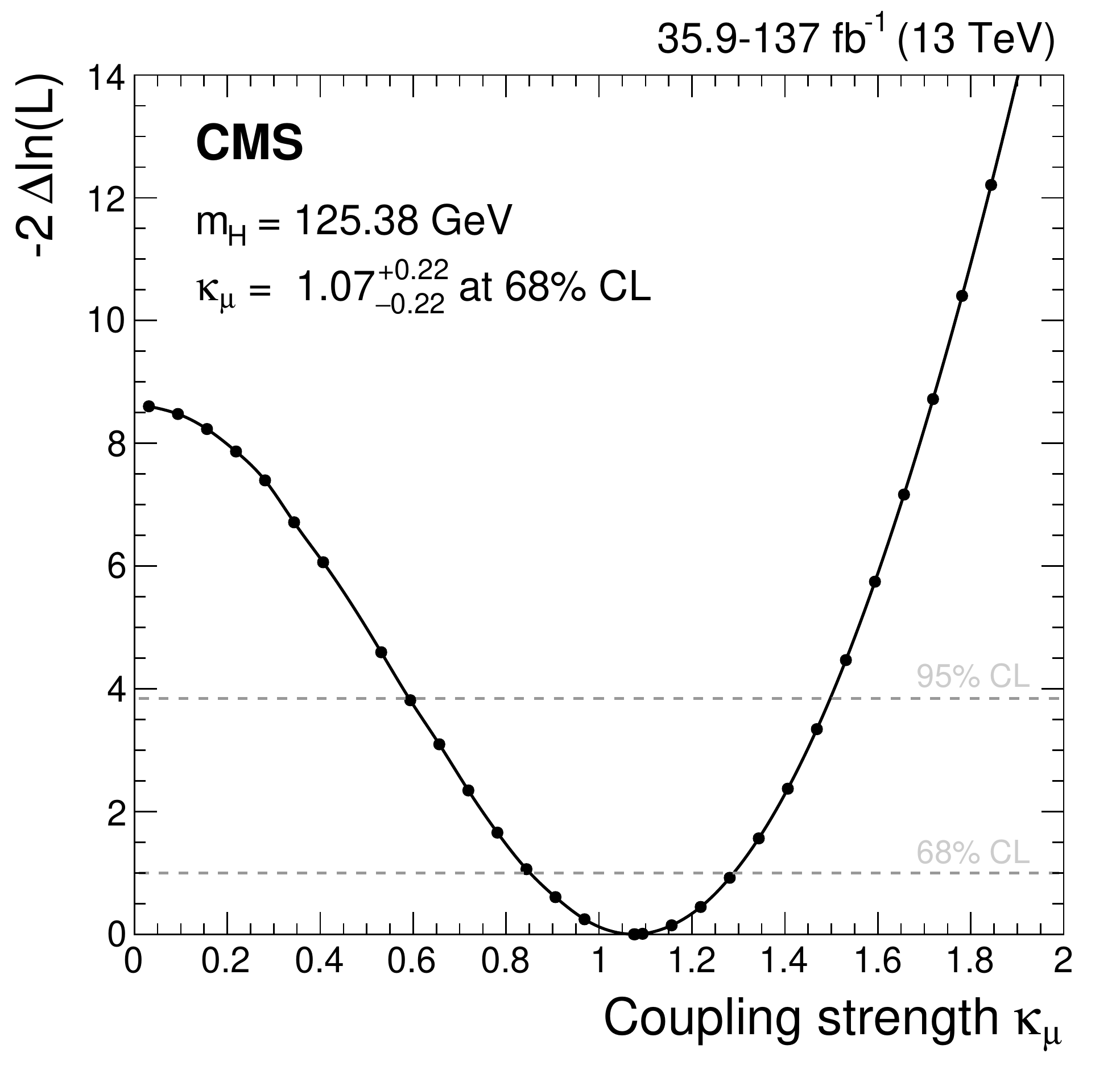}
  \includegraphics[width=0.45\textwidth]{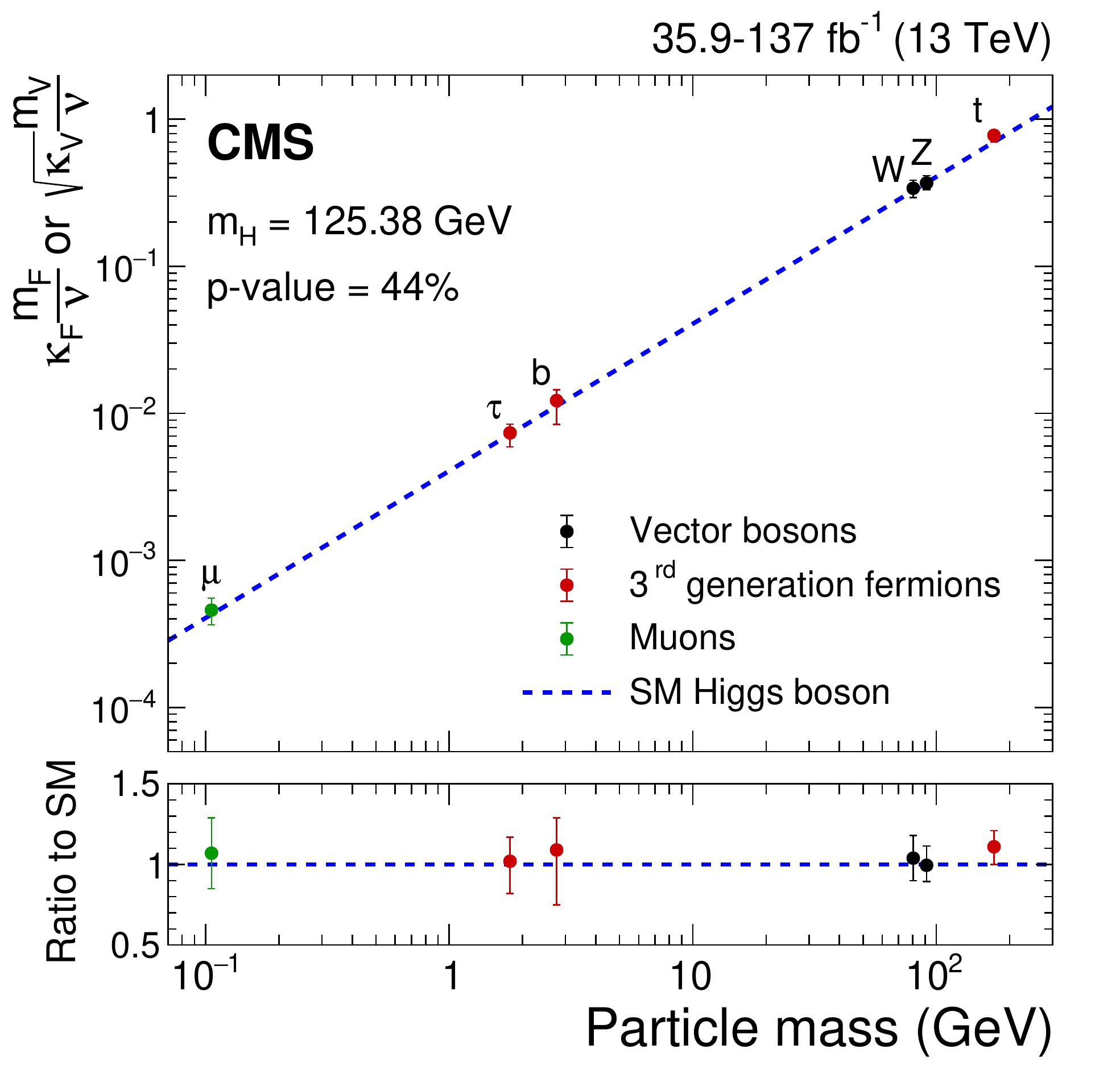}
  \caption{Left: the observed profile likelihood ratio as a function of $\kappa_{\mu}$ for ${\mh = 125.38\GeV}$, obtained from a combined fit with Ref.~\cite{Sirunyan:2018koj} in the $\kappa$-framework. The best fit value for $\kappa_{\mu}$ is 1.07 and the corresponding observed 68\% \CL interval is $0.85 < \kappa_{\mu} < 1.29$. Right: the best fit estimates for the reduced coupling modifiers extracted for fermions and weak bosons from the resolved $\kappa$-framework compared to their corresponding prediction from the SM. The error bars represent 68\% \CL intervals for the measured parameters. In the lower panel, the ratios of the measured coupling modifiers values to their SM predictions are shown.}
  \label{fig:combo_coupling_kappamu}
\end{figure*}

The results presented in this paper are the most precise measurement of the \hmm decay rate reported to date, and provide the best constraint of the coupling between the Higgs boson and the muon. The signal strength measured in the \hmm analysis cannot be translated directly into a measurement of the Higgs boson coupling to muons because it is also sensitive to the interactions between the Higgs boson and several SM particles involved in the production processes considered, primarily the top quark and vector boson couplings. These Higgs boson couplings to other particles are constrained by combining the result of this analysis with those presented in Ref.~\cite{Sirunyan:2018koj}, based on $\Pp\Pp$ collision data recorded by the {CMS} experiment at ${\sqrt{s}=13\TeV}$ in 2016 corresponding to an integrated luminosity of 35.9\fbinv. Under the assumption that there are no new particles contributing to the Higgs boson total width, Higgs boson production and decay rates in each category are expressed in terms of coupling modifiers within the $\kappa$-framework~\cite{Heinemeyer:2013tqa}. Six free coupling parameters are introduced in the likelihood function ($\kappa_{\PW}$, $\kappa_{\PZ}$, $\kappa_{\cPqt}$, $\kappa_{\tau}$, $\kappa_{\cPqb}$, and $\kappa_{\mu}$) and are extracted from a simultaneous fit across all event categories. In the combined fit, the event categories of the ${\sqrt{s} = 13\TeV}$ \hmm analysis described in this paper supersede those considered in Ref.~\cite{Sirunyan:2018koj}. Figure~\ref{fig:combo_coupling_kappamu} (\cmsLeft) shows the observed profile likelihood ratio as a function of $\kappa_{\mu}$ for ${\mh=125.38\GeV}$. The best fit value for $\kappa_{\mu}$ ($\kappa_{\mu}=1.07$), as well as those for the other couplings, are compatible with the SM prediction. The corresponding 68 and 95\% \CL intervals for the $\kappa_{\mu}$ parameter are ${0.85 < \kappa_{\mu} < 1.29}$ and ${0.59 < \kappa_{\mu} < 1.50}$, respectively. Note that the observed (expected) significances reported in Table~\ref{tab:combo_signif_table} and Fig.~\ref{fig:pvalue_vs_mass} are computed assuming SM production cross sections and decay rates, constrained within the corresponding theoretical uncertainties. In the result presented in Fig.~\ref{fig:combo_coupling_kappamu} (left), the freely floating coupling modifiers are allowed to simultaneously modify both Higgs boson production cross sections and decay rates within the constraint of keeping the total Higgs boson width fixed to the SM value.

In the SM, the Yukawa coupling between the Higgs boson and the fermions ($\lambda_{\mathrm{F}}$) is proportional to the fermion mass ($m_{\mathrm{F}}$), while the coupling to weak bosons ($g_{\PV}$) is proportional to the square of the vector boson masses ($m_{\PV}$). The results from the $\kappa$-framework fit can therefore be translated in terms of reduced coupling strength modifiers, defined as ${y_{\PV} = \sqrt{\smash[b]{\kappa_{\PV}}} \, m_{\PV}/\nu}$ for weak bosons and ${y_{\mathrm{F}} = \kappa_{\mathrm{F}}\,m_{\mathrm{F}}/\nu}$ for fermions, where $\nu$ is the vacuum expectation value of the Higgs field of 246.22\GeV~\cite{pdg2020}. Figure~\ref{fig:combo_coupling_kappamu} (\cmsRight) shows the best fit estimates for the six reduced coupling strength modifiers as a function of particle mass, where lepton, vector boson, and quark masses are taken from Ref.~\cite{pdg2020}. The compatibility between the measured coupling strength modifiers and their SM expectation is derived from the ${-2 \, \Delta\ln(\mathcal{L})}$ separation between the best fit and an alternative one, performed by fixing the six coupling modifiers to the SM prediction ($\kappa_{\PW}=\kappa_{\PZ}=\kappa_{\cPqt}=\kappa_{\tau}=\kappa_{\cPqb}=\kappa_{\mu}=1$), yielding a $p$-value of 44\%.

\section{Summary}\label{sec:summary}

Evidence for Higgs boson decay to a pair of muons is presented. This result combines searches in four exclusive categories targeting the production of the Higgs boson via gluon fusion, via vector boson fusion, in association with a vector boson, and in association with a top quark-antiquark pair. The analysis is performed using proton-proton collision data at ${\sqrt{s}=13\TeV}$, corresponding to an integrated luminosity of ${137\fbinv}$, recorded by the {CMS} experiment at the {CERN} {LHC}. An excess of events over the background expectation is observed in data with a significance of 3.0 standard deviations, where the expectation for the standard model (SM) Higgs boson with mass of 125.38\GeV is 2.5. The combination of this result with that from data recorded at $\sqrt{s} = 7$ and 8\TeV, corresponding to integrated luminosities of 5.1 and 19.7\fbinv, respectively, increases both the expected and observed significances by 1\%. The measured signal strength, relative to the SM prediction, is ${1.19^{+0.40}_{-0.39}\stat^{+0.15}_{-0.14}\syst}$. This result constitutes the first evidence for the decay of the Higgs boson to second generation fermions and is the most precise measurement of the Higgs boson coupling to muons reported to date.

\begin{acknowledgments}

  We congratulate our colleagues in the CERN accelerator departments for the excellent performance of the LHC and thank the technical and administrative staffs at CERN and at other CMS institutes for their contributions to the success of the CMS effort. In addition, we gratefully acknowledge the computing centres and personnel of the Worldwide LHC Computing Grid for delivering so effectively the computing infrastructure essential to our analyses. Finally, we acknowledge the enduring support for the construction and operation of the LHC and the CMS detector provided by the following funding agencies: BMBWF and FWF (Austria); FNRS and FWO (Belgium); CNPq, CAPES, FAPERJ, FAPERGS, and FAPESP (Brazil); MES (Bulgaria); CERN; CAS, MoST, and NSFC (China); COLCIENCIAS (Colombia); MSES and CSF (Croatia); RIF (Cyprus); SENESCYT (Ecuador); MoER, ERC IUT, PUT and ERDF (Estonia); Academy of Finland, MEC, and HIP (Finland); CEA and CNRS/IN2P3 (France); BMBF, DFG, and HGF (Germany); GSRT (Greece); NKFIA (Hungary); DAE and DST (India); IPM (Iran); SFI (Ireland); INFN (Italy); MSIP and NRF (Republic of Korea); MES (Latvia); LAS (Lithuania); MOE and UM (Malaysia); BUAP, CINVESTAV, CONACYT, LNS, SEP, and UASLP-FAI (Mexico); MOS (Montenegro); MBIE (New Zealand); PAEC (Pakistan); MSHE and NSC (Poland); FCT (Portugal); JINR (Dubna); MON, RosAtom, RAS, RFBR, and NRC KI (Russia); MESTD (Serbia); SEIDI, CPAN, PCTI, and FEDER (Spain); MOSTR (Sri Lanka); Swiss Funding Agencies (Switzerland); MST (Taipei); ThEPCenter, IPST, STAR, and NSTDA (Thailand); TUBITAK and TAEK (Turkey); NASU (Ukraine); STFC (United Kingdom); DOE and NSF (USA).
  
  \hyphenation{Rachada-pisek} Individuals have received support from the Marie-Curie programme and the European Research Council and Horizon 2020 Grant, contract Nos.\ 675440, 752730, and 765710 (European Union); the Leventis Foundation; the A.P.\ Sloan Foundation; the Alexander von Humboldt Foundation; the Belgian Federal Science Policy Office; the Fonds pour la Formation \`a la Recherche dans l'Industrie et dans l'Agriculture (FRIA-Belgium); the Agentschap voor Innovatie door Wetenschap en Technologie (IWT-Belgium); the F.R.S.-FNRS and FWO (Belgium) under the ``Excellence of Science -- EOS" -- be.h project n.\ 30820817; the Beijing Municipal Science \& Technology Commission, No. Z191100007219010; the Ministry of Education, Youth and Sports (MEYS) of the Czech Republic; the Deutsche Forschungsgemeinschaft (DFG) under Germany's Excellence Strategy -- EXC 2121 ``Quantum Universe" -- 390833306; the Lend\"ulet (``Momentum") Programme and the J\'anos Bolyai Research Scholarship of the Hungarian Academy of Sciences, the New National Excellence Program \'UNKP, the NKFIA research grants 123842, 123959, 124845, 124850, 125105, 128713, 128786, and 129058 (Hungary); the Council of Science and Industrial Research, India; the HOMING PLUS programme of the Foundation for Polish Science, cofinanced from European Union, Regional Development Fund, the Mobility Plus programme of the Ministry of Science and Higher Education, the National Science Center (Poland), contracts Harmonia 2014/14/M/ST2/00428, Opus 2014/13/B/ST2/02543, 2014/15/B/ST2/03998, and 2015/19/B/ST2/02861, Sonata-bis 2012/07/E/ST2/01406; the National Priorities Research Program by Qatar National Research Fund; the Ministry of Science and Higher Education, project no. 02.a03.21.0005 (Russia); the Programa Estatal de Fomento de la Investigaci{\'o}n Cient{\'i}fica y T{\'e}cnica de Excelencia Mar\'{\i}a de Maeztu, grant MDM-2015-0509 and the Programa Severo Ochoa del Principado de Asturias; the Thalis and Aristeia programmes cofinanced by EU-ESF and the Greek NSRF; the Rachadapisek Sompot Fund for Postdoctoral Fellowship, Chulalongkorn University and the Chulalongkorn Academic into Its 2nd Century Project Advancement Project (Thailand); the Kavli Foundation; the Nvidia Corporation; the SuperMicro Corporation; the Welch Foundation, contract C-1845; and the Weston Havens Foundation (USA).
\end{acknowledgments}

\bibliography{auto_generated}
\cleardoublepage \appendix\section{The CMS Collaboration \label{app:collab}}\begin{sloppypar}\hyphenpenalty=5000\widowpenalty=500\clubpenalty=5000\vskip\cmsinstskip
\textbf{Yerevan Physics Institute, Yerevan, Armenia}\\*[0pt]
A.M.~Sirunyan$^{\textrm{\dag}}$, A.~Tumasyan
\vskip\cmsinstskip
\textbf{Institut f\"{u}r Hochenergiephysik, Wien, Austria}\\*[0pt]
W.~Adam, T.~Bergauer, M.~Dragicevic, J.~Er\"{o}, A.~Escalante~Del~Valle, R.~Fr\"{u}hwirth\cmsAuthorMark{1}, M.~Jeitler\cmsAuthorMark{1}, N.~Krammer, L.~Lechner, D.~Liko, I.~Mikulec, F.M.~Pitters, N.~Rad, J.~Schieck\cmsAuthorMark{1}, R.~Sch\"{o}fbeck, M.~Spanring, S.~Templ, W.~Waltenberger, C.-E.~Wulz\cmsAuthorMark{1}, M.~Zarucki
% ...
\vskip\cmsinstskip
\dag: Deceased\\
1:  Also at Vienna University of Technology, Vienna, Austria\\
% ...
\end{sloppypar}
\end{document}